\documentclass[12pt]{article}
\usepackage{amsmath}
\usepackage{graphicx}
\usepackage{enumerate}
\usepackage{natbib} 
\usepackage{url} 

\usepackage{mathtools} 
\usepackage{marvosym} 
\usepackage{amssymb, amsmath, amsfonts, amsthm, bm}

\usepackage{hyperref}

\newcommand{\mbbR}{\mathbb{R}}

\newcommand{\mbfY}{\mathbf{Y}}
\newcommand{\bfmu}{\bm{\mu}}
\newcommand{\bfbeta}{\bm{\beta}}
\newcommand{\mcalN}{\mathcal{N}}
\newcommand{\matern}{{Mat\'{e}rn}}
\newcommand{\bmy}{\bm{y}}

\newcommand{\bmY}{\bm{Y}}
\newcommand{\HPi}{\scalebox{1.5}{$\Pi$}}
\newcommand{\HSigma}{\scalebox{1.5}{$\Sigma$}}
\newcommand{\mcalG}{\mathcal{G}}

\newcommand{\bmX}{\bm{X}}
\newcommand{\bmmu}{\bm{\mu}}
\newcommand{\bmB}{\bm{B}}
\newcommand{\bmS}{\bm{S}}
\newcommand{\bmZ}{\bm{Z}}
\newcommand{\mbbE}{\mathbb{E}}
\newcommand{\bmSIGMA}{\bm{\Sigma}}
\newcommand{\bmD}{\bm{D}}

\newcommand{\bmR}{\bm{R}}
\newcommand{\bmC}{\bm{C}}
\newcommand{\bmBT}{\bm{BT}}
\newcommand{\bmU}{\bm{U}}
\newcommand{\bmE}{\bm{E}}

\newcommand{\lglftbr}{\scalebox{1.5}[1.5]{$[$}}
\newcommand{\lgrgtbr}{\scalebox{1.5}[1.5]{$]$}}
\newcommand{\bmzero}{\boldsymbol{0}}

\newcommand{\mcalO}{\mathcal{O}}

\newcommand{\mcalD}{\mathcal{D}}

\newcommand{\mcalE}{\mathcal{E}}

\newcommand{\mcalY}{\mathcal{Y}}

\newtheorem{definition}{Definition}
\newtheorem{theorem}{Theorem} 

\newtheorem{lemma}{Lemma}
\newtheorem{observation}{Observation}

\newtheorem{proposition}{Proposition}

\usepackage[linesnumbered,ruled,vlined]{algorithm2e}
\usepackage{algpseudocode} 

\usepackage{caption} 

\usepackage{lipsum} 

\graphicspath{{Figures/}} 
\usepackage{subcaption} 

\usepackage[titletoc]{appendix} 

\usepackage{booktabs} 

\usepackage{breqn}

\usepackage[linesnumbered,ruled,vlined]{algorithm2e}
\usepackage{setspace} 

\usepackage{array} 

\usepackage{threeparttable} 

\usepackage{enumitem} 

\usepackage{authblk}
\date{}

\title{Highly Multivariate Large-scale Spatial Stochastic Processes -- A Cross-Markov Random Field Approach}

\author[1]{\small Xiaoqing Chen}
\author[2]{\small Peter Diggle}
\author[3]{\small James V. Zidek}
\author[4]{\small Gavin Shaddick}

\affil[1]{The Alan Turing Institute \\ British Library, 96 Euston Rd., London NW1 2DB, London, UK}
\affil[2]{\small CHICAS, Lancaster Medical School, Lancaster University,
Lancaster, LA1 4YB, U.K.}
\affil[3]{\small Department of Statistics, University of British Columbia, Vancouver, BC V6T 1Z4, Canada}
\affil[4]{\small College of Physical Sciences and Engineering, Cardiff University \\ Cardiff, CF10 3AT, U.K.}
\affil[*]{Corresponding author: Xiaoqing Chen; email: xiaoqing.a.chen@gmail.com}

\begin{document}
\maketitle


\begin{abstract}
Key challenges in the analysis of highly multivariate large-scale spatial stochastic processes, where both the number of components ($p$) and spatial locations ($n$) can be large, include achieving maximal sparsity in the joint precision matrix $\HSigma_{np \times np}^{-1}$, ensuring efficient computation for its generation, accommodating asymmetric cross-covariance in the joint covariance matrix $\HSigma_{np \times np}$, and delivering scientific interpretability. We propose a cross-MRF model class, consisting of a mixed spatial graphical model framework and cross-MRF theory, to collectively address these challenges in one unified framework across two modelling stages. The first stage exploits scientifically informed conditional independence (CI) among $p$ component fields and allows for a step-wise parallel generation of 
$\HSigma_{np \times np}$ and $\HSigma_{np \times np}^{-1}$, enabling a simultaneous accommodation of asymmetric cross-covariance in $\HSigma_{np \times np}$ and sparsity in $\HSigma_{np \times np}^{-1}$. The second stage further extends the first-stage CI to doubly CI among both $p$ and $n$ and unearths the cross-MRF via an extended Hammersley-Clifford theorem for multivariate spatial stochastic processes. 
This results in the sparsest possible representation of $\HSigma_{np \times np}^{-1}$ with the highest percentage of exact-zero entries and ensures the lowest generation complexity of  $\HSigma_{np \times np}^{-1}$. We demonstrate with 1D simulated comparative studies and 2D real-world data.
\end{abstract}

\noindent%
\vfill

\def\spacingset#1{\renewcommand{\baselinestretch}%
{#1}\small\normalsize} \spacingset{1}
\spacingset{1.65} 

\section{Introduction}
The interaction of multiple quantities of interest across large spatial domains spans diverse disciplines, including climate (e.g., temperature, precipitation, wind speed), pandemics (e.g., protein mutation rate, UV radiation intensity, regional vaccination coverage), 
air quality (e.g., PM2.5, NO$_2$, O$_3$), and social economy (e.g., crime rates, housing prices, income levels), among others.

Modelling such data jointly presents challenges in model characterisation, computation and scientific interpretability.
We propose a cross-Markov Random Field (cross-MRF) model class, consisting of a
mixed spatial graphical model framework and cross-MRF theory, to address the above challenges collectively in one unified framework.


One example dataset in environmental science is the ECMWF CAMS \citep{inness2019cams}. It provides re-analysed concentrations for various pollutants at a horizontal resolution of $0.75^{\circ} \times 0.75^{\circ} $ grid, covering 27384 grid cells for all land regions worldwide. In particular, it includes concentrations for different components of PM2.5, i.e., Black Carbon (BC), Dust (DU), Sulfate (SU), Organic Matter (OM), and Sea Salt (SS), whose accurate quantification is crucial for advancing environmental health research and guiding policy formulation.

\begin{figure}[h]
  \centering
  \begin{subfigure}[b]{0.32\textwidth}
    \caption{Auto-correlation of DU}
    \includegraphics[width=0.75\textwidth]{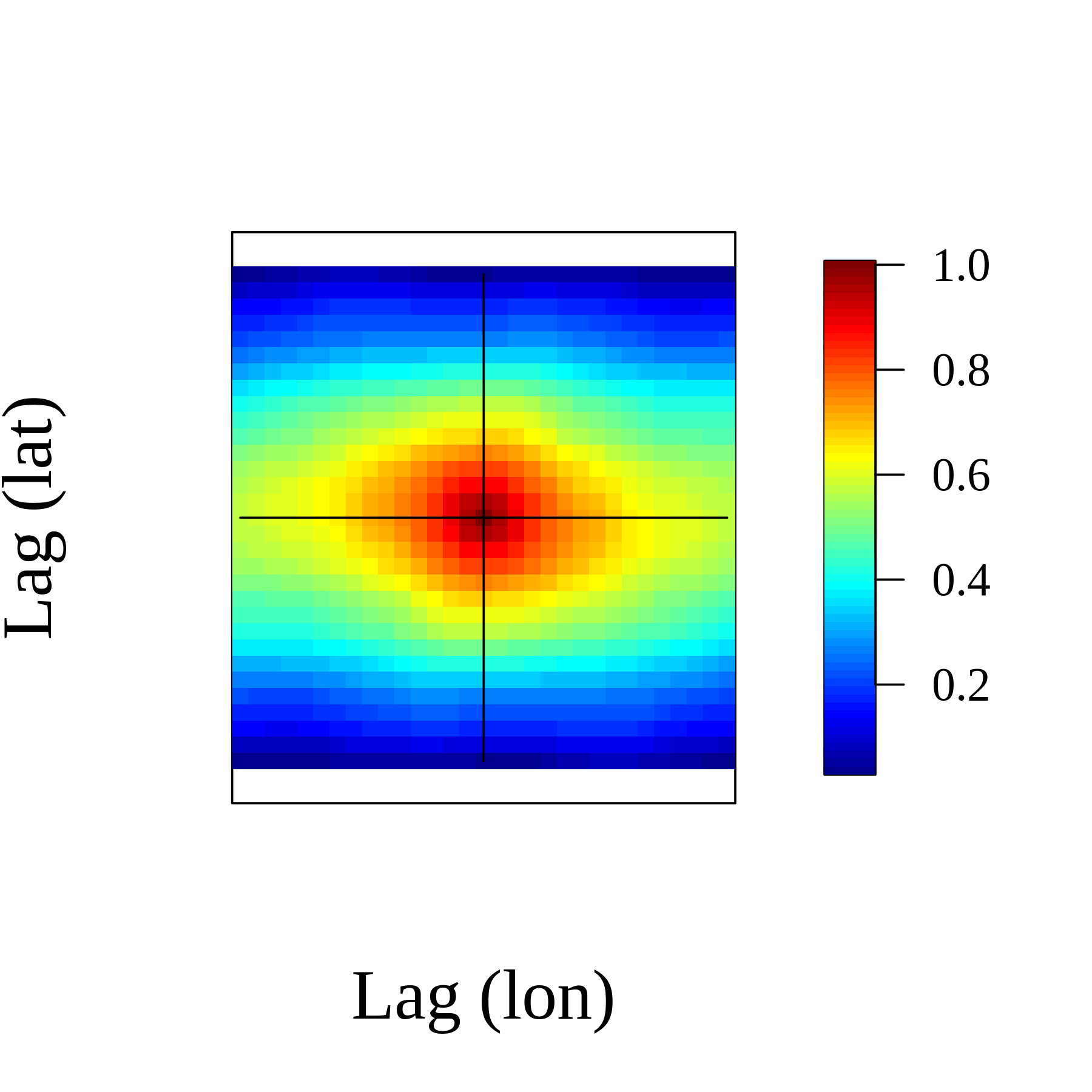}
    \label{fig:auto_du}
  \end{subfigure}
  \begin{subfigure}[b]{0.32\textwidth}
    \caption{Auto-correlation of SU}
    \includegraphics[width=0.75\textwidth]{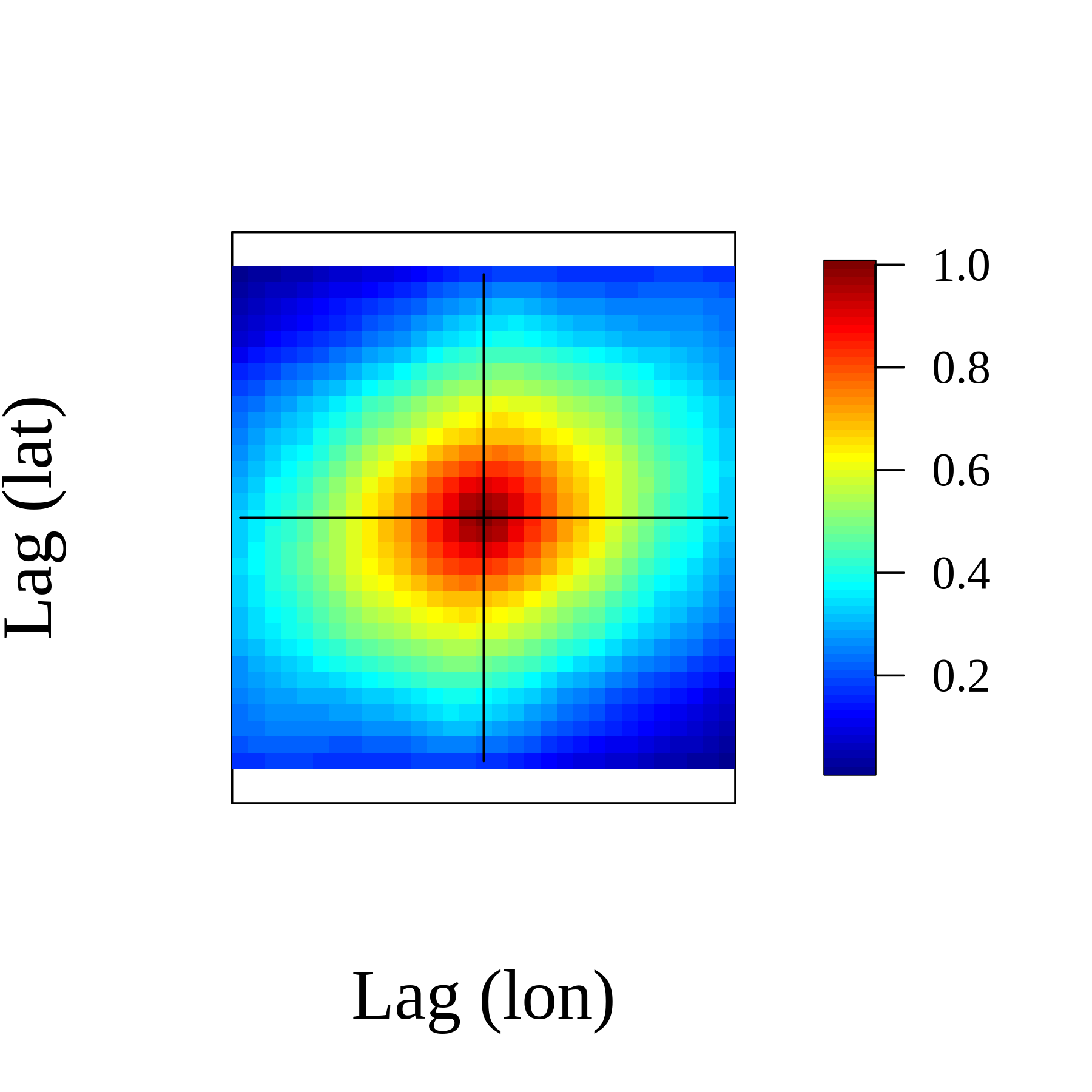}
    \label{fig:auto_su}
  \end{subfigure}
  \begin{subfigure}[b]{0.32\textwidth}
    \caption{Cross-correlation of DU, SU}
    \includegraphics[width=0.75\textwidth]{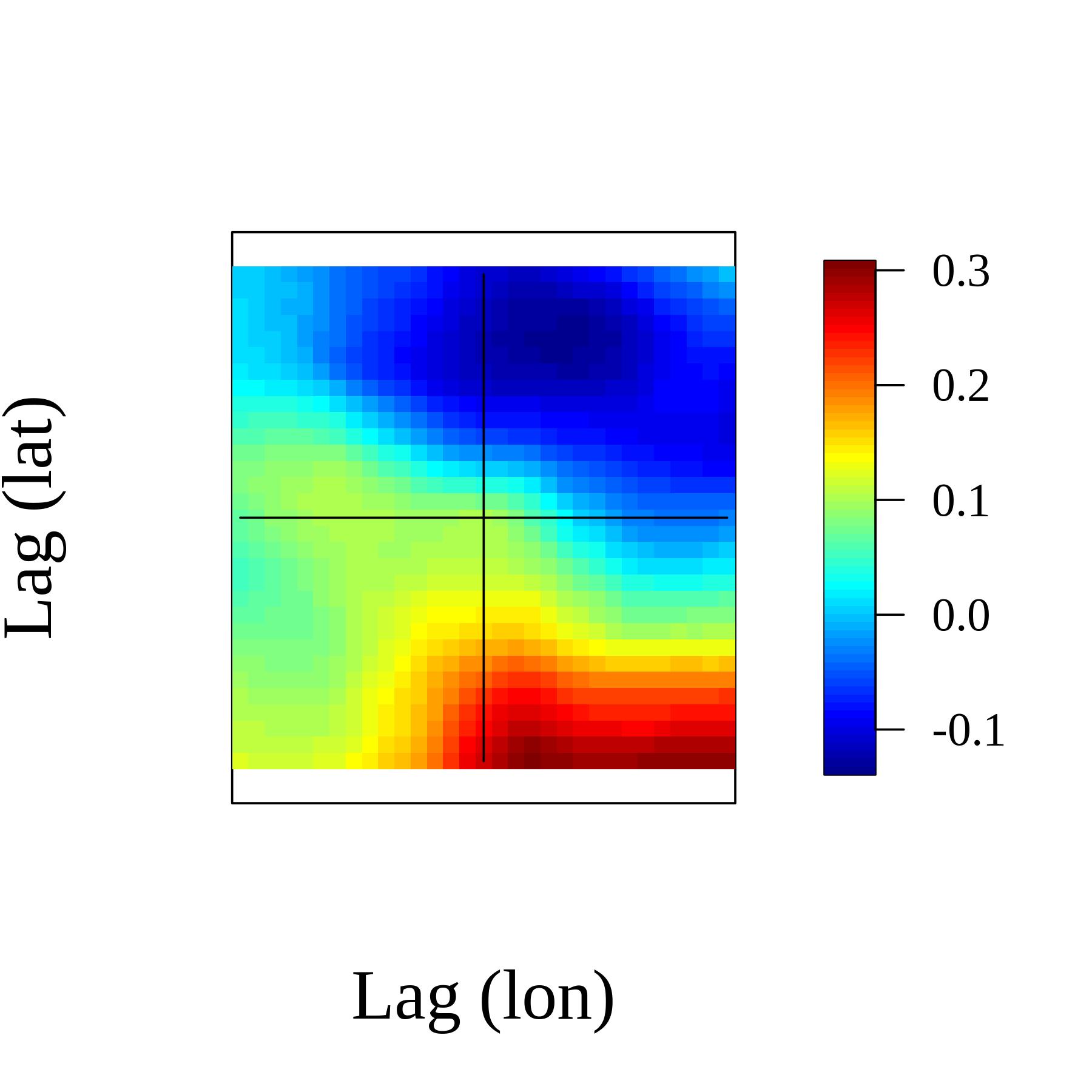}
    \label{fig:cross_dusu}
  \end{subfigure}
  \caption{(a) and (b) show the auto-correlation of DU and SU residuals after removing a quadratic trend surface; both are symmetric. (c) shows the cross-correlation between DU and SU residuals, which is asymmetric. Each plot spans $-15$ to $+15$ degrees in both coordinate directions.}
  \label{fig:cross_corr}
\end{figure}

Each pollutant not only interacts with itself within its nearby region in the spatial domain $\mcalD$
(reflected by \textit{auto-correlation}), but also interacts with other pollutants across certain ranges in $\mcalD$ (captured by \textit{cross-correlation}). Auto-correlation is symmetric, e.g., $corr(DU(s_1), DU(s_2)) = corr(DU(s_2), DU(s_1))$, see Figure \ref{fig:auto_du} and \ref{fig:auto_su}.
In contrast, the cross-correlation is \textit{asymmetric}, e.g., $corr(DU(s_1), SU(s_2)) \neq corr(DU(s_2), SU(s_1))$, see Figure \ref{fig:cross_dusu}. For more properties of the asymmetric cross-correlation, see \citet{Chen2025ICSTA150}.


A series of scientific studies (Supplementary Material \ref{app:sec_evidence}) indicates that ordering exists among different pollutants when they interact, especially among the five components of PM2.5. For example, DU enhances the mass concentration of SU by more than an order of magnitude \citep{acp-10-365-2010}, whilst DU frequently mixes into SS during their transport in the marine boundary layer \citep{ZHANGSSDU}. 
Moreover, each pollutant interacts only with a subset of the other pollutants, implying conditional independence (CI) among pollutants over the spatial domain $\mcalD$.

Formally, we consider the data to arise from a real-valued, discretely indexed
spatial stochastic process
$\{ (Y_1(s_i), Y_2(s_i), \ldots, Y_p(s_i)): i = 1, 2, \ldots, n \}$, 
where both the number of components (e.g. pollutants) $p$ and the number of spatial locations $n$ can be large, whilst multiple components interact only across a
certain spatial range.
We call a process of this kind a \textit{highly multivariate large-scale (HMLS) spatial stochastic process}. 
In what follows, we consider the set of spatial locations to be fixed, forming a regular grid.

Let $\bmY = [\bmY^{T}_1(\cdot), \ldots, \bmY^{T}_l(\cdot), \ldots, \bmY^{T}_p(\cdot)]^{T} = [\bmY^{T}_1, \ldots, \bmY^{T}_i, \ldots, \bmY^{T}_n]^{T}$ represent a vector of $np$ random variables whose joint probability distribution $pr(\bmY)$ is supported on a product space $\Omega = \Omega_1 \times \ldots \times \Omega_{np}$, where each $\Omega_{*}$ is the support for one random variable $Y_l(s_i)$, $l = 1, \ldots, p$, and $i = 1, \ldots, n$. 
The ``$\cdot$'' denotes all $n$ locations in $\mcalD$. Here, $\bmY_l(\cdot)= [Y_l(s_1), \ldots, Y_l(s_n)]^T  \in \mbbR^n $ 
spanning across all $n$ locations for a particular component $l$, 
while $\bmY_i  = [Y_1(s_i), \ldots, Y_p(s_i)]^T \in \mbbR^p$ 
collecting all $p$ components at a particular location $s_i$.

Under the Gaussian process assumption, the goals are to (1) construct a valid joint covariance $\HSigma_{np \times np}$ and precision matrix $\HSigma_{np \times np}^{-1}$, (2) ensure maximal sparsity in the joint precision matrix $\HSigma_{np \times np}^{-1}$, (3) derive it with efficient computation, 
and (4) accommodate asymmetric cross-covariance in the joint covariance $\HSigma_{np \times np}$ for accurate data feature characterisation.


Much of the existing literature either focuses on the univariate large-scale spatial problem (large $n$ but $p = 1$), 
see \citet{heaton2019case} and the reference therein,  
or on multivariate (mostly bivariate, $p = 2$) spatial problems within either small spatial domains (small $n$), see \citet{ver1998constructing}, \citet{gneiting2010matern}, \citet[p.~155]{wackernagel2013multivariate}, or large spatial domain (large $n$), see \citet{kleiber2019model} and \citet{guinness2022nonparametric}. 
\citet{dey2022graphical} provided a framework for highly multivariate ($p \gg 2$) settings using undirected graphical structure, while \citet{taylor2019spatial} and \citet{krock2023modeling} addressed the highly multivariate and large-scale spatial setting (both $p$ and $n$ are large) via direct construction of joint precision matrix $\HSigma_{np \times np}^{-1}$ or orthogonal basis.
However, orthogonal basis methods restrict the joint covariance matrix $\HSigma_{np \times np}$ to be symmetric, while direct construction of joint precision $\HSigma_{np \times np}^{-1}$ has no guarantee that the joint covariance $\HSigma_{np \times np}$ inverted from $\HSigma_{np \times np}^{-1}$ has asymmetric cross-covariance in it off-diagonal blocks or that asymmetry is objectively learned from data.
\citet{li2011approach} does accommodate the asymmetry in $\HSigma_{np \times np}$, but
lacks sparsity in the joint precision matrix $\HSigma_{np \times np}^{-1}$, computational scalability is hence limited.

From a conditional methods perspective, \citet{mardia1988multi} proposed a Gaussian model for constructing the joint precision matrix $\HSigma_{np \times np}^{-1}$. 
Let $\mathbf{Y}_i$ represent the vector of $p$ components at location $s_i$, where $\mbfY_i \in \mbbR^p$.
Model each $\mbfY_i$ conditionally as
    $E[\mbfY_i \mid \mbfY_{-i}] = \bfmu_i + \sum_{j \in \mcalN(i)} \bfbeta_{ij} (\mbfY_j - \bfmu_j)$ and $
    Var[\mbfY_i \mid \mbfY_{-i}] = \mathbf{\Gamma}_i$.
Then, the multivariate Gaussian joint distribution for $\mbfY = [\mbfY_1^{T}, \cdots, \mbfY_n^{T}]^{T} \in \mbbR^{np}$ has precision matrix 
    $\HSigma^{-1} = \{ \mbox{block diag}(\mathbf{\Gamma}_i)^{-1} \} \{ \mbox{block} (- \bfbeta_{ij}) \}$, 
where 
$\HSigma^{-1} \in \mbbR^{np \times np}$, 
provided the symmetric 
and the positive definite condition 
are satisfied.

\begin{figure}[ht]
    \centering
    \begin{subfigure}[b]{0.4\textwidth}
        \centering
        \caption{Row-wise conditional}
        \includegraphics[width=0.68\textwidth]{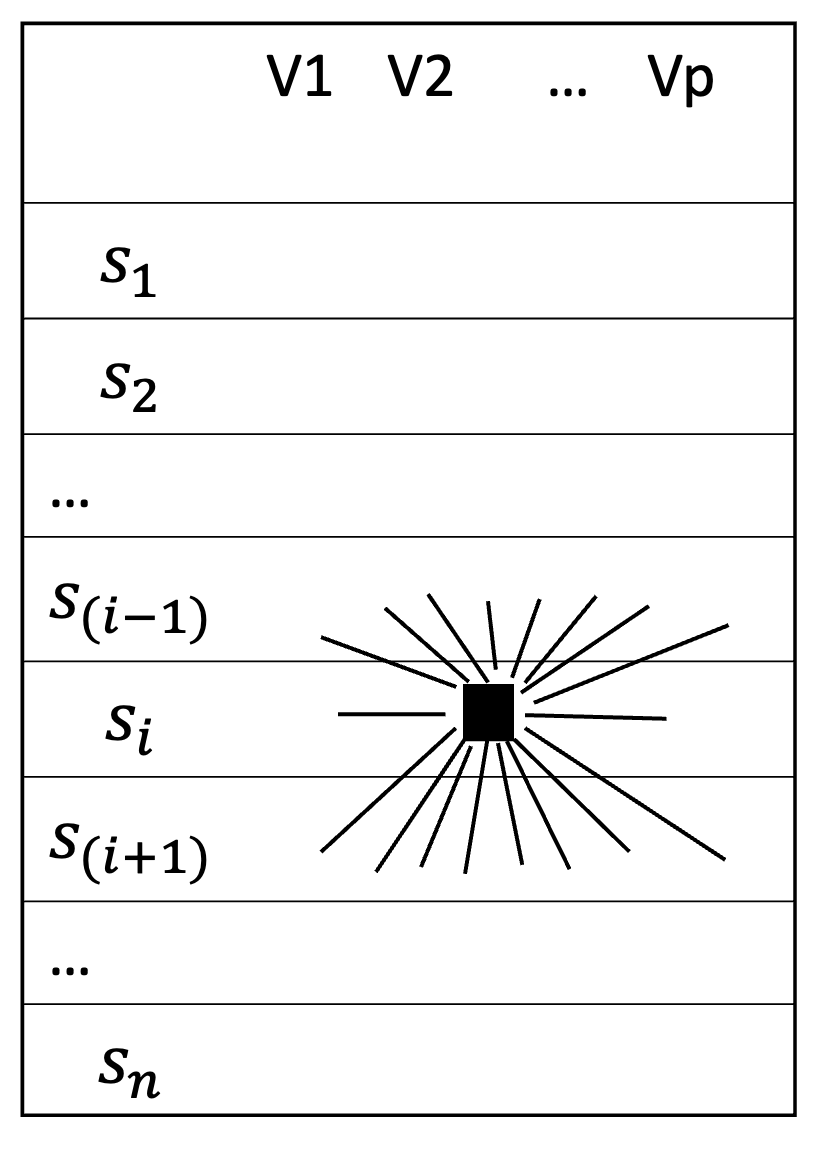}
        \label{fig:row_cond}
    \end{subfigure}
    \hfill
    \begin{subfigure}[b]{0.4\textwidth}
        \centering
        \caption{Column-wise conditional}
        \includegraphics[width=0.68\textwidth]{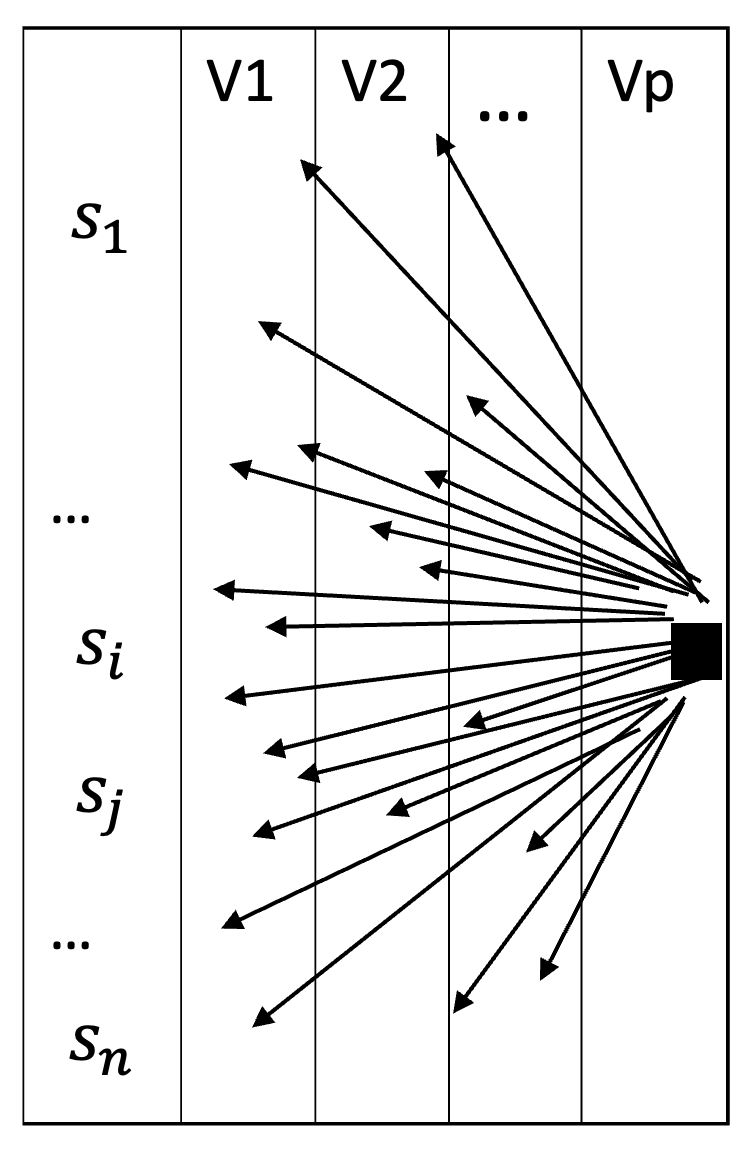}
        \label{fig:col_cond}
    \end{subfigure}
    \caption{Schematic representations of the row-wise conditional and column-wise conditional method. Row-wise conditional only regresses on values at neighbourhood locations but across all components, while column-wise conditional regresses on all the previous components across all spatial locations.}
    \label{fig:row-col}
\end{figure}

If organising the $np$ random variables 
into a $p$-column by $n$-row table, $\mbfY_i \in \mbbR^p$ represents an entire row of the table. Accordingly, we refer to this conditional approach as \textit{row-wise conditional}. See Figure \ref{fig:row_cond} for a schematic representation.

The advantage of this conditional method is that 
one directly obtains the desired joint precision matrix $\HSigma^{-1}_{np \times np}$, providing convenience for further inference of the Gaussian likelihood
without the need to invert $\HSigma_{np \times np}$.
Additionally, since the conditional mean depends only on values at neighbourhood locations, the obtained $\HSigma^{-1}_{np \times np}$ naturally embodies structural sparsity due to the Markovian property of the conditional independence (CI) among $n$ spatial locations (row-wise direction). 

The disadvantage is that the $p$ components have to be treated as a whole, with no exploitation of CI among them. Moreover, the accommodation of the asymmetric cross-covariance in $\HSigma_{np \times np}$ is absent.

\citet{cressie2016multivariate} proposed an alternative conditional modelling scheme. Instead of obtaining the joint precision matrix $\HSigma^{-1}_{np \times np}$, 
it constructs the joint covariance matrix $\HSigma_{np \times np}$ by modelling the conditional mean and covariance of the $q^{th}$ component process given all the preceding $(q - 1)$ components. Specifically, 
\begin{align}
    \label{eq:col-wise_multi1}
    E(Y_q(s) \mid \{\bmY_r(\cdot): r = 1, 2, \cdots, (q-1)\}) &= \sum_{r = 1} ^{(q-1)} \int_D b_{qr} (s, v) Y_r(v) dv; \quad s \in \mcalD, \\[3pt]
    \label{eq:col-wise_multi2}
    cov(Y_q(s), Y_q(u) \mid \{\bmY_r(\cdot): r = 1, 2, \cdots, (q-1) \}) &= C_{q \mid (r < q)}(s, u); \quad s, u \in \mcalD ,    
\end{align} 
The idea of obtaining the joint covariance from conditional mean and conditional covariance dates back to \citet{yule1907theory}; see also 
\citet[pp. 370-372]{bishop2007pattern}.

Since $\bmY_r(\cdot) \in \mbbR^n$ is an entire column of the table, we refer to this conditional approach as \textit{column-wise conditional}, see Figure \ref{fig:col_cond}.

The advantage of this method is that it can accommodate asymmetric cross-covariance in the joint $\HSigma_{np \times np}$ directly via the real-valued integrable $b$ functions in equation \eqref{eq:col-wise_multi1}.

However, summing over all the preceding $(q-1)$ processes when modelling the $q^{th}$ component induces several limitations. First, it creates arbitrariness in the absence of a natural ordering among the $p$ components. Second, it fails to exploit CI among the $p$ components, lacking scalability in highly multivariate settings where $q$ is large. Additionally, it does not account for CI among the $n$ spatial locations, as was done with the ``row-wise conditional'' method of \citet{mardia1988multi}.
Finally, its reliance on Cholesky inversion incurs cubic computational costs ($\mcalO(n^3p^3)$), and the resulting precision matrix has no sparse structure at all, making computations prohibitively expensive in practice.

The {\it cross-MRF} model class proposed in this paper retains the advantages and overcomes the disadvantages of both \citet{mardia1988multi} and \citet{cressie2016multivariate}. It (1) achieves the maximal sparsity in $\HSigma^{-1}_{np \times np}$ via \textit{doubly} CI over both $p$ component fields and $n$ locations, (2) ensures $\HSigma^{-1}_{np \times np}$ is obtained with the lowest generation complexity, (3) circumvents the dilemma of choosing between constructing the joint covariance matrix $\HSigma_{np \times np}$ (capturing asymmetric cross-covariance) and the joint precision matrix $\HSigma^{-1}_{np \times np}$ (embodying structural sparsity), and (4) allows for scientific interpretability.



Concretely, the first stage proposes hybrid probabilistic spatial graphs, a mixed spatial graphical model framework and the corresponding algorithm that exploits the scientifically informed CI among $p$ component fields, 
scaling to any $p$ and any set of customised component relationships.
In particular, the step-wise parallel generation of $\HSigma_{np \times np}$ and $\HSigma^{-1}_{np \times np}$ enables a simultaneous accommodation of asymmetric cross-covariance in $\HSigma_{np \times np}$ and sparsity in $\HSigma^{-1}_{np \times np}$, while reducing the generation complexity of $\HSigma^{-1}_{np \times np}$ to be linear in $p$.

The second stage extends the first-stage CI to \textit{doubly} CI among both $p$ components and $n$ spatial locations via the development of various classes of neighbourhoods (auto-/cross-neighbourhood) 
and the cross-MRF theory.
This results in $\HSigma^{-1}_{np \times np}$ achieving the maximal sparsity with the highest percentage of exact-zero entries. Additionally, it further reduces the generation complexity of $\HSigma^{-1}_{np \times np}$ to $\mcalO(pn^2)$, yielding even faster $\HSigma^{-1}_{np \times np}$ generation time. 

The remainder of the paper is organised as follows. 
Section \ref{sec:def_mix_G} defines the \textit{mixed spatial graph}. 
Section \ref{sec:Mix_model} presents the first-stage framework exploiting the CI among $p$ component fields, along with the
theorem that underpins the step-wise parallel generation of $\HSigma_{np \times np}$ and $\HSigma_{np \times np}^{-1}$. 
The corresponding algorithm is in Section \ref{sec:algo}. 
Section \ref{sec:simulation} presents the 1D simulation with ten component fields 
($p$ = 10) demonstrating the simultaneous accommodation of asymmetric cross-covariance
in $\HSigma_{np \times np}$ and initial-stage sparsity in $\HSigma_{np \times np}^{-1}$.
Section \ref{sec:cross_MRF} explores the existence of a doubly CI structure among both the $p$ components and the $n$ spatial locations. This includes the development of auto-/cross-neighbourhoods, the cross-MRF and its conditions. 
Section \ref{sec:embed_CrossMRF} provides linking strategies to unify the cross-MRF theory with the proposed mixed spatial graphical framework, realising the desired doubly CI structure.
Section \ref{sec:1D_compare_study} conducts 1D simulated studies to compare various aspects of different CI scenarios and conditional modelling strategies, showing that the doubly CI strategy achieves the highest percentage of exact-zero entries in $\HSigma_{np \times np}^{-1}$ and the lowest generation complexity of $\HSigma_{np \times np}^{-1}$. 
Section \ref{sec:2D_illustration} illustrates the derived method using 2D real-world data. 
Section \ref{sec:conclude_discussion} presents conclusions and
discusses limitations and future work.

\section{Definitions of Mixed Probabilistic Spatial Graphs}
\label{sec:def_mix_G}

Before proposing the mixed spatial graphical model framework, we first define the \textit{mixed probabilistic spatial graph},  
including the corresponding nodes, edges, and graphs. Standard graph theory concepts (such as non-descendants) follow \citet[p.~34-37, 57, 118, 150]{koller2009probabilistic}.

\begin{definition}[Nodes of a probabilistic spatial graph]
A node in a probabilistic spatial graph $\mcalG$ is a random quantity associated with a component $q$ at a spatial location $s_i$, denoted as $Y_q(s_i)$, $q = 1, \ldots, p$, $i = 1, \ldots, n$. 
The collection $\{ Y_q(s_i): q = 1,  \ldots, p; i = 1, \dots, n \}$ is a node set $\mcalY$ consisting of the probabilistic spatial graph $\mcalG$. 
\end{definition}

\begin{definition}[component field]
Given a component $q$, the collection $\{Y_q(s_i): i = 1, \ldots , n  \}$ of random nodes across $n$ spatial locations is a component field, denoted as $\bm{Y}_q(\cdot)$, $ q = 1, \ldots, p$.    
\end{definition}

\begin{definition}[Probabilistic spatial graph -- undirected]
An undirected probabilistic spatial graph $\mcalG^{UD}= (\mcalY, \mcalE^{UD})$ consists of a node set $\mcalY =\{Y_q(s_i): i = 1, \ldots , n  \}$ and an undirected edge set $\mcalE^{UD}$. 

Within a given component field, edges between neighbouring nodes are undirected (``---''), i.e.,  $Y_k(s_i)$ --- $Y_k(s_j) \in \mcalE^{UD}$.

Local independence (``$\perp$'') of $Y_k(s_i)$ in
$\mcalG^{UD}$ is that given its neighbouring nodes, it is conditionally independent of the rest of the nodes in $\mcalY$, i.e., 
$Y_k(s_i) \perp (\mcalY - \{Y_k(s_i)\} - \mcalN(Y_k(s_i)))  \mid  \mcalN(Y_k(s_i)) $.
\end{definition}

\begin{definition}[Component field spatial graph -- directed]
The directed component field spatial graph $\mcalG^{D}= (\mcalY, \mcalE^{D})$ consists of a node set $\mcalY = \{ Y_q(s_i): q = 1,  \ldots, p; i = 1, \dots, n \}$ and a directed edge set $\mcalE^{D}$.

Across two component fields, e.g., $\bm{Y}_k(\cdot)$ and $\bm{Y}_l(\cdot)$, edges connecting them are directed (``$\rightarrow$''), pointing from parent component $k$ to child component $l$. That is, $\bm{Y}_k(\cdot) \rightarrow \bm{Y}_l(\cdot) \in \mcalE^{D}$. The parent and child relationships are denoted as $k \in Pa(l)$. Here, $k$, $l$ follow topological order and are acyclic.

Local independence of component field $\bm{Y}_l(\cdot)$ in $\mcalG^{D}$ is that given its parent fields, it is conditionally independent of its non-descendant fields, i.e.,
$\bm{Y}_l(\cdot) \perp NonDescendants (\bm{Y}_l(\cdot)) \mid  Pa(\bm{Y}_l(\cdot))$.
\end{definition}

\begin{figure}[htbp]
    \centering
    \includegraphics[width=0.26\textwidth]{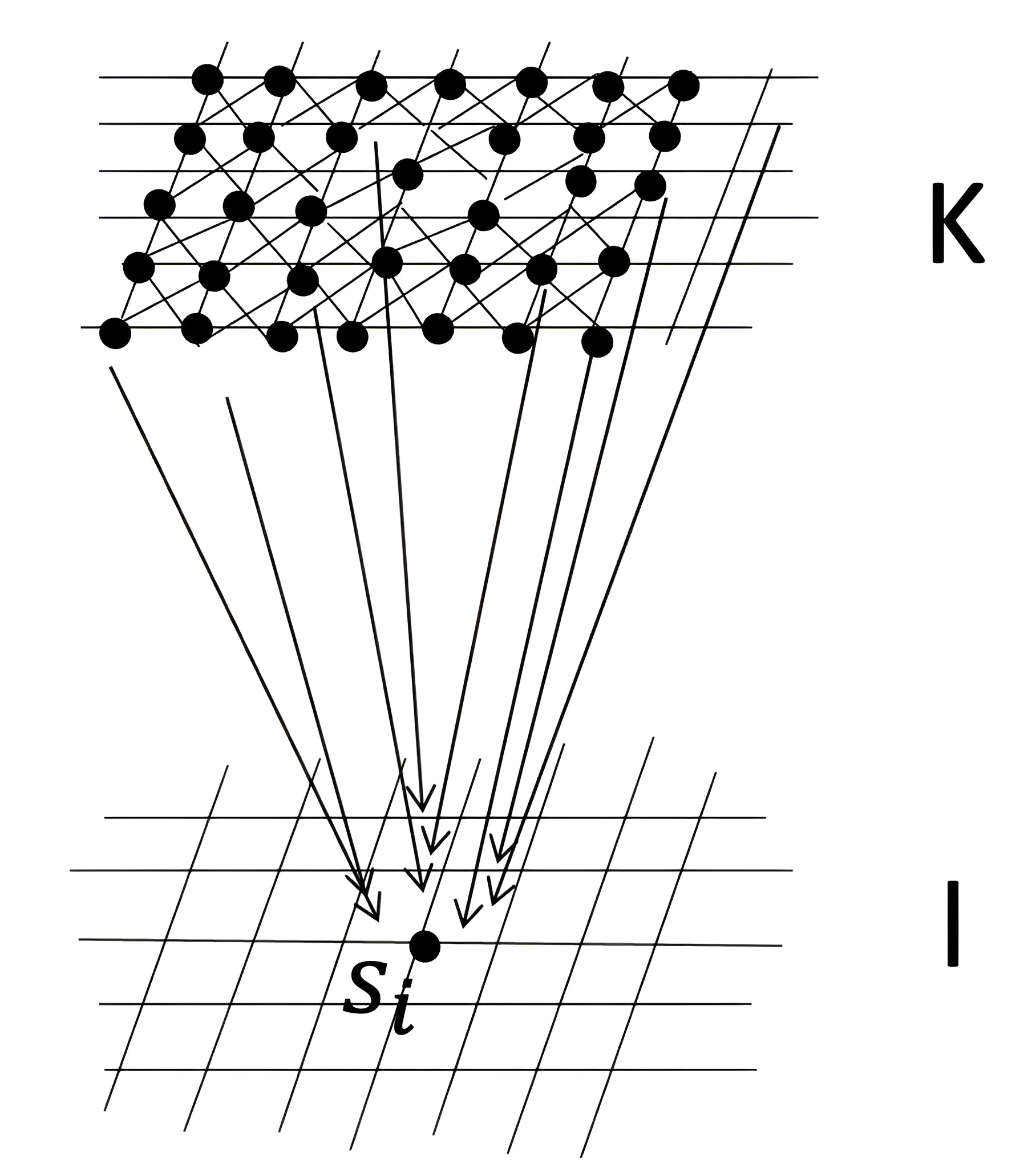}
    \caption{Illustration of a mixed probabilistic spatial graph. Across different component fields $\bm{Y}_k(\cdot)$ and $\bm{Y}_l(\cdot)$, nodes are connected using directed edges. Within a component field $\bm{Y}_k(\cdot)$, nodes are connected by undirected edges. Each node is a random quantity. Altogether, they consist of a mixed probabilistic spatial graph.}
    \label{fig:mixed_sp_graph}
\end{figure}

\begin{definition}[Mixed probabilistic spatial graph]
A mixed probabilistic spatial graph $\mcalG^{MX} = (\mcalY, \mcalE^{MX})$ consists of a node set $\mcalY = \{ Y_q(s_i): q = 1,  \ldots, p; i = 1, \dots, n \}$ and an edge set $\mcalE^{MX}$. 

Connections between nodes across different
component fields (e.g., $\bm{Y}_k(\cdot)$ and $\bm{Y}_l(\cdot)$) are represented by directed edges, i.e., $Y_k(s_i) \rightarrow Y_l(s_j) \in \mcalE^{MX}$.
Connections between nodes within each component field (e.g., $\bm{Y}_k(\cdot)$) are represented by undirected edges, i.e., $Y_k(s_i)$---$Y_k(s_j) \in \mcalE^{MX}$.
Altogether, they form a mixed probabilistic spatial graph.  
See Fig. \ref{fig:mixed_sp_graph}. 
\end{definition}

\section{Mixed Spatial Graphical Model Framework}
\label{sec:Mix_model}

\subsection{The first-stage model: CI among $p$ component fields}
\label{sec:model}
Our objective is to construct the joint covariance $\HSigma_{np \times np}$
and joint precision matrix $\HSigma^{-1}_{np \times np}$. The fundamental idea is to obtain the desired matrices step-wisely from the first and second central moments of conditional distributions; obtaining the joint covariance $\HSigma_{np \times np}$ step-wise dates back to \citet{yule1907theory} and has also been studied in \citet{shachter1989gaussian, bishop2007pattern}.

We start by embedding the mixed probabilistic spatial graph, specifically the directed component field graph at this stage, into the column-wise representation of the conditional mean and covariance in equations \eqref{eq:col-wise_multi1} and \eqref{eq:col-wise_multi2}
\citep{cressie2016multivariate} to exploit the CI among $p$ component fields. 
Theorem \ref{THRM:INDUCTION} generalises an updating formula for computing the joint covariance $\HSigma_{np \times np}$ for any $p$ under any customised component field graph structure.
Theorem \ref{thrm:Sigma_inv_oder} details the step-wise parallel generation of the joint precision $\HSigma_{np \times np}^{-1}$.

Let $ k = \{ 1, 2, \ldots, (p-1) \}$, $l = \{p \}$, 
$k^c$ denotes an arbitrary element of $k$ (i.e., $k^c \in k$), and 
denote the collection $\{k^c: k^c \in Pa(l) \}$ as $k^{sub} \subseteq k$.

A mean-zero process  $Y_l(s_i)$
can be represented as a regression equation,
\begin{align}
\label{reg-eq-uni}
    Y_l(s_i) = \sum_{k^c \in Pa(l)} \sum_{j = 1}^n b_{lk^c}(s_i, s_j) y_{k^{c}}(s_j) + \varphi_{l\mid k^{sub}}^{\frac{1}{2}}(s_i, s_j) Z_l(s_i), \quad Z_l(s_i) \sim N(0, 1).
\end{align}    
Here, $\sum_{k^c \in Pa(l)} \sum_{j = 1}^n b_{lk^c}(s_i, s_j) y_{k^{c}}(s_j)$ is the conditional mean $\mbbE[Y_l(s_i) \mid \bmY_{k^{sub}} (\cdot) =  \bmy_{k^{sub}} (\cdot)]$, where $ b_{lk^c}(s_i, s_j)$ is a real-valued function describing how $Y_l(s_i)$ depends on $Y_{k^{c}}(s_j)$,
and 
$\varphi_{l\mid k^{sub}}(s_i, s_j)$ is the conditional covariance between the values of component $l$ at locations $s_i$ and $s_j$, i.e., $cov[Y_l(s_i), Y_l(s_j) \mid \bmY_{k^{sub}}(\cdot) = \bmy_{k^{sub}}(\cdot)]$.

The corresponding matrix form of equation \eqref{reg-eq-uni} is 
\begin{align}
\label{reg-eq-block}
    \bmY_{np \times 1} = \mathcal{B}_{np \times np} \bmY_{np \times 1} + \bmS_{np \times np} \bmZ_{np \times 1}, \quad \bmZ \sim \mcalN(\bm{0}, \bm{I}_{np \times np}),
\end{align}
where 
$\mathcal{B}$ has block matrix $ \lglftbr block[b_{ll}(\cdot, \cdot)]\lgrgtbr_{n \times n}$ on the main diagonal 
and $\lglftbr block [b_{lk^c}(\cdot, \cdot)]\lgrgtbr_{n \times n} \triangleq \lglftbr \bmB_{lk^c}(\cdot, \cdot)\lgrgtbr$ on the lower off-diagonals, where $\bmB_{lk^c}(\cdot, \cdot)$ is an $n \times n$ matrix of $b_{lk^c}$ function evaluated at all pairs of locations.
$\bmS = \bmD^{\frac{1}{2}}$, where $\bmD = diag \lglftbr block[\varphi_{l \mid k^{sub}}(\cdot , \cdot)] \lgrgtbr$, 
is an $np \times np$ diagonal block matrix, with each main diagonal block being $block[\varphi_{l \mid k^{sub}}(\cdot , \cdot)]$.


\begin{theorem}[Graph-structure-guided Updating Formula for $\HSigma_{np \times np}$]
\label{THRM:INDUCTION}
For any customised structure among component fields in the mixed spatial graph, 
the joint covariance matrix $\HSigma_{np \times np}$ for any $p$, $p \triangleq j+1$,  
can be obtained using the updating formula:
$\HSigma_{np \times np} \triangleq \HSigma_{(j+1)n \times (j+1)n} = $    
 \begin{align*} 
     \left[ \begin{array}{cc}
        \bmSIGMA_{\{1, \cdots, j \}\{1, \cdots, j \}} (\cdot,\cdot) & 
        \bmSIGMA_{\{1, \cdots, j \}\{1, \cdots, j \}}(\cdot,\cdot) \bmB_{ \{ j+1\} \{1, \cdots, j \}}^T (\cdot,\cdot) \\
        \bmB_{ \{ j+1\} \{1, \cdots, j \}} (\cdot,\cdot) \bmSIGMA_{\{1, \cdots, j \}\{1, \cdots, j \}} (\cdot,\cdot) &  \bmB_{ \{ j+1\} \{1, \cdots, j \}} \bmSIGMA_{\{1, \cdots, j \}\{1, \cdots, j \}} \bmB_{ \{ j+1\}\{1, \cdots, j \}}^T  + \bmD_{ \{ j+1\}\{ j+1\}} (\cdot,\cdot)
    \end{array} \right],   
\end{align*} 
where $\bmB_{(j+1)k^c}(\cdot, \cdot) \neq \bm{0}$ for $k^c \in Pa(j+1)$. $k = \{1, \cdots, j \}$, $k^c \in k$.
\end{theorem}

\begin{proof}
    See Supplementary Material \ref{app:proof_induction}.
\end{proof}

Here, $\bmSIGMA_{\{1, \cdots, j\}\{1, \cdots, j\}}(\cdot, \cdot)$ is the joint covariance matrix for the first $j$ component fields $\bmY_1(\cdot), \ldots, \bmY_{j}(\cdot)$. 
The first and second subscripts denote row and column indices respectively, spanning from 
1 to $j$.
$(\cdot, \cdot)$ means spanning over all pairs of spatial locations.

$\HSigma_{np \times np}$ can be computed for any $p$ by setting $j$. For example, setting $j = 99$ yields $\HSigma_{100n \times 100n}$ (hundred-variate processes).

Beyond bivariate, the $\HSigma_{np \times np}$ structure varies with different graph structures of the $p$ component fields. For example, the structure of $\HSigma_{3n \times 3n}$ of a tri-variate process varies depending on whether $1 \in Pa(3)$ or $2 \in Pa(3)$ or $1, 2 \in Pa(3)$. Only when both $1, 2 \in Pa(3)$, do we sum over all the previous component fields, as in \citet{cressie2016multivariate}.

\begin{proposition}[Positive Definite Condition for $\HSigma_{np \times np}$]
\label{prop:pd}
$\HSigma_{np \times np}$ is positive definite (PD) if and only if $\bmSIGMA_{\{1, \cdots, j \}\{1, \cdots, j \}}(\cdot, \cdot)$ is PD and its Schur complement $\bmD_{\{j+1 \}\{j+1 \}}(\cdot, \cdot)$ is PD. 
\end{proposition}

\begin{proof}
 See Supplementary Material \ref{app:proof_prop1}.
 \end{proof}

To reveal the sparsity induced by the CI among $p$ component fields, we also need to construct the corresponding joint precision matrix $\HSigma_{np \times np}^{-1}$. 

The induction formula in Theorem \ref{THRM:INDUCTION} implies that $\HSigma_{np \times np}$ of any size can always be divided into four different blocks: the leading diagonal block SG,  the row block $\bmR$ beneath SG, the column block $\bmC$ to the right of SG, and the bottom-right block $\bmSIGMA_{rr}$.
That is,
\begin{scriptsize}
$   \left[ \begin{array}{cc} 
  \left[SG\right]   & \left[ \bmC \right] \\
  \left[ \bmR\right]   & \left[ \bmSIGMA_{rr} \right] 
\end{array} \right] $
\end{scriptsize}, where here we use $r$ to denote $j + 1$ for convenience.

The following theorem presents a step-wise computing method for the $\HSigma_{np \times np}^{-1}$ at a much reduced computational cost (linear in $p$) compared to the $\mcalO(p^3n^3)$ complexity of Cholesky inversion.

\begin{theorem}[Step-wise Construction of $\HSigma^{-1}_{np \times np}$]
\label{thrm:Sigma_inv_oder}
At the $r^{th}$ step, the joint precision matrix $\HSigma^{-1}_{np \times np}$ for the first $r$ component fields,
 denoted as $\HSigma^{-1}_{rn \times rn}$,
 can be obtained using the updating formula 
    \begin{align*}
        \HSigma^{-1}_{rn \times rn} = 
        \left[ \begin{array}{cc}
           \Sigma_{(r-1)n \times (r-1)n}^{-1}[\Sigma_{(r-1)n \times (r-1)n} + \bmC \bmD_{rr}^{-1} \bmR]\Sigma_{(r-1)n \times (r-1)n}^{-1}  & -\Sigma_{(r-1)n \times (r-1)n}^{-1} \bmC \bmD_{rr}^{-1}  \\
           - \bmD_{rr}^{-1} \bmR \Sigma_{(r-1)n \times (r-1)n}^{-1}  &  \bmD_{rr}^{-1}
        \end{array}  \right],
    \end{align*}     
in parallel to the construction of the joint covariance matrix $\HSigma_{rn \times rn}$ at this step, where $r = 2, \dots, p$.
This step-wise construction has dominant computational complexity $p* \mcalO(n^3)$.
\end{theorem}

\begin{proof}
   
See Supplementary Material \ref{app:proof_thrm2}.
\end{proof}

\begin{proposition}[Positive Definite Condition for $\HSigma^{-1}_{np \times np}$]
\label{prop:cond_sigma_inv}
The matrix $\HSigma^{-1}_{np \times np}$ constructed at the $r^{th}$ step above is 
positive definite if and only if
$\bmD_{rr}^{-1}$ and $\HSigma^{-1}_{(r-1)n \times (r-1)n}$ at the $r^{th}$ step are both positive definite. 
\end{proposition}

\begin{proof}
See Supplementary Material \ref{app: proof_prop2}.
\end{proof}

\subsection{The algorithm}
\label{sec:algo}
Following the rules revealed by Theorems \ref{THRM:INDUCTION} and \ref{thrm:Sigma_inv_oder}, together with the specified directed acyclic graphical (DAG) for $p$ component fields, $p \gg2$, Algorithm \ref{alg:my-algorithm} (Supplementary Material \ref{app:algo1}) generates the $\HSigma_{np \times np}$ for any customised CI structure, involving only component fields that exhibit conditional dependence (parent-child relationships), see steps 9 to 12. 

Meanwhile, $\HSigma_{np \times np}^{-1}$ is generated step-wise in parallel at a significantly reduced computational complexity (i.e., $p * \mcalO( n^3)$) compared to Cholesky inversion ($\mcalO(p^3n^3)$).


\subsection{1D Simulation: Ten component Fields}
\label{sec:simulation}
The spatial domain $\mathcal{D}$ for this simulation is [-1, 1], with a grid spacing of 0.05. 

\begin{figure}[htpb]
    \centering
    \includegraphics[width = 0.29 \textwidth]{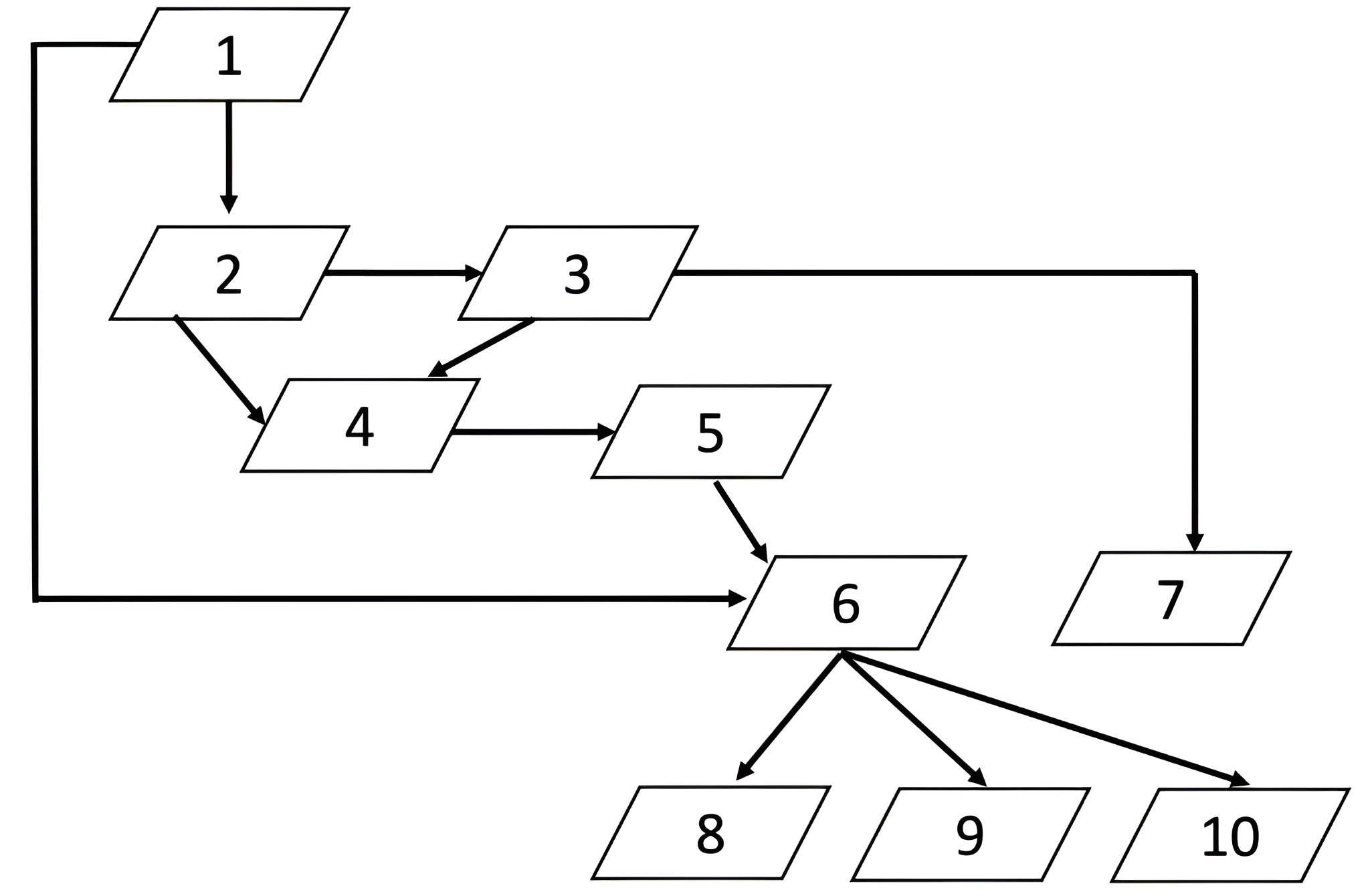}
    \caption{A hypothetically organised acyclic graphical structure for ten fields. Each parallelogram represents a component field.}
    \label{fig:ten_fields}
\end{figure}

Figure \ref{fig:ten_fields} is a hypothetically organised DAG for ten component fields. 

\begin{figure}[ht]
    \centering
    \includegraphics[width=0.7\linewidth]{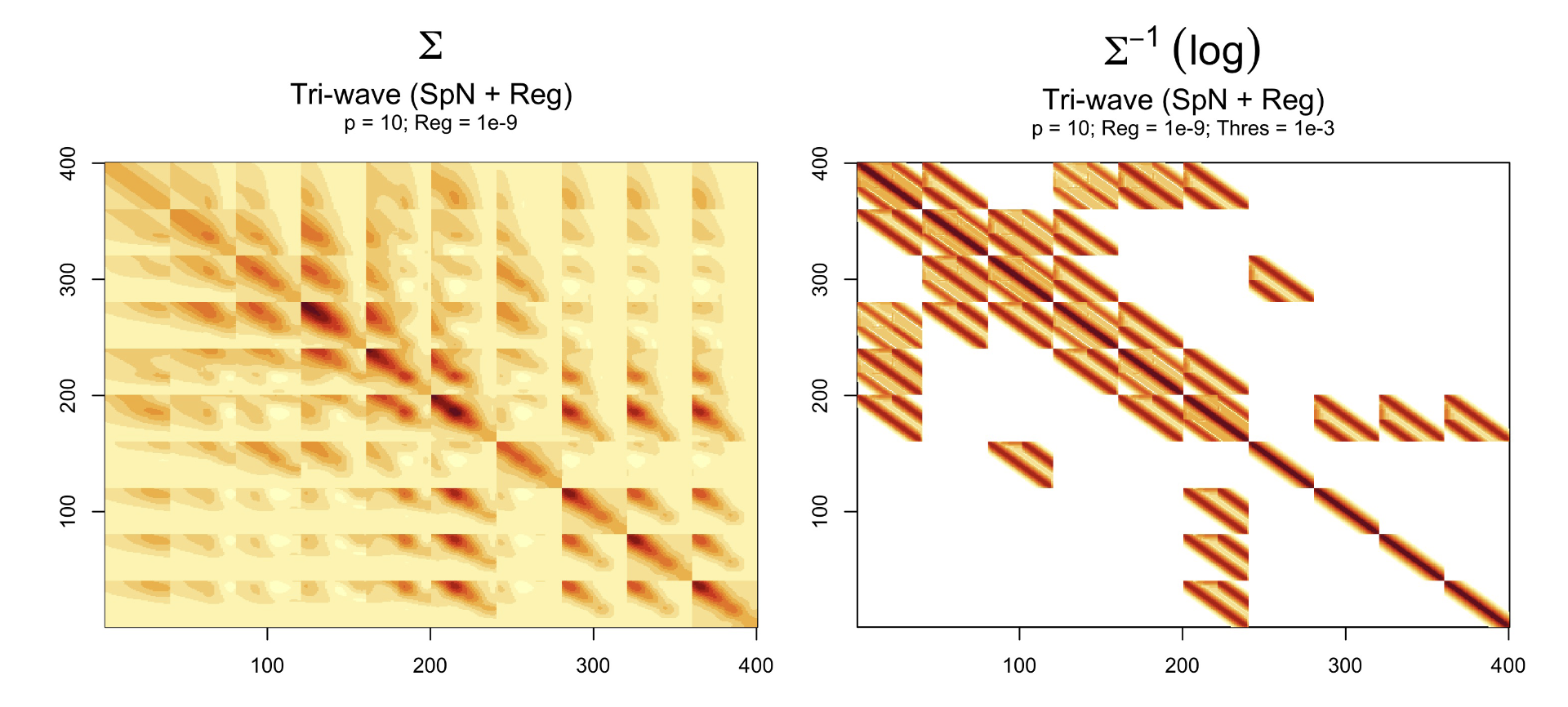}
    \caption{1D simulation of \text{$\Sigma$} and \text{$\Sigma^{-1}$} for ten component fields ($p = 10$). $b(\cdot, \cdot)$ function is modelled using a modified tri-wave function.
    The left figure is the joint covariance matrix \text{$\Sigma$} and the right one is the joint precision matrix \text{$\Sigma^{-1}$}, on a log scale. Both $\textbf{B}_{rt}$ are under a combination of spectral normalisation (SpN) and regularisation (Reg) transformation. Parameter A = 0.1, \text{$\Delta$} = 0.5. The largest possible threshold for \text{$\Sigma^{-1}$} is $1e^{-3}$, the smallest possible regularisation number is $1e^{-9}$.}
     \label{fig:SG_SG_inv_SBS}
\end{figure}

Figure \ref{fig:SG_SG_inv_SBS} shows the simultaneously obtained joint covariance $\HSigma_{np \times np}$ (left) and the joint precision matrix $\HSigma^{-1}_{np \times np}$ (right) for ten component fields ($p$ = 10) organised in Figure \ref{fig:ten_fields}.

The left figure shows the accommodation of asymmetric cross-covariances in the off-diagonal blocks of $\HSigma_{np \times np}$, while all the diagonal blocks, representing auto-covariances, are symmetric. The right figure displays the
structural sparsity in $\HSigma^{-1}_{np \times np}$ at the current stage.

For full details of the simulation, 
see Supplementary Material \ref{app:1dsimu}.



\section{Beyond CI among p-component Fields: Cross-MRF}
\label{sec:cross_MRF}
The method derived so far generated $\HSigma_{np \times np}$ and $\HSigma_{np \times np}^{-1}$ step-wise in parallel with a much reduced $\HSigma_{np \times np}^{-1}$ generation order ($p* \mcalO(n^3)$), accommodated asymmetric cross-covariance in $\HSigma_{np \times np}$ and achieved
structural sparsity in $\HSigma_{np \times np}^{-1}$ via CI among $p$ component fields.

However, generating $\HSigma_{np \times np}^{-1}$ still remains \textit{cubic} in $n$.
The possibility of its further reduction and enhanced sparsity in the precision $\HSigma_{np \times np}^{-1}$ is worth probing. 

We explore incorporating CI among $n$ spatial locations on top of the first-stage CI among $p$ component fields,
aiming to achieve \textit{doubly} CI 
structure among both $p$ and $n$.

The fundamental question is
whether this conditional distribution $[Y_l(s_i) | \bmY_{k^{sub}} (\mcalN(s_i)) =  \bmy_{k^{sub}} (\mcalN(s_i))]$
rather than this conditional $[Y_l(s_i) | \bmY_{k^{sub}} (\cdot) =  \bmy_{k^{sub}} (\cdot)] $ 
can still define $pr(\bmY)$, where $\bmY \in \mbbR^{np}$. 
We appeal to the Hammersley-Clifford (H-C) theorem \citep{clifford1971markov}. 

In the remainder of this section, we temporally put aside the parent-child relationships among component fields. Section \ref{sec:embed_CrossMRF} will link the theory derived in this section with the first-stage framework for CI structure between component fields.


\subsection{H-C theorem for multivariate spatial stochastic processes}
Besag (1974) proved the univariate version ($p = 1$) of the H-C theorem
\citep{clifford1971markov}, which provides a foundation for the ``\textit{construction of a valid spatial stochastic process through conditional probabilities}''.

We first explore the extension of the H-C theorem to multivariate spatial stochastic processes ($p \gg 2$) and obtain the following observation.
\begin{observation}[Multivariate Extension of the Hammersley-Clifford Theorem]
\label{lemma_mv_HC}
For a given 
$l \in \{ 1, \ldots, p \}$, let $k = \{1, \ldots, p  \} \backslash \{ l\}$, $k^c \in k$, 
$l \cup k = \{ 1, \ldots, p \}$,  $i, j = 1, \ldots, n$, $i \neq j$. 
Partition the collection of $p$ random vectors 
$(\bmY_1(\cdot), \ldots, \bmY_p(\cdot))$ into $(\bmY_k(\cdot), \bmY_l(\cdot))$, where 
$\bmY_l(\cdot) \in \mbbR^n$ indicates one random vector corresponding to index $l$, and $\bmY_k(\cdot)$ indicates a collection of random vectors corresponding to the remaining indices except $l$. The ``$\cdot$'' represents all locations in $\mcalD$.

The desired joint distribution $pr(\bmY)$ is denoted as $pr(\bmY_k(\cdot), \bmY_l(\cdot))$ and 
\begin{small}
\begin{align}
\label{HC-multi}
    \frac{pr(\bmY_k(\cdot), \bmY_l(\cdot))}{pr(\bmY_k(\cdot), \bmX_l(\cdot))} = 
    \HPi_{i = 1} ^n \frac{pr(Y_l(s_i) \mid Y_l(s_1), \cdots, Y_l(s_{i-1}), X_l(s_{i+1}), \ldots, X_l(s_n), \bmY_k(\cdot))}{pr(X_l(s_i) \mid Y_l(s_1), \cdots, Y_l(s_{i-1}), X_l(s_{i+1}), \ldots, X_l(s_n), \bmY_k(\cdot))},
\end{align}      
\end{small}
where $(\bmY_k(\cdot), \bmY_l(\cdot)) \in \Omega = \Omega_1 \times \ldots \times \Omega_{np}$.
\end{observation}

 \begin{proof}
    See Supplementary Material \ref{app:Proof_HC_col}.
\end{proof}

The main problem with such a multivariate version is that it does not provide much detailed information about other component fields in $\bmY_k(\cdot)$.
An alternative perspective on the multivariate H-C theorem, derived below, may provide more useful insights.

Following \citet[Sec.~3]{besag1974spatial}, 
we define
\begin{small}
\begin{align}
            \label{def:Q}
                Q(\bmY) = log \frac{pr(\bmY)}{pr(\bm{0})}, \quad \mbox{where } pr(\bm{0}) > 0, \bmY \in \mbbR^{np}.
            \end{align}
            \end{small}
Then, for any $pr(\bmY)$, $Q(\bmY)$ has a unique expansion for all $\bmY \in \Omega$ 
expressed as a sum of various \textit{G-functions}.
\begin{small}
\begin{dmath}
\label{Q-G}
    Q(\bmY) = \sum_{l=1}^p \sum_{i=1}^n Y_l(s_i)G_i^l(Y_l(s_i)) + 
    \sum_{l=1}^p \sum_{1 \leq i, j \leq n} Y_l(s_i) Y_l(s_j) G_{ij}^{ll} (Y_l(s_i), Y_l(s_j)) + \cdots + \sum_{l = 1}^p Y_l(s_1) \cdots Y_l(s_i) \cdots Y_l(s_n)G_{1 \cdots i \cdots n}^{l \quad l \;\;\; l} (\cdot) 
    + \sum_{1 \leq k^c, l \leq p} \sum_{1 \leq i \leq n} Y_l(s_i) Y_{k^c}(s_i) G_{ii}^{l k^{c}}(Y_l(s_i), Y_{k^c}(s_i)) + \cdots + \sum_{i = 1}^n Y_1(s_i) \cdots Y_l(s_i) \cdots Y_{k^c}(s_i) \cdots Y_p(s_i)G_{i \cdots i \cdots i \cdots i}^{1 \;\; l \;\;\; k^{c}\;\; p}(\cdot) 
     +  \sum_{1 \leq k^c, l \leq p} \sum_{1 \leq i, j \leq n} Y_l(s_i) Y_{k^c}(s_j)G_{i j}^{l {k}^c} (Y_l(s_i), Y_{k^c}(s_j)) + \cdots 
     + Y_1(s_1) \cdots Y_l(s_i) \cdots Y_{k^c}(s_j) \cdots Y_p(s_n) G_{1 \cdots i \cdots j \cdots n}^{1 \;\; l \;\;\; k^c \;\; p} (\cdot) 
     + Y_p(s_1) \cdots Y_{k^c}(s_i) \cdots Y_l(s_j) \cdots Y_1(s_n) G_{1 \cdots i \cdots j \cdots, n}^{p \;\; k^c \;\; l \quad 1} (\cdot). 
\end{dmath}       
\end{small}
Such an expansion follows a $p$-column-$n$-row table, expanding column-wisely, row-wisely, and
then diagonally, from singleton to pairwise and higher-order interaction terms.

For the \textit{G-functions}, see \citet[Sec.~3]{besag1974spatial} and \citet[pp.~178-179]{cressie2011statistics}.

For a particular component $l$ at a particular site $s_i$,
let $\bmY_{-(li)}$ denote a vector $\bmY$ with a particular element $Y_l(s_i)$
set to zero, i.e., 
$\bmY_{-(li)} = (Y_l(s_1), \cdots, Y_l(s_{i-1}), 0, Y_l(s_{i+1}), Y_l(s_n), \bmY_k(\cdot))$. This leads to a quantity $Q(\bmY) - Q(\bmY_{-(li)})$, where
\begin{small}
\begin{align}
\label{Q-Qil-log}
    Q(\bmY) - Q(\bmY_{-(li)}) &= log \frac{pr(\bmY)}{pr(\bmY_{-(li)})}  \quad \mbox{(by definition of $Q(\bmY)$ in equation \eqref{def:Q}})  \nonumber \\ 
    &= log \frac{pr(Y_l(s_i) \mid \{Y_l(s_1), \cdots, Y_l(s_{i-1}), Y_l(s_{i+1}), \cdots, Y_l(s_n), \bmY_k(\cdot) \})}{pr(0_l(s_i) \mid \{Y_l(s_1), \cdots, Y_l(s_{i-1}),  Y_l(s_{i+1}), \cdots, Y_l(s_n), \bmY_k(\cdot) \} )}. \;
\end{align}   
\end{small}

The quantity $Q(\bmY) - Q(\bmY_{-(li)})$ includes only the terms related to the specific $Y_l(s_i)$, as all other terms not involving this particular $Y_l(s_i)$ are cancelled out. This implies that the expansion of $Q(\bmY)$ in equation \eqref{Q-G} can be modified to the
expansion of $Q(\bmY) - Q(\bmY_{-(li)})$.
\begin{small}
\begin{dmath}
\label{Q-Qil}
    Q(\bmY) - Q(\bmY_{-(li)}) =  Y_l(s_i)G_i^l(Y_l(s_i)) + \\
     \sum_{1 \leq j \leq n} Y_l(s_i) Y_l(s_j) G_{ij}^{ll} (Y_l(s_i), Y_l(s_j)) + \cdots + 
     Y_l(s_1) \cdots Y_l(s_i) \cdots Y_l(s_n)G_{1 \cdots i \cdots n}^{l \quad l \;\;\; l} (\cdot) \\
    + \sum_{1 \leq k^c \leq p}  Y_l(s_i) Y_{k^c}(s_i) G_{ii}^{l k^{c}}(Y_l(s_i), Y_{k^c}(s_i)) + \cdots + Y_1(s_i) \cdots  Y_l(s_i) \cdots Y_{k^c}(s_i) \cdots Y_p(s_i)G_{i \cdots i \cdots i \cdots i}^{1 \;\; l \;\;\; k^{c} \;\; p}(\cdot) \\
     +  \sum_{1 \leq k^c \leq p} \sum_{1 \leq  j \leq n} Y_l(s_i) Y_{k^c}(s_j)G_{i j}^{l {k}^c} (Y_l(s_i), Y_{k^c}(s_j)) + \cdots \\
     + Y_1(s_1) \cdots Y_l(s_i) \cdots Y_{k^c}(s_j) \cdots Y_p(s_n) G_{1 \cdots i \cdots j \cdots n}^{1 \;\; l \;\;\; k^c \;\; p} (\cdot). 
\end{dmath}    
\end{small}

Some G-functions in the expansion \eqref{Q-Qil} may be null, hence the expansion \eqref{Q-Qil} can be simplified if three additional neighbourhood classes are introduced.

\begin{figure}[htbp]
    \centering
    \begin{subfigure}[b]{0.31\textwidth}
        \centering
        \caption{Same-component \\ auto-neighbourhood of $Y_l(s_i)$}
        \includegraphics[width=0.63\textwidth]{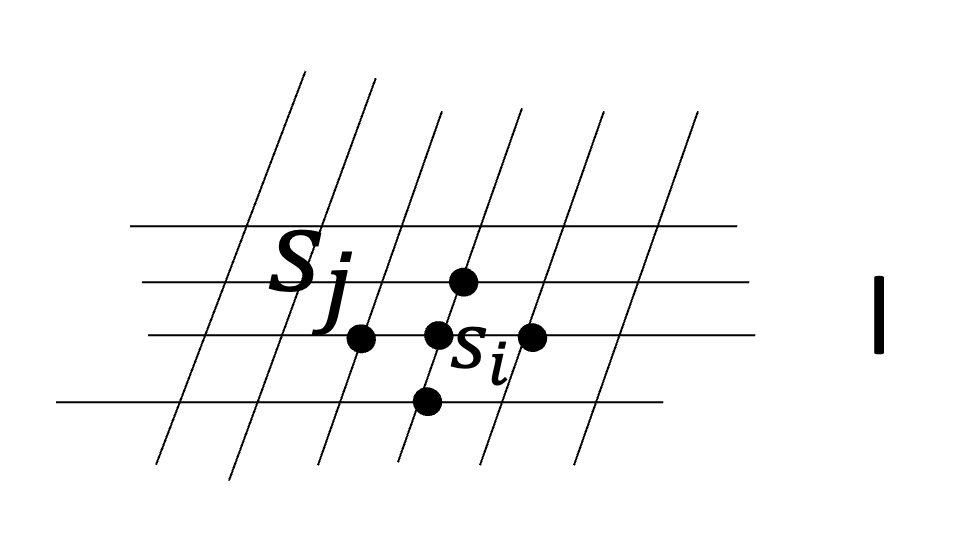}   
        \label{fig:samevar-atuo-N}
    \end{subfigure}
    \hfill
    \begin{subfigure}[b]{0.31\textwidth}
        \centering
        \caption{Same-location \\ auto-neighbourhood of $Y_l(s_i)$}
        \includegraphics[width=0.53\textwidth]{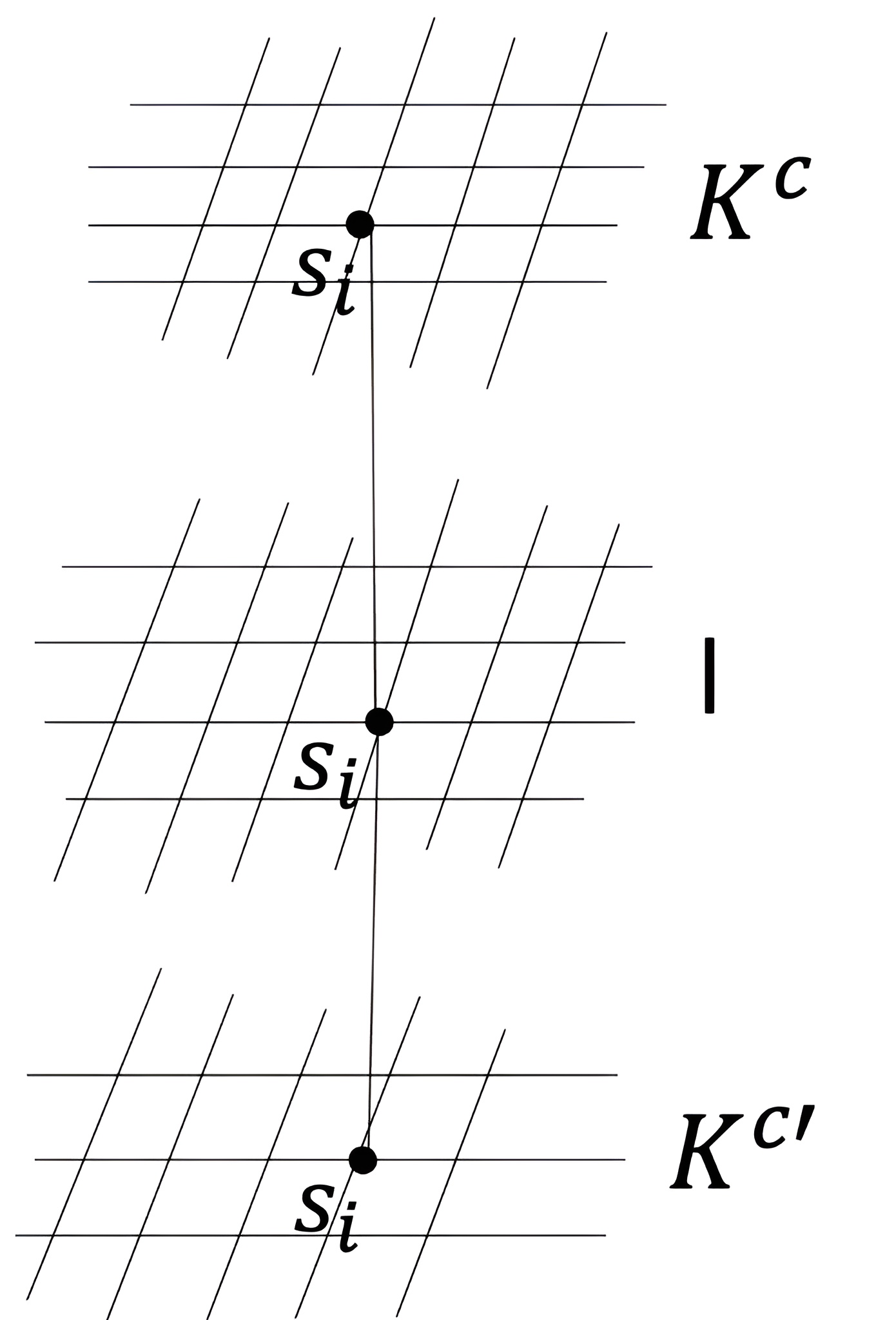}
        \label{fig:sameloc-auto-N}
    \end{subfigure}
    \hfill
    \begin{subfigure}[b]{0.31\textwidth}
        \centering
        \caption{Cross-neighbourhood of $Y_l(s_i)$}
        \includegraphics[width=0.63\textwidth]{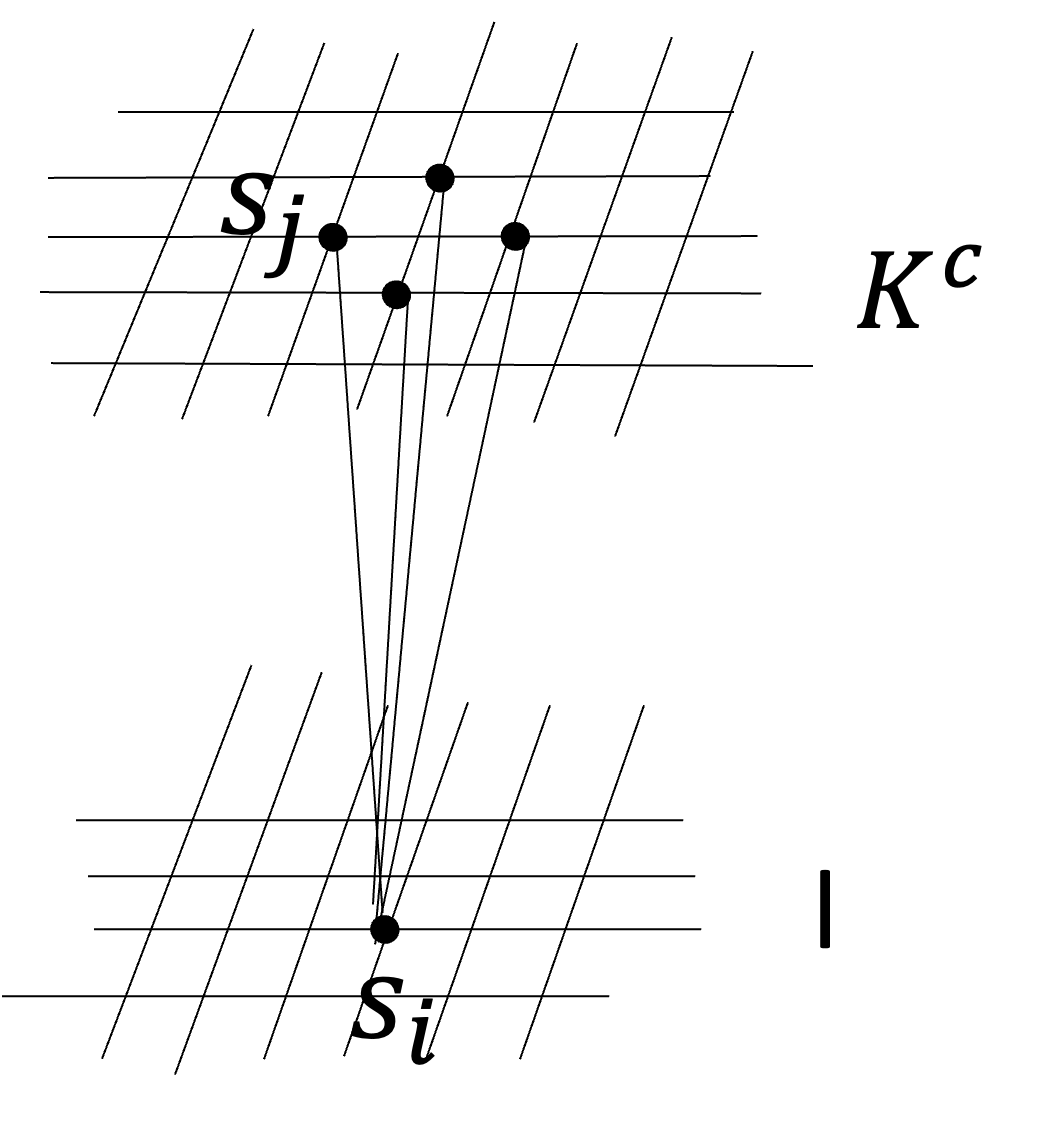}
        \label{fig:cross-N}
    \end{subfigure}
    \caption{Illustration of the same-component auto-neighbourhood $\{ Y_l(s_j): j \in \mcalN(i)\}$, the same-location auto-neighbourhood $\{Y_{k^c}(s_i): k^c \text{ --- } l , k^c \in \{1, \ldots, p\} \backslash \{ l\} \}$, and the 
    cross-neighbourhood $\{Y_{k^c}(s_j): k^c \text{ --- } l , k^c \in \{1, \ldots, p\} \backslash \{ l\}, j \in \mcalN(i) \}$ of $Y_l(s_i)$.}
\end{figure}

\begin{definition}[Same-component auto-neighbour, same-component auto-neighbourhood]
$Y_l(s_i)$ and $Y_l(s_j)$ are same-component auto-neighbours if for a given component $l$, $s_j \in \mcalN(s_i)$ (denoted as $j \in \mcalN(i)$ for convenience). 
The collection $\{ Y_l(s_j): j \in \mcalN(i) \}$ is the same-component auto-neighbourhood of $Y_l(s_i)$. 
$l = 1, \ldots, p$, $i = 1, \ldots, n$. See Fig. \ref{fig:samevar-atuo-N}.
\end{definition}

\begin{definition}[Same-location auto-neighbour, same-location auto-neighbourhood]
$Y_l(s_i)$ and $Y_{k^c}(s_i)$ are same-location auto-neighbours if at a given location $s_i$, component $l$ and $k^c$ are directly connected by an undirected edge, denoted as 
$l \text{ --- } k^c$.
The collection $\{Y_{k^c}(s_i): k^c \text{ --- } l, k^c \in \{1, \ldots, p\} \backslash \{ l\}\}$ is the same-location auto-neighbourhood of $Y_l(s_i)$. $l = 1, \ldots, p$, $i = 1, \ldots, n$. See Fig. \ref{fig:sameloc-auto-N}. 
\end{definition}

\begin{definition}[Cross-neighbour, cross-neighbourhood]
$Y_l(s_i)$ and $Y_{k^c}(s_j)$ are cross-neighbours if component $l$ and $k^c$ are directly connected by an undirected edge 
($l \text{ --- } k^c$) meanwhile $s_j \in \mcalN(s_i)$ ($j \in \mcalN(i)$).
The collection $\{ Y_{k^c}(s_j):  k^c \text{ --- } l , k^c \in \{1, \ldots, p\} \backslash \{ l\}, j \in \mcalN(i) \}$ is the cross-neighbourhood of $Y_l(s_i)$. $l = 1, \ldots, p$, $i = 1, \ldots, n$. Figure \ref{fig:cross-N}.
\end{definition}

With the above definitions, 
$Q(\bmY) - Q(\bmY_{-(li)})$ in equation \eqref{Q-Qil} will depend on 
$Y_l(s_j)$ only when $Y_l(s_i)$ and $Y_l(s_j)$ are same-component auto-neighbours, 
or $Y_l(s_j)$ is an element of $Y_l(s_i)$'s same-component auto-neighbourhood $\{Y_l(s_j): j \in \mcalN(i) \}$. 

Similarly, the quantity will depend on 
$Y_{k^c}(s_i)$ only when $Y_l(s_i)$ and $Y_{k^c}(s_i)$ are same-location auto-neighbours, 
and will depend on $Y_{k^c}(s_j)$ only when $Y_l(s_i)$ and $Y_{k^c}(s_j)$ are cross-neighbours. 

These are equivalent to say $Q(\bmY) - Q(\bmY_{-(li)})$ in equation \eqref{Q-Qil} will depend on 
$Y_l(s_j)$, $Y_{k^c}(s_i)$, and $Y_{k^c}(s_j)$
only when the corresponding \textit{same-component auto G-function} $Y_l(s_i) Y_l(s_j)G_{ij}^{ll}$, or the \textit{same-location auto G-function} $Y_l(s_i)Y_{k^c}(s_i)G_{ii}^{l k^c}$, or the 
\textit{cross G-function} $Y_l(s_i) Y_{k^c}(s_j) G_{i j}^{l k^c}$ in the expansion \eqref{Q-Qil} 
is non-null. 

Removing all null auto and cross G-functions simplifies the equation \eqref{Q-Qil-log} to 
\begin{small}
\begin{dmath}
\label{eq:QQ_HC}
    Q(\bmY) - Q(\bmY_{-(li)}) = log \frac{pr(\bmY)}{pr(\bmY_{-(li)})} \nonumber \\
    = log \frac{pr(Y_l(s_i) \mid \{Y_l(s_j) \}, \{Y_{k^c}(s_i)\}, 
    \{Y_{k^c}(s_j)\}, \{ 0_r(s_h) : h \neq i, j, r \neq k^c, l \} )}{pr(0_l(s_i) \mid
    \{Y_l(s_j) \}, \{Y_{k^c}(s_i) \}, \{Y_{k^c}(s_j) \}, \{ 0_r(s_h) : h \neq i, j; r \neq k^c, l \} )}, 
\end{dmath}    
\end{small}
where $j \in \mcalN(i)$, $k^c \text{ --- } l , k^c \in \{1, \ldots, p\} \backslash \{ l\}$, and $0_l(s_i)$ denotes $Y_l(s_i) = 0$.

Equation \eqref{eq:QQ_HC} shows that the change in the joint probability between the case with a particular $Y_l(s_i)$ and the case without this $Y_l(s_i)$ only depends on the auto-neighbourhoods and cross-neighbourhood of this particular $Y_l(s_i)$.
See supplementary material \ref{app:verification_prop_product} for details.

More generally, for any $l = 1, \ldots, p $ and any $i =  1, \ldots, n$, the following lemma provides an alternative perspective,
offering more refined insights than Observation \ref{lemma_mv_HC}.

\begin{lemma}[Alternative Perspective on the Multivariate Hammersley-Clifford Theorem]
\label{thrm:alt_H-C}
When certain conditions are satisfied, each of the $np$ random quantities of the multivariate spatial stochastic process together with their associated auto-neighbourhoods (same-component, same-location) and cross-neighbourhood is sufficient to define the desired joint probability distribution $pr(\bmY)$, where $\bmY \in \mbbR^{np}$.    
\end{lemma}

\begin{proof}[Proof sketch]
Each G-function can be written in terms of $Q(\cdot)$ (by expansion \eqref{Q-G}), and $Q(\cdot)$ can be expressed as local conditional probabilities (by definition of $Q(\cdot)$ in equation \eqref{def:Q}). For example, the cross-G function $Y_l(s_i)Y_{k^c}(s_j) G_{ij}^{lk^c}(Y_l(s_i), Y_{k^c}(s_j))$ can be written as
\begin{small}
\begin{dmath*}
    Y_l(s_i)Y_{k^c}(s_j) G_{ij}^{lk^c}(Y_l(s_i), Y_{k^c}(s_j))
= Q(0, \cdots, 0, Y_l(s_i), 0, \cdots, 0, \underline{0}, 0, \cdots, 0, \underline{0}, 0, \cdots, 0, Y_{k^c}(s_j), 0, \cdots, 0) \nonumber \\
- Q(0, \cdots, 0, \underline{0}, 0, \cdots, 0, \underline{0}, 0, \cdots, 0, \underline{0}, 0, \cdots, 0, Y_{k^c}(s_j), 0, \cdots, 0) \nonumber \\
- Q(0, \cdots, 0, Y_l(s_i), 0, \cdots, 0, \underline{0}, 0, \cdots, 0, \underline{0}, 0, \cdots, 0, \underline{0}, 0, \cdots, 0) \\
= log \frac{pr(Y_l(s_i) \mid \{Y_{k^c}(s_j): k^c \text{ --- } l , k^c \in \{1, \ldots, p\} \backslash \{ l\}, j \in \mcalN(i) \}, \{ 0_r(s_h): r \neq k^c, l, h \neq i,j  \})}
{ pr(0_l(s_i) \mid \{Y_{k^c}(s_j): k^c \text{ --- } l , k^c \in \{1, \ldots, p\} \backslash \{ l\}, j \in \mcalN(i) \}, \{ 0_r(s_h): r \neq k^c, l, h \neq i,j  \})}
- log \frac{pr(Y_l(s_i) \mid \{0_r(s_h):  r \neq l, h \neq i \})}{pr(0_l(s_i) \mid \{0_r(s_h):  r \neq l , h \neq i\})}.
\end{dmath*}   
\end{small}

For the other G-functions and a full proof, see Supplementary Material \ref{app:proof_lemma1}. 

The above equalities indicate that from local conditional distributions reflecting auto-/cross-neighbourhood structures, we can obtain the corresponding auto-G and cross-G functions; from these G-functions, we could obtain $Q(\bmY)$ (by equation \eqref{Q-G}); and from the $Q(\bmY)$, we can obtain  $pr(\bmY)$, where $\bmY \in \mbbR^{np}$ (by equation \eqref{def:Q}). 

\end{proof}

The fact revealed by Lemma \ref{thrm:alt_H-C} that we could arrive at our desired joint $pr(\bmY)$ from a set of local conditional distributions
indicates that this conditional distribution $[Y_l(s_i) \mid \bmY_{k^{sub}} (\mcalN(s_i))]$, 
rather than this conditional distribution $[Y_l(s_i) \mid \bmY_{k^{sub}} (\cdot)]$,
can still define $pr(\bmY)$, where $\bmY \in \mbbR^{np}$, provided certain conditions are satisfied.

\subsection{Conditions}
\label{conditions&verify}
In general, the conditions are 
\begin{itemize}[itemsep=0.2pt]
    \item strict positivity condition: $pr(\bmY) > 0$ for all $\bmY \in \Omega$, $\bmY \in \mbbR^{np}$; particularly,  $pr(\boldsymbol{0}) > 0$;
    \item summability of $exp(Q(\bmY))$: $\sum_{\bmY \in \Omega} exp(Q(\bmY)) < \infty$;
    \item symmetry of auto G-functions and cross G-functions: 
(a) $Y_l(s_i) Y_l(s_j) G_{ij}^{ll}(\cdot) = Y_l(s_j)Y_l(s_i) G_{j i}^{ll}(\cdot)$; 
(b) $Y_l(s_i) Y_{k^c}(s_i) G_{ii}^{l k^c}(\cdot) = Y_{k^c}(s_i) Y_l(s_i) G_{ii}^{k^c l}(\cdot)$; (c) $Y_l(s_i) Y_{k^c}(s_j) G_{i j}^{l k^c}(\cdot) = Y_{k^c}(s_j) Y_l(s_i) G_{ji}^{k^c l}(\cdot)$
\end{itemize}


The first condition indicates that an entire zero realisation is possible. This ensures that the definition of $Q(\bmY)$ in equation \eqref{def:Q} is valid. 
The second condition ensures that the desired joint distribution $pr(\bmY)$ exists and is a valid probability distribution.

The third condition ensures that the $\HSigma_{np \times np}$ and $\HSigma_{np \times np}^{-1}$ for $pr(\bmY)$ are symmetric. 
It implies a requirement of symmetry between components $k^c$ and $l$, as spatial locations are symmetric by nature already.
The symmetry between components $k^c$ and $l$ is usually reflected via an undirected graph. We detail the linking strategy between this requirement and the first-stage framework in Section \ref{sec:embed_CrossMRF}.

\subsection{Cross-MRF}
\label{sec_cross-MRF}
In a univariate spatial setting ($p=1$), a
univariate spatial stochastic process $\{Y(s_i): i = 1, 2, \ldots, n \}$ is
called a Markov Random Field when the 
joint distribution $\bmY(\cdot)$, where $\bmY \in \mbbR^n$, is defined by
a set of univariate conditional distributions $\{ [Y(s_i) \mid \{ Y(s_j): j \in \mcalN(i) \}]: i = 1,2, \ldots, n \}$  \citep[p.~176]{cressie2011statistics}. 

For the multivariate case,
the \textit{cross-Markov Random Field} is presented below. 
\begin{definition}[Cross-Markov Random Field]
\label{def_cross_MRF}
When the conditions in Section \ref{conditions&verify} are satisfied, and
the cross-conditional distributions $\{ [Y_l(s_i) \mid \{ Y_{k^c}(s_j): k^c \text{ --- } l , k^c \in \{1, \ldots, p\} \backslash \{ l\},  j \in \mcalN(i) \}]: 
l = 1, \ldots, p; i = 1, \ldots, n \} $
together with two types of auto-conditional distributions, i.e., the same-component auto $\{ [Y_l(s_i) \mid \{ Y_l(s_j): j\in \mcalN(i) \}]: l = 1, \ldots, p; i = 1, \ldots, n\}$ and the same-location auto
$\{ [Y_l(s_i) \mid \{ Y_{k^c}(s_i): k^c \text{ --- } l, k^c \in \{1, \ldots, p\} \backslash \{ l\} \}]: 
l = 1, \ldots, p; i = 1, \ldots, n \} $
can define the joint distribution $pr(\bmY)$, where $\bmY \in \mbbR^{np}$, then
the multivariate spatial stochastic process $\{(Y_1(s_i),  \ldots, Y_p(s_i)):  i = 1, \ldots, n \}$
is a \textit{cross Markov Random Field} (cross-MRF). 
\end{definition}

When the multivariate spatial stochastic process is a cross-MRF, it
features directly connected components $k^c$, $l$ being in a symmetric relationship, while $s_j \in \mcalN(s_i)$. 
The ``cross'' here refers to the \textit{doubly} conditional independence (CI) among
both the $n$ spatial locations (row-wisely) and $p$ components (column-wisely).

In fact, 
it is straightforward to verify that $pr (\bmY)$ is directly proportionate to the product of two types of auto-conditional distributions and a cross-conditional distribution. 
See Supplementary Material \ref{app:verify_p(y)} for the verification.


Table~\ref{Tab:compare_MRF_crossMRF} in Supplementary Material \ref{app:compare_tab} compares different facets between the MRF for univariate spatial
processes and the cross-MRF for multivariate spatial processes.

\section{Linking the Cross-MRF and Mixed Spatial Graphical Model Framework}
\label{sec:embed_CrossMRF}
A cross-MRF requires directly connected components $k^c$, $l$ to be symmetric and $s_j \in \mcalN(s_i)$. 

Therefore, linking the cross-MRF and the mixed spatial graphical model framework requires the realisation of the symmetry between component fields $Y_l(\cdot)$ and $Y_{k^c}(\cdot)$ (in which spatial locations are symmetric by nature already), and $s_j \in \mcalN(s_i)$ in the framework.

\subsection{Realising symmetry between component fields}

To link the cross-MRF theory and the framework, we need to transform the directed component graph in the first-stage model into an undirected graph via \textit{moralisation}.
That is connecting parent fields that share the same child field using undirected edges and dropping the arrows of the remaining edges, see \citet[p.~135]{koller2009probabilistic}.

This realises one of the features of the cross-MRF, namely, directly connected components $l, k^c$ are symmetric.

\subsection{Realising $s_j \in \mcalN(s_i)$}
\label{sec:realising_sjsi}
The undirected spatial graph within each component field in the mixed spatial graph (see Fig. \ref{fig:mixed_sp_graph}) reminds us to
use the conditional autoregressive
(CAR) formulation \citep[Section 4.2]{besag1974spatial}
to model the \textit{inverse} of the univariate
conditional covariance matrices  $\bmSIGMA_{11}^{-1}$, $\bmD_{22}^{-1}$, $\ldots, \bmD_{pp}^{-1}$, each of dimension $n \times n$. 
See the updated Algorithm \ref{alg:my-algorithm_update} in Supplementary Material \ref{app:updated_algo}. 

The advantage is that it further reduces the $\HSigma_{np \times np}^{-1}$ generation complexity to $ \mcalO(pn^2)$, due to the banded structure in $\bmSIGMA_{11}^{-1}$, $\bmD_{22}^{-1}$, $\ldots, \bmD_{pp}^{-1}$, induced by (univariate) MRF from the CAR. 
This, in turn, reduces the overall 
generation time (see Table \ref{tab:elapsed_time} in Section \ref{sec:1D_compare_study}). 

$s_j \in \mcalN(s_i)$ can also be realised by tapering a univariate \matern{} covariance function, as in \citet{furrer2006covariance}. Simulation results in 
Supplementary Material \ref{app:car_vs_matern} shows that the univariate CAR strategy achieves a faster generation time than the tapering \matern{}.

\section{1D Comparative Study}
\label{sec:1D_compare_study}
To compare different CI structures and conditional modelling strategies, 
we conduct simulations under two sizes of one-dimensional spatial domain with a grid spacing of 0.05. One is $\mathcal{D}_1 = [-10, 10]$, totalling 400 locations per component field ($n = 400$), and 
the other is $\mathcal{D}_2 = [-15, 15]$, totalling 600 locations per component field ($n = 600$).

\begin{figure}[htbp]
    \centering
    \begin{subfigure}[b]{0.49\textwidth}
        \centering
        \caption{Original structure}
    \includegraphics[width=0.55\textwidth]{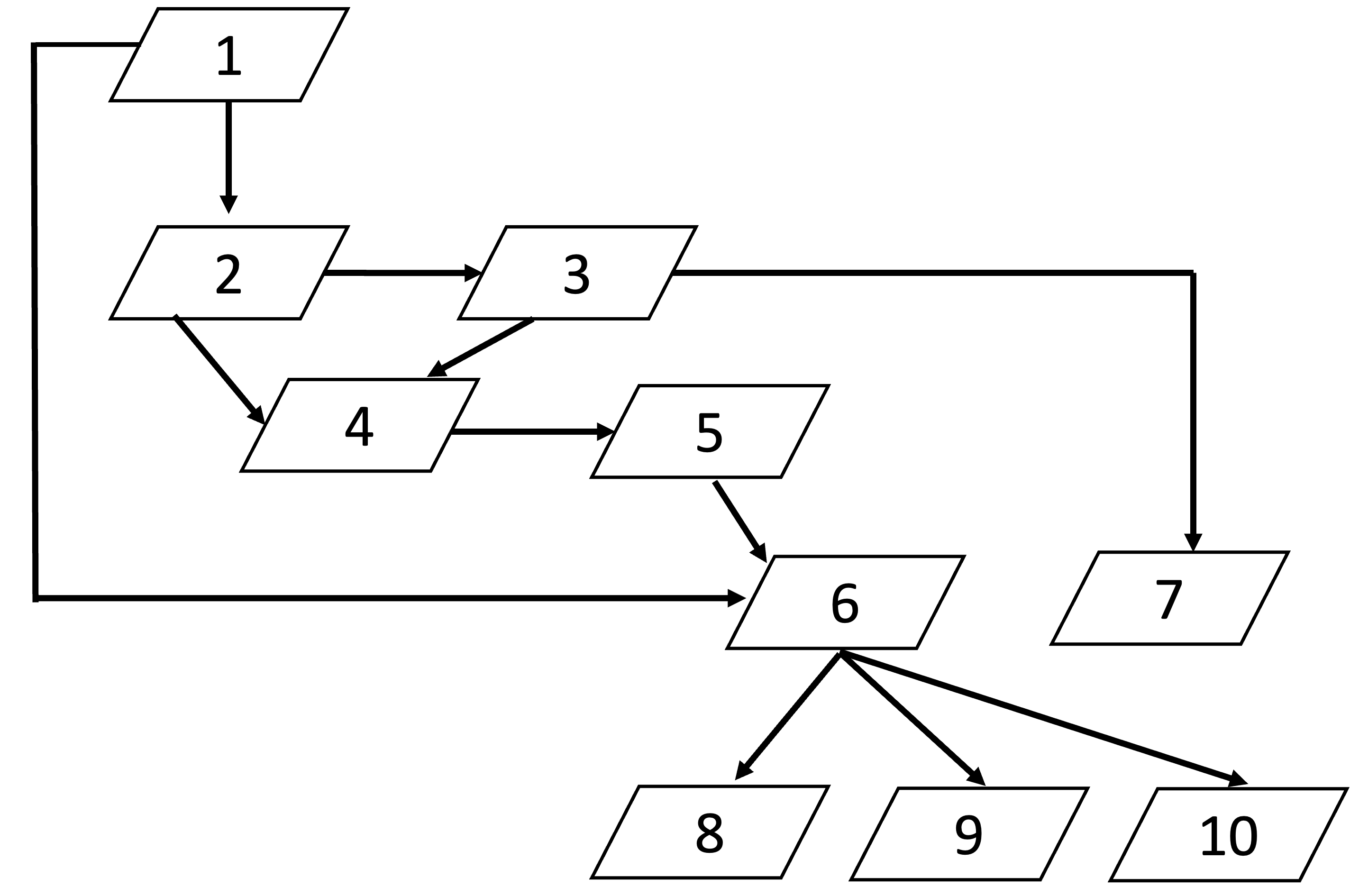}
        \label{fig: original_ten}
    \end{subfigure}
    \hfill
    \begin{subfigure}[b]{0.49\textwidth}
        \centering
        \caption{Moralised structure}
        \includegraphics[width=0.55\textwidth]{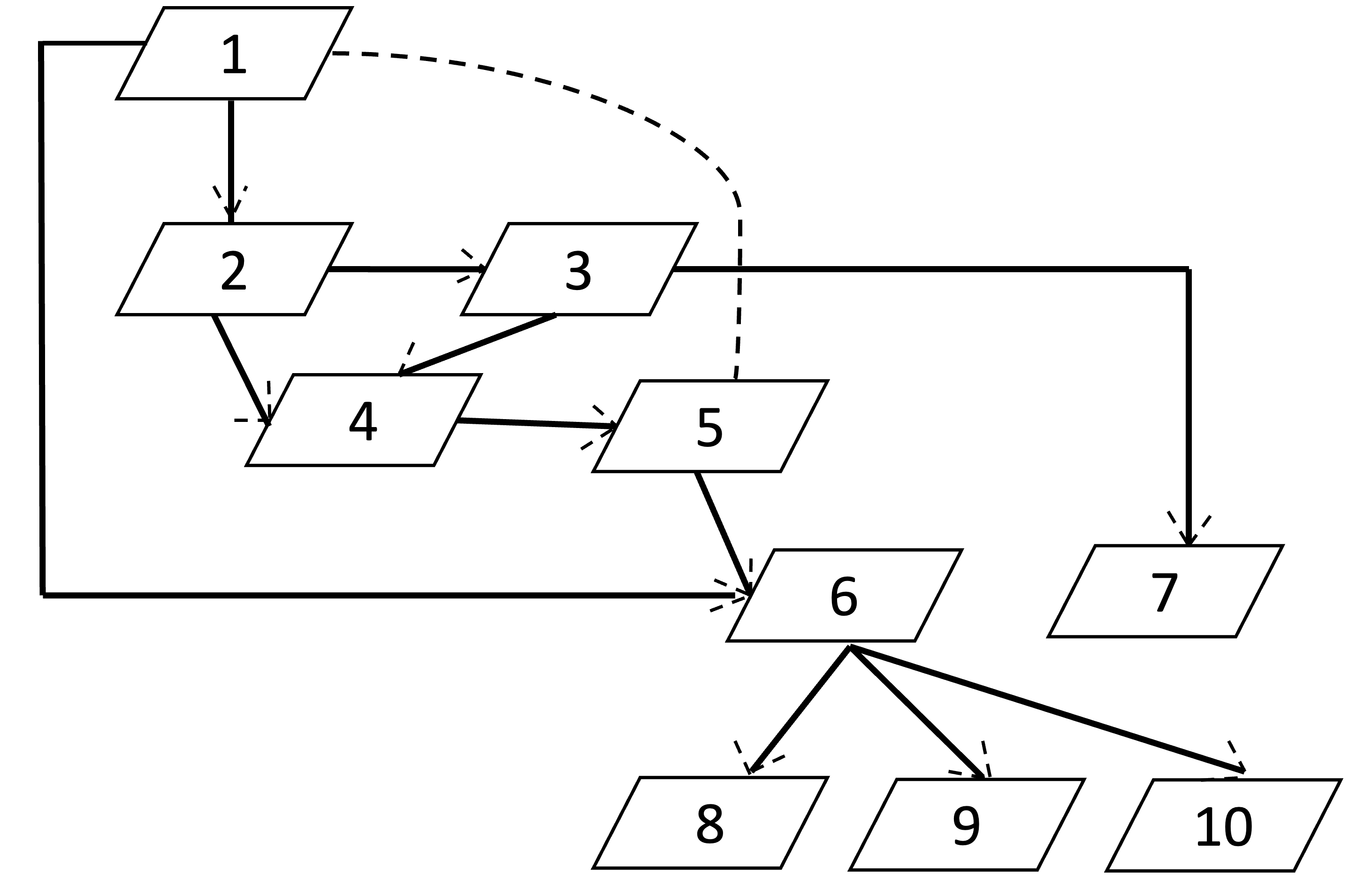}
        \label{fig:moral_ten}
    \end{subfigure}
    \caption{The left figure (a) is the original graph structure for ten component fields, while the right one (b) displays the corresponding moralised structure. The dotted line indicates the moralisation, or ``marriage", of parent fields (1 and 5) that share a common child field (6). 
    The dotted arrows indicate the directed edges that become undirected after moralisation.}
\end{figure}

The interactions among ten component fields ($p = 10$)
were specified by the DAG shown in Fig. \ref{fig: original_ten}; the corresponding moralised graph is shown in Fig. \ref{fig:moral_ten}.


The doubly CI among $p$ components and $n$ locations is realised by combining the mixed spatial graphical model framework with cross-MRF theory introduced in Section \ref{sec:embed_CrossMRF}. 
For convenience, we call this a \textit{cross-conditional} strategy.
The \textit{inverse} of the univariate conditional covariance matrices are all modelled using CAR models with an equally weighted lag-3 neighbourhood structure, i.e., $\mcalN(s_i) = \{s_j: j \pm 1, j \pm2, j \pm 3 \}$. Thus, the same-location and same-component auto-neighbourhood and the cross-neighbourhood of a certain point, for example $Y_2(s_i)$, are $\{Y_1(s_i), Y_3(s_i), Y_4(s_i) \}$, $\{Y_2(s_j): j\pm 1, j \pm 2, j \pm 3, j \neq i \}$, and $\{Y_1(s_j), Y_3(s_j), Y_4(s_j): j\pm 1, j \pm 2, j \pm 3, j \neq i \}$, respectively.


The CI among $p$ components only is the first-stage model, where each univariate conditional covariance matrix is modelled using special Mat\'{e}rn ($\nu = 3/2$).

We also adopt a lag-3 neighbourhood structure for constructing $\HSigma_{np \times np}^{-1}$ using row-wise conditional 
\citep{mardia1988multi}, which considers CI among $n$ locations only. 

\begin{figure}[ht]
    \centering
    \includegraphics[width=0.65\textwidth]{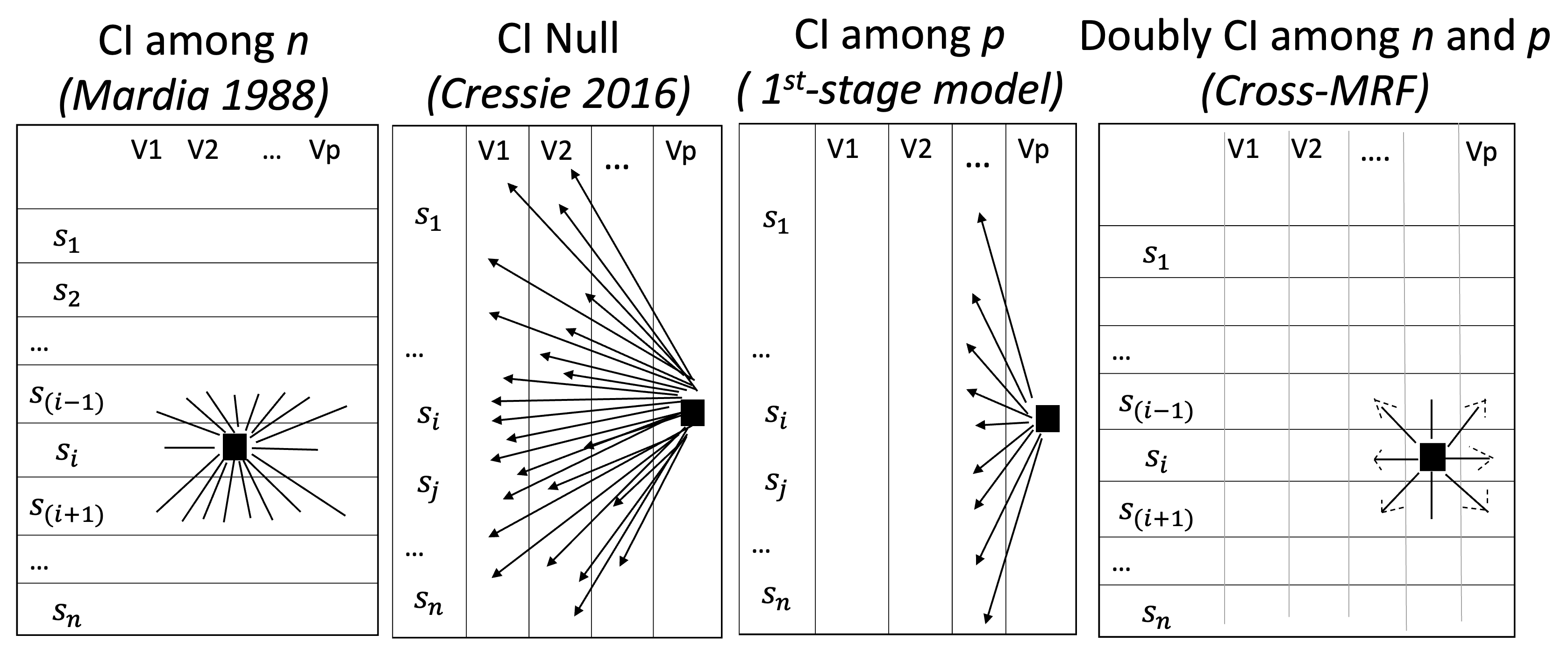}  
    \caption{The description for the first two CI scenarios see Fig. \ref{fig:row-col}.
    CI among $p$ only conditions on the conditionally dependent component fields across the whole spatial domain. Cross-MRF conditions on conditionally dependent components across a neighbourhood spatial domain.
    ``$\mid$'' reflects same-component auto-conditionals, the horizontal ``---'' reflects same-location auto-conditionals, and the diagonal ``---'' reflect cross-conditionals. Dotted arrows indicate those arrows that are dropped after moralisation.}
    \label{fig:scheme_4CI}
\end{figure}

Figure \ref{fig:scheme_4CI} displays the schematic representation of the various CI structures.

Table~\ref{tab:generation_order_compare} compares the generation complexity of $\HSigma_{np \times np}^{-1}$ under four different CI scenarios.
\begin{table}[ht]
    \centering
    \begin{threeparttable}
        \caption{$\HSigma_{np \times np}^{-1}$ Generation Complexity Under Different Conditional Independence (CI) Scenarios} 
    \label{tab:generation_order_compare}
    \begin{tabular}{ccccc}
    \toprule
     &\begin{tabular}[c]{@{}l@{}}  \quad CI among $n$ \\ \small{\citep{mardia1988multi}} \end{tabular} 
    &\begin{tabular}[c]{@{}l@{}}  \quad CI Null \\ \small{(Cressie, 2016)} \end{tabular} 
    &\begin{tabular}[c]{@{}l@{}} \quad CI among $p$ \\ \small{(1st-stage model)} \end{tabular} & \begin{tabular}[c]{@{}l@{}} \quad Doubly CI among $n$, $p$ \\ \quad \small{(Cross conditional)}  \end{tabular}  \\ \hline
    &$\mcalO(p^3 n^2)^{\dag}$ & $\mcalO(p^3n^3)^{\ddag}$ & $\mcalO(pn^3)$ & $\mcalO(pn^2)$   \\
    \bottomrule
    \end{tabular}
    \begin{tablenotes}
        \begin{scriptsize} \item[\dag] By $\HSigma^{-1} = \{ \mbox{block diag}(\mathbf{\Gamma}_i)^{-1} \} \{ \mbox{block} (- \bfbeta_{ij}) \}$ \citep{mardia1988multi}  \end{scriptsize}
        \begin{scriptsize} \item[\ddag] By Cholesky inversion. \end{scriptsize}
    \end{tablenotes}
    \end{threeparttable}
\end{table}

The reduced $\HSigma_{np \times np}^{-1}$ generation order from $\mcalO(pn^3)$ (CI among $p$ only) to $\mcalO(pn^2)$ (doubly CI) is reflected in the elapsed wall time on a local machine (macOS Sonoma V14.6, 16 GB).
Table \ref{tab:elapsed_time} shows the elapsed wall time for the generation of $\HSigma^{-1}_{np \times np}$ and $\HSigma_{np \times np}$ step-wise in parallel under two CI scenarios (CI among $p$ and doubly CI). 
\begin{table}[ht]
    \centering
    \begin{threeparttable}
        \caption{Elapsed Wall Time of Step-wise Parallel Generation of $\HSigma_{np \times np}$ and $\HSigma_{np \times np}^{-1}$ Under Two CI Scenarios on Local Machine (Unit: seconds) \\ 
        \hspace*{\fill} (Number of components \(p = 10\); Spatial locations \(n = 600\); $np = 6000$) \hspace*{\fill}} 
    \label{tab:elapsed_time}
    \begin{tabular}{ccccc}
    \toprule
     &\begin{tabular}[c]{@{}l@{}}  \quad CI among $n$ \\ \small{\citep{mardia1988multi}} \end{tabular} 
    &\begin{tabular}[c]{@{}l@{}}  \quad CI Null \\ \small{(Cressie, 2016)} \end{tabular} 
    &\begin{tabular}[c]{@{}l@{}} \quad CI among $p$ \\ \small{(1st-stage model)} \end{tabular} & \begin{tabular}[c]{@{}l@{}} \quad Doubly CI among $n$, $p$ \\ \quad \small{(Cross conditional)}  \end{tabular}  \\ \hline
    &\begin{tabular}[c]{@{}l@{}}965.44 + 42.37 = 1007.814  $\dag$ \\ ($\approx$ 16.80 min)\end{tabular} & $-^{\ddag}$ & \begin{tabular}[c]{@{}l@{}}1077.513\\ ($\approx$ 17.58 min)\end{tabular} &  \begin{tabular}[c]{@{}l@{}}591.652\\ ($\approx$ 9.86 min) \end{tabular}  \\
    \bottomrule
    \end{tabular}
    \begin{tablenotes}
        \begin{scriptsize}
            \item[\dag] Only construct a single object $\HSigma_{np \times np}^{-1}$ or $\HSigma_{np \times np}$, lacking the rigour to make comparisons here.
            \item[$\ddag$] Exceeds laptop's available local memory capacity.
        \end{scriptsize}
    \end{tablenotes} 
    \end{threeparttable}
\end{table}

To reveal sparsity, 
Table~\ref{tab:exat_perct} presents the percentage of exact-zero entries in the joint precision matrix $\HSigma_{np \times np}^{-1}$ for ten component fields over the two sizes of the spatial domain ($n = 400, 600$).
\begin{table}[htbp]
    \centering
        \caption{Percentage of Exact-zero Entries in $\HSigma_{np \times np}^{-1}$ Under Different Conditional Independence (CI) Scenarios \\ 
        \hspace*{\fill} (Number of components \(p = 10\); Spatial locations \(n = 400, 600\)) \hspace*{\fill}}
    \label{tab:exat_perct}
    \begin{tabular}{ccccc}
    \toprule
    & \begin{tabular}[c]{@{}l@{}}  \quad CI among $n$ \\ \small{\citep{mardia1988multi}} \end{tabular} 
    & \begin{tabular}[c]{@{}l@{}}  \quad CI Null \\ \small{(Cressie, 2016)} \end{tabular} 
    & \begin{tabular}[c]{@{}l@{}} \quad CI among $p$ \\ \small{(1st-stage model)} \end{tabular} & \begin{tabular}[c]{@{}l@{}} \quad Doubly CI among $n$, $p$ \\ \quad \small{(Cross conditional)} \end{tabular}  \\ \hline
    $n = 400$  &89.36 \% & 0\% & 82.23 \% & 97.19 \%  \\
    $n = 600$   &92.77 \% & 0\% & 90.42 \% & 98.08 \%  \\
    \bottomrule
    \end{tabular}
\end{table}

Table~\ref{tab:order_compare} shows the computational cost of $\bmy^{T} \HSigma^{-1}_{np \times np} \bmy$.
$m $ is the number of non-zero locations after imposing the CI among spatial locations, $m \ll n$.
\begin{table}[htbp]
    \centering
        \caption{Computational Complexity of $\bmy^{T} \HSigma^{-1}_{np \times np} \bmy $ Under Different Conditional Independence (CI) Scenarios}
    \label{tab:order_compare}
    \begin{tabular}{ccccc}
    \toprule
    & \begin{tabular}[c]{@{}l@{}} \quad CI among $n$ \\ \small{\citep{mardia1988multi}}\end{tabular} 
    & \begin{tabular}[c]{@{}l@{}} \quad CI Null \\ \small{(Cressie, 2016)} \end{tabular} 
  & \begin{tabular}[c]{@{}l@{}} \quad CI among $p$ \\ \small{(1st-stage model)}  \end{tabular} & \begin{tabular}[c]{@{}l@{}} \quad Doubly CI among $n$, $p$ \\ \quad \small{(Cross conditional)} \end{tabular}  \\ \hline
    &$\mcalO(n p^2)$ & $\mcalO(n^2 p^2) $ & $\mcalO(pn^2)$ & $\mcalO(pmn)$  \\
    \bottomrule
    \end{tabular}
\end{table}

We end this section with Table \ref{tab:condi_mod_compare}, summarising three conditional modelling strategies. 
\begin{table}[htpb]
    \centering
        \caption{Comparison of Three Conditional Model Strategies}
    \label{tab:condi_mod_compare}
    \begin{tabular}{lcccc}
    & \multicolumn{3}{c}{} &  \\ \toprule
    & \begin{tabular}[c]{@{}l@{}}  Row-wise Condit' \\ \small{\citep{mardia1988multi}} \end{tabular} & \begin{tabular}[c]{@{}l@{}} Col-wise Condit' \\ \small{(Cressie, 2016)} \end{tabular} 
  & \begin{tabular}[c]{@{}l@{}} Cross Conditional \\ \small{(Cross-MRF)}  \end{tabular}  \\ \hline
    Joint $\HSigma$, $\HSigma^{-1}$ valid & \checkmark & \checkmark & \checkmark & \\
    Asymmetric cross-covariance & $\times$ & \checkmark & \checkmark & \\
    CI among $n$ locations & \checkmark & $\times$ & \checkmark & \\
    CI among $p$ components & $\times$ & $\times$ & \checkmark  &\\
    Joint $\HSigma^{-1}$ structural sparse  & \checkmark  & $\times$ & \checkmark & \\
    Obtain $\HSigma^{-1}$ in real-time   & \checkmark  & $\times$ & \checkmark &\\
    Obtain $\HSigma$, $\HSigma^{-1}$ together & $\times$ & $\times$ & \checkmark & \\
    \bottomrule
    \end{tabular}
\end{table}

\section{2D Illustration: CAMS Reanalysis Data}
\label{sec:2D_illustration}
We illustrate the step-wise parallel construction of $\HSigma_{np \times np}^{-1}$ and $\HSigma_{np \times np}$ under the cross-conditional modelling strategy using five PM2.5 components (BC, DU, OM, SS, SU) of the CAMS data across a two-dimensional spatial domain, where $p = 5$. 

\begin{figure}[ht]
    \centering
    \begin{subfigure}[b]{0.49\textwidth}
        \centering
        \caption{Moralised CAMS graph structure}
        \includegraphics[width=0.75\textwidth]{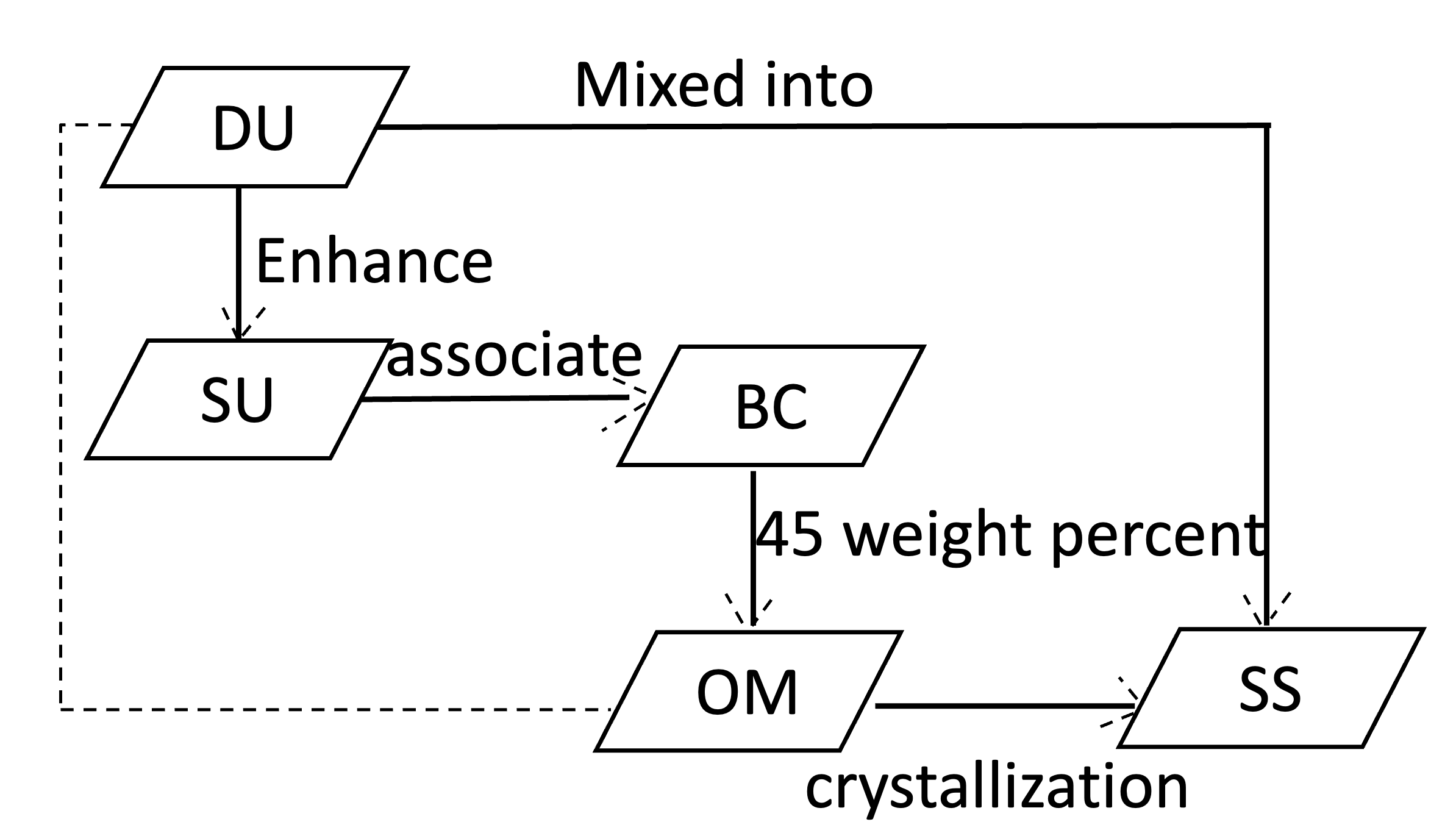}
        \label{fig:cams_moral}
    \end{subfigure}
    \hfill
    \begin{subfigure}[b]{0.49\textwidth}
        \centering
        \caption{Local CI structure of OM($s_i$)}
        \includegraphics[width=0.78\textwidth]{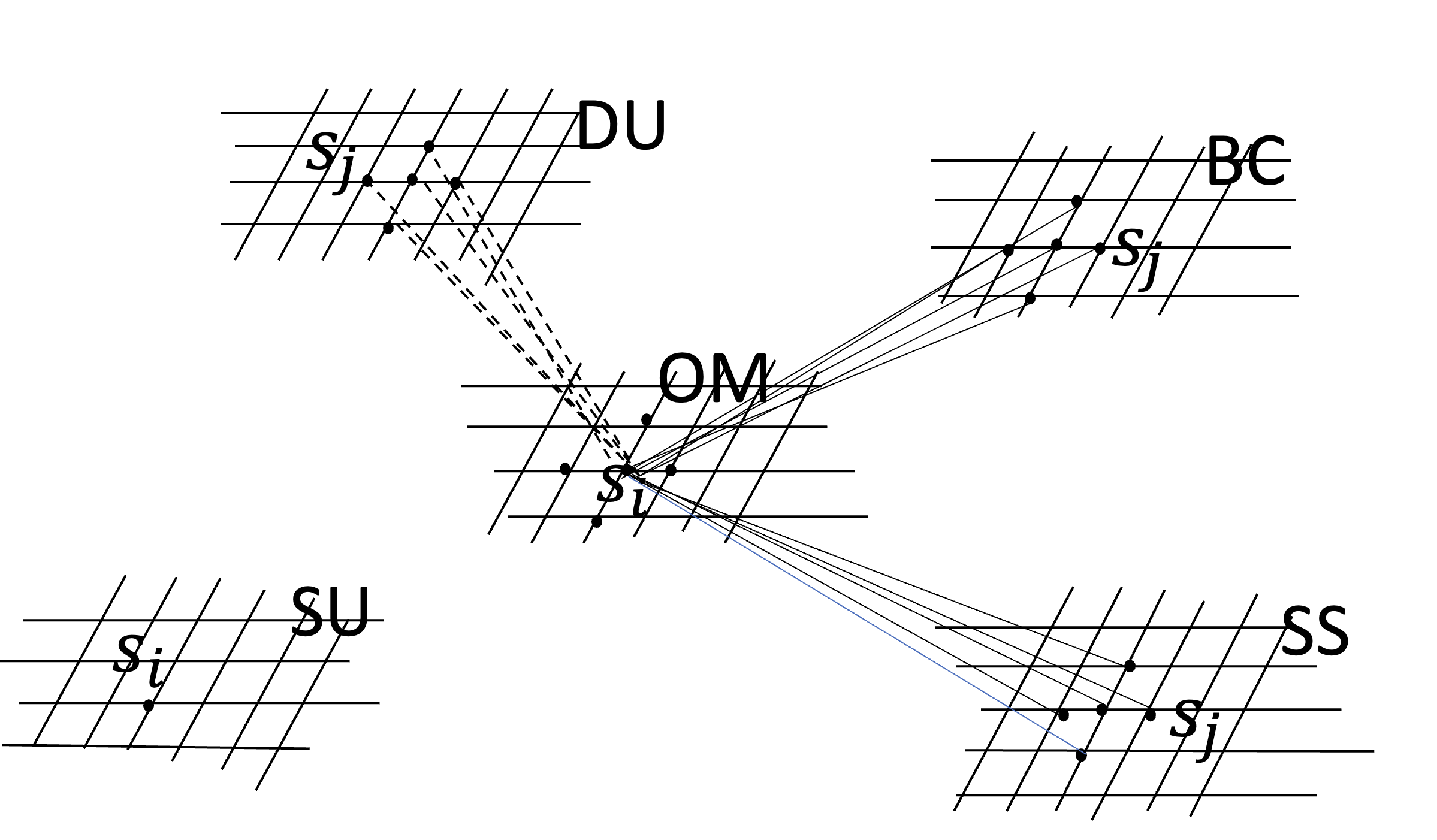}
        \label{fig:Local_CI_CAMS}
    \end{subfigure}
    \caption{Schematic representations of (a) moralised graph structure of five component fields of PM2.5. The dotted line indicates the moralisation of parent fields that share a common child field. The dotted arrows indicate the edges that become undirected after moralisation; (b) The same-location auto neighbourhood, same-component auto neighbourhood and cross-neighbourhood of OM($s_i$) in the mixed spatial graph after moralisation.}
    \label{fig:cams_fields_moral_app}
\end{figure}

Following the scientific evidence (Supplementary Material \ref{app:sec_evidence}), the moralised graph structure is shown in Fig. \ref{fig:cams_moral}. The local conditional independence structure of the component OM at location $s_i$ given its auto-/cross-neighbourhoods
is shown in Fig. \ref{fig:Local_CI_CAMS}.

The local conditional independence of OM($s_i$) indicates that conditional on the rest of the pollutant fields, OM($s_i$) will produce crystallisation only with SS at the same location $s_i$ and its neighbouring location $\mcalN(s_i)$; only values of BC at the same $s_i$ and $\mcalN(s_i)$ will contribute to 45 weight per cent of OM at $s_i$; only values of DU at $s_i$ and $\mcalN(s_i)$ will have some underlying physical processes associated with OM at $s_i$; only OM within $\mcalN(s_i)$
will associate with OM at $s_i$ during these physical processes; finally, OM($s_i$) is conditionally independent of the entire SU field and the rest of the pollutants at locations other than $s_i$ and $\mcalN(s_i)$.

Computationally, with the CAMS data spanning $n = 27384$ locations, the joint $\HSigma_{np \times np}^{-1}$ and $\HSigma_{np \times np}$ are of dimension $(27384*5)\times (27384*5)$, totalling 139.68GB in memory. 

The 80 GB memory limit of a single NVIDIA A100 GPU node on the Baskerville HPC necessitates domain segmentation. We divided the data into four
equal-width longitude strips, with the first strip containing $3793$ spatial locations.


Thus, the dimensions of $\HSigma_{np \times np}$ and $\HSigma_{np \times np}^{-1}$ are $(3793*5 = 18965)\times (3793*5 = 18965)$, corresponding to 359671225 entries in $\HSigma_{np \times np}$, capturing asymmetric cross-covariance in the off-diagonal blocks, while $\HSigma_{np \times np}^{-1}$ remains sparse.

To speed up the computation, large matrix multiplications in steps 24-30 in Algorithm \ref{alg:my-algorithm_update} were offloaded to GPU for the construction of $\HSigma^{-1}_{np \times np}$, while steps 5-23 for  $\HSigma_{np \times np}$ remained in the CPU to avoid communication overheads between CPU and GPU during iteration of \textit{for} loops.

\begin{figure}
    \centering
    \includegraphics[width=0.5\linewidth]{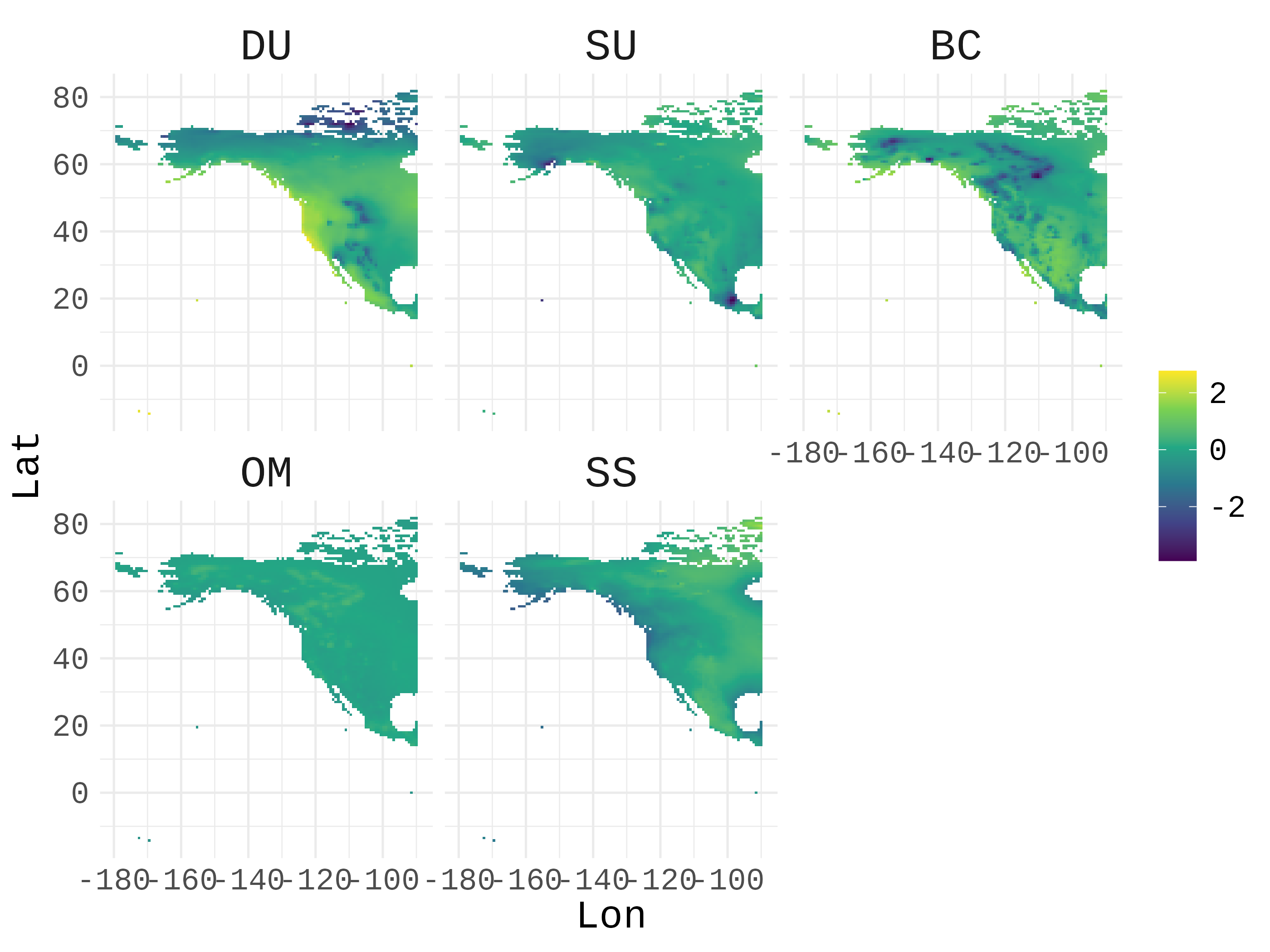}
    \caption{Jointly predicted residuals of true processes of five PM2.5 components (BC, DU, OM, SS, SU) of the CAMS data in the first longitude strip.}
    \label{fig:10}
\end{figure}

Figure \ref{fig:10} displays the jointly predicted residuals of true processes of five PM2.5 components (BC, DU, OM, SS, SU) of the CAMS data in the first longitude strip.

\section{Conclusions and Discussions}
\label{sec:conclude_discussion}
\subsection{Conclusions}
\label{sec:conclusions}
The proposed framework allows step-wise parallel generation of the joint precision $\HSigma_{np \times np}^{-1}$ and joint covariance matrix $\HSigma_{np \times np}$. 
This not only reduces the $\HSigma_{np \times np}^{-1}$ generation complexity (linear in $p$, see Table \ref{tab:generation_order_compare}), but also 
enables a simultaneous accommodation of asymmetric cross-covariance in the off-diagonal blocks of $\HSigma_{np \times np}$ and sparsity in $\HSigma_{np \times np}^{-1}$ (see Fig. \ref{fig:SG_SG_inv_SBS}).


Maximal sparsity in $\HSigma_{np \times np}^{-1}$ is achieved through two modelling stages. The first stage accommodates CI among the $p$ component fields, while
the second stage derives cross-MRF to establish the 
\textit{doubly} CI structure among both $p$ components and $n$ locations. 
This results in the sparsest possible representation of $\HSigma_{np \times np}^{-1}$, with the highest percentage of exact-zero entries (see Table \ref{tab:exat_perct}), leading to the Gaussian likelihood evaluation scaling linearly in $p$ and $n$ (see Table \ref{tab:order_compare}). It also further reduces the generation complexity of $\HSigma_{np \times np}^{-1}$ to $\mcalO(pn^2)$ (see Table \ref{tab:generation_order_compare}), significantly lowering computation time (see Table \ref{tab:elapsed_time}).
The scientific interpretability is also evident (see Sec.\ref{sec:2D_illustration}).

Hence, the challenges of the HMLS spatial data class (i.e., maximal sparsity in $\HSigma_{np \times np}^{-1}$, scalable generation efficiency of $\HSigma_{np \times np}^{-1}$, asymmetric cross-covariance in $\HSigma_{np \times np}$, and scientific interpretability) are addressed collectively in one unified framework. 

\subsection{Discussion}
\label{sec:discussion}


Computational memory storage remains a rigid constraint and a challenge. 
Most methods addressing the large $n$ large $p$ problem discussed in Section \ref{sec1} circumvent this by leaving the asymmetric cross-covariance in $\HSigma_{np \times np}$ unaddressed and focusing only on the sparse $\HSigma_{np \times np}^{-1}$. 
However, both \citet[Sec. 3.3]{cressie2016multivariate} and \citet[Sec.5]{Chen2025ICSTA150} have demonstrated the prominent difference in the prediction accuracy between the models with and without characterising the asymmetric cross-covariance in $\HSigma_{np \times np}$.

To overcome this memory constraint without neglecting asymmetry in $\HSigma_{np \times np}$, 
parallelisation schemes and associated inference methods that would allow spatially segmented data to be distributed and inferred across multiple GPU nodes, while accounting for spatial boundary effects, 
represent a natural next step. 
Distributed inference \citep{4102537, varshney2012distributed} that allows global inference to be performed on local subsets of the dataset is one possible option. The complication here is ensuring consistency of global estimates when local subset datasets are spatially correlated rather than independent and identical.

The proposed framework can be extended to geostatistical data, in which the locations $s_i$ are discretely sampled from a spatially continuous domain, by exploiting the link between Gaussian MRFs and spatially continuous Gaussian processes established in \citet{lindgren2011explicit}.

The doubly CI structure of the cross-MRF advances the CI structure in \citet{mardia1988multi} and \citet{cressie2016multivariate} by
extending the neighbourhood structure from 2D spatial planes to 3D (e.g., cross neighbourhood), 
allowing for scientific interpretability without violating the H-C theorem.
The further generalisation of the cross-MRF theory to a spatio-temporal setting can be another future work.

\section*{Author contributions statement}
XC developed the method and theory, implemented the code, and drafted the manuscript. PD contributed to methodology modification, manuscript writing and editing. JVZ provided broader insights and manuscript editing. GS supplied the dataset and theme lead.

\section*{Competing interests} 
No competing interest is declared.

\section*{Acknowledgments}
The first author thanks Noel Cressie for participation in designing the DAG in Fig.\ref{fig:ten_fields},
for many beneficial discussions,
and for his kind encouragement in exploring the cross-MRF. 

Thanks also extend to Daniel Falbel 
who solved the problem of the HPC apptainer, 
and to Turing's RSE team members 
who answered numerous questions regarding the HPC.

The first author was supported by the Alan Turing Institute under EPSRC grant EP/N510129/1. 
The computations described in this research were performed on Baskerville Tier 2 HPC service (https://www.baskerville.ac.uk/), 
funded by the EPSRC and UKRI under EP/T022221/1 and EP/W032244/1.

\bibliography{main}
\clearpage

\bigskip

\pagenumbering{arabic} 
\setcounter{page}{1}   

\begin{center}
{\large\bf Supplementary Materials For Highly Multivariate Large-scale Spatial Stochastic Processes -- A Cross-Markov Random Field Approach}
\end{center}

\begin{center}
\author{Xiaoqing Chen \hspace{.2cm}\\
    Department of Mathematics and Statistics, University of Exeter, Exeter EX4 4PY,  U.K.\\
    Peter Diggle \\ CHICAS, Lancaster Medical School, Lancaster University, Lancaster, LA1 4YB, U.K. \\
    James V. Zidek \\
    Department of Statistics, University of British Columbia, Vancouver, BC V6T 1Z4, Canada\\
    Gavin Shaddick \\
    College of Physical Sciences and Engineering, Cardiff University, Cardiff, CF10 3AT, U.K.
    }
\end{center}

\begin{appendices}

\section{Scientific Evidence}
\label{app:sec_evidence}
\citet{acp-10-365-2010} shows that DU enhances the mass concentration of coarse SU by more than an order of magnitude; BC is the result of incomplete combustion of fossil fuels, which is usually associated with SU \citep{NOAA_SOSSite}; 45 to 55 weight per cent of OM is carbon \citep{NOM_Website}; OM tends to mix into SS and influences the final result of SS during crystallisation processes \citep{Silva2014AnalysisOT}; and DU frequently becomes mixed into SS during their transport in the marine boundary layer \citep{ZHANGSSDU}.

\section{Proof of Theorem \ref{THRM:INDUCTION}}
\label{app:proof_induction}

\subsection{Lemma \ref{b_ll}}
\begin{lemma}
\label{b_ll}
    $b_{ll}(s_i, s_i) = 0$ due to no self-node regression.
\end{lemma}

\subsection{Lemma \ref{lemma:b=UU}}
\begin{lemma}
\label{lemma:b=UU}
    When $l = \{ p\}, k = \{1, \cdots, (p-1) \}$, $k^c$ is an arbitrary element of $k$ (i.e., $k^c \in k$),
    $\bmU_{lk} \bmU_{kk}^{-1} = \bmB_{lk}$.
\end{lemma}

\begin{proof}
    Without loss of generality, we set $p = 5$ and assume a directed chain structure among p components. 
$\bmB = [b_{lk^c}]$, where $b_{lk^c} = 0$ if $k^c \notin Pa(l)$, and $b_{ll} = 0$ as there's no self-node regression. 

So, $\bmB$ is 
$
\left[ \begin{array}{ccccc}
0 &  & & &  \\ b_{21} & 0 & & & \\ & b_{32} & 0 & &  \\ & & b_{43} & 0 &  \\ & & & b_{54} & 0 \end{array} \right],
$
then $(\bm{I} - \bmB)$ is 
$
\left[ \begin{array}{ccccc}
1 &  & & &  \\ -b_{21} & 1 & & & \\ & -b_{32} & 1 & &  \\ & & -b_{43} & 1 &  \\ & & & -b_{54} & 1 \end{array} \right], 
$
and by the formula of the inverse of the lower bi-diagonal matrix given in  \citet{kilicc2013inverse}, 

$
(\bm{I} - \bmB)^{-1} \triangleq \bmU = 
\left[ \begin{array}{ccccc}
1 &  & & &  \\ b_{21} & 1 & & & \\ b_{32}b_{21}& b_{32} & 1 &  &  \\ b_{43}b_{32}b_{21} & b_{43}b_{32} & b_{43} & 1 &  \\ b_{54}b_{43}b_{32}b_{21} & b_{54}b_{43}b_{32} & b_{54}b_{43} & b_{54} & 1 \end{array} \right].
$

$\bmU_{lk} = \bmU_{\{5\}\{ 1,2,3,4\} } =
\left[ \begin{array}{cccc}
    b_{54}b_{43}b_{32}b_{21} & b_{54}b_{43}b_{32} & b_{54}b_{43} & b_{54}
\end{array} \right]$

$\bmU_{kk}^{-1} = \bmU_{ \{1,2,3,4 \}\{1,2,3,4 \} }^{-1} = \left[ \begin{array}{cccc}
   1  & & &  \\
    -b_{21} & 1 & &  \\ 
    &  -b_{32} & 1 &  \\
    & & -b_{43} & 1 \\
\end{array} \right],
$
therefore, $ \bmU_{lk} \bmU_{kk}^{-1} = \left[ \begin{array}{cccc}
    0 & 0 & 0 & b_{54} 
\end{array} \right] = \bmB_{ \{ 5\} \{ 1,2,3,4 \}  } = \bmB_{lk}$, which is right the 5th row and the first four columns of matrix $\bmB$.
\end{proof}

\subsection{Lemma \ref{lemma:general=univariate}}
When $k = \{1, 2, \ldots, r \}$, $l = \{(r+1), \ldots, p \}$, \citet[Appendix.~B, Theorem 7]{shachter1989gaussian} states a general form for non-spatial multivariate multiple regression, as below
\begin{align*}
    \bmY_l = \bmmu_l + \bmU_{lk} \bmE_{k} + \bmU_{ll} \bmE_{l},
\end{align*}
in which $\bmU_{lk} \triangleq (\bm{I} - \bmB)^{-1}$, $\bmE \triangleq \bmS \bmZ$, 
$\bmS = \bmD^{1/2}$. 

\begin{lemma}
\label{lemma:general=univariate} 
\begin{align}
\label{eq:reg_nonsp}
    \bmY_l = \bmmu_l + \bmU_{lk} \bmE_{k} + \bmU_{ll} \bmE_{l},
\end{align} 
is the general multivariate multiple regression form of 
\begin{align}
    \bmY_l = \bmmu_l + \bmB_{lk} (\bmY_k - \bmmu_k) + \bmE_l
\end{align}
\end{lemma}

\begin{proof}
We know $\bmS \bmZ \triangleq  \bmE  = (\bm{I} - \bmB)(\bmY - \bmmu)$, 
so, 
\begin{align}
    \bmE_k &= \bmU_{kk}^{-1}(\bmY_k - \bmmu_k) \\
    \bmU_{lk}\bmE_k &= \bmU_{lk} \bmU_{kk}^{-1}(\bmY_k - \bmmu_k),
\end{align}
and $\bmU_{ll} = (\bm{I} - \bmB)_{ll}^{-1} = 1$.

So, re-write equation \eqref{eq:reg_nonsp} as 
\begin{align}
\label{eq:reg_general}
    \bmY_l = \bmmu_l + \bmU_{lk} \bmU_{kk}^{-1}(\bmY_k - \bmmu_k) + \bmE_l
\end{align}

By Lemma \ref{lemma:b=UU}, when $l = \{ p\}$, $k=\{1, 2, \ldots, (p-1) \}$,
equation \eqref{eq:reg_general} degenerates to 
one component (outcome) multiple regression equation: 
\begin{align*}
    Y_l = \mu_l + \bmB_{lk} (\bmY_k - \bmmu_k) + E_l
\end{align*}

\end{proof}

\subsection{Proof of Theorem \ref{THRM:INDUCTION}}
\begin{proof}
We start from the general equation \eqref{eq:reg_nonsp} and expand it to spatial settings.

Let the components $N = \{1, 2, \ldots, p \}$, and divide it into two parts, one is $k = \{1, 2, \ldots, r \}$, and the other is $l = \{(r+1), \ldots, p \}$

For any given location $s_i$, we have 
\begin{align*}
    \bmY_l(s_i) =  \sum_{j =1}^n \bmU_{lk}(s_i, s_j) \bmU^{-1}_{kk} (s_j, s_j)\bmY_k(s_j)  + \bmU_{ll}(s_i, s_i) \bmE_{l}(s_i)
\end{align*}

The covariance between a pair of locations $s_i$ and $s_j$ across all components is
    \begin{align*}
    \HSigma(s_i, s_j) &= cov(\bmY_{\cdot}(s_i), \bmY_{\cdot}(s_j) ) \\
    &= \mbbE \left[ 
        Var \left[  \begin{array}{cc}
        \bmY_k (s_i) | \bmY_k (\cdot)  \\
         \bmY_l (s_j) | \bmY_k (\cdot)
    \end{array}  \right] 
    \right]  +  
    Var \left[  \mbbE \left[ 
    \begin{array}{cc}
        \bmY_k(s_i) | \bmY_k(\cdot) \\
        \bmY_l(s_j) | \bmY_k(\cdot)
    \end{array} \right]   
    \right]
\end{align*}

in which, 
\begin{align*}
    \mbbE \left[ 
        Var \left[  \begin{array}{cc}
        \bmY_k (s_i) | \bmY_k (\cdot)  \\
         \bmY_l (s_j) | \bmY_k (\cdot)
    \end{array}  \right] 
    \right]   \nonumber \\ 
    &= \left[ \begin{array}{cc}
        \bmzero & \bmzero \\
        \bmzero & cov[(\bmY_l(s_j), \bmY_l(s_i) | \bmY_k(\cdot))]
    \end{array}  \right]   \nonumber  \\  
     &= \left[ \begin{array}{cc}
        \bmzero & \bmzero \\
        \bmzero & cov[\bmU_{ll}(s_j, s_j) \bmE_l(s_j), \bmU_{ll}(s_i, s_i) \bmE_l(s_i)]
    \end{array} \right]    \nonumber  \\
    &= \left[ \begin{array}{cc}
        \bmzero & \bmzero \\
        \bmzero & \bmU_{ll}(s_j, s_j) \bmD_{ll}(s_j, s_i) \bmU_{ll}^T(s_i, s_i)
    \end{array}
    \right] \\ 
    \mbox{(by $\bmE_l(s_i)$  $\triangleq$ $\bmS_{ll}(s_i, s_i)\bmZ_l(s_i)$ )},  
\end{align*}
and 
\begin{align*}
    \mbbE \left[ \begin{array}{c}
         \bmY_k(s_i) | \bmY_k(\cdot) \\
         \bmY_l(s_j) | \bmY_k(\cdot)
    \end{array}
    \right] 
    &= \mbbE \left[ \begin{array}{c}
         \bmY_k(s_i) \\
         \mbbE[\bmY_l(s_j) | \bmY_k(\cdot)]
    \end{array} \right] \\
    &= \left[ \begin{array}{c}
         \bmY_k(s_i) \\
         \sum_{m = 1}^n \bmU_{lk}(s_j, s_m) \bmU_{kk}^{-1} \bmY_k(s_m) 
    \end{array} \right] 
\end{align*}

$
Var \left[ \begin{array}{c}
         \bmY_k(s_i) \\
         \sum_{m=1}^n \bmU_{lk}(s_j, s_m) \bmU_{kk}^{-1} \bmY_k(s_m) 
    \end{array} \right] =  
$
\small
\begin{align*}
    \left[ \begin{array}{cc}
        \bmSIGMA_{kk}(s_i, s_j) & \sum_{m=1}^n \bmSIGMA_{kk}(s_i, s_m) (\bmU_{lk}(s_j, s_m) \bmU_{kk}^{-1}(s_m, s_m) )^T  \\
         \sum_{m=1}^n \bmU_{lk}(s_j, s_m) \bmU_{kk}^{-1}(s_m, s_m) \bmSIGMA_{kk}(s_m, s_i)  & 
         \sum_{q,m }  \bmU_{lk}(s_j, s_m) \bmU_{kk}^{-1}(s_m, s_m) \bmSIGMA_{kk}(s_m, s_q) (\bmU_{lk}(s_i, s_q) \bmU_{kk}^{-1} (s_q, s_q) )^T 
    \end{array} \right],    
\end{align*}
When $l = \{ p\}$, $k = \{1, 2, \ldots, (p-1) \}$, $k^c \in k$, by Lemma \ref{lemma:b=UU}
\begin{align*}
    \bmU_{lk}(s_j, s_m) \bmU_{kk}^{-1}(s_m, s_m)  \triangleq \bmB_{lk}(s_j, s_m) &= [b_{lk^c}(s_j, s_m)] \\
    \bmU_{lk}(s_i, s_q) \bmU_{kk}^{-1}(s_q, s_q) \triangleq \bmB_{lk}(s_i, s_q) &= [b_{lk^c}(s_i, s_q)],   
\end{align*}
where $b_{lk^c}(s_j, s_m) \neq 0$ if $k^c \in Pa(l)$ and $b_{ll}(s_j, s_m) = 0$ when $s_m = s_j$ (by Lemma \ref{b_ll}).
So,
\begin{align*}
    \HSigma(s_i, s_j) &=
    \left[ 
        \begin{array}{cc}
          \bmSIGMA_{kk}(s_i, s_j) &  \sum_{m=1}^n \bmSIGMA_{kk}(s_i, s_m) \bmB_{lk}^T(s_j, s_m)  \\
          \sum_{m=1}^n \bmB_{lk}(s_j, s_m) \bmSIGMA_{kk}(s_m, s_i)   & \sum_{q } \sum_{m} \bmB_{lk}(s_j, s_m) \bmSIGMA_{kk}(s_m, s_q) \bmB_{lk}^T(s_i, s_q) + \bmD_{ll}(s_j, s_i)
        \end{array}
    \right]   
\end{align*}
From one pair of location $(s_i, s_j)$ to all locations in $\mcalD$, we get
\begin{align*}
    \HSigma_{np \times np} =
    \left[ 
    \begin{array}{cc}
      \bmSIGMA_{kk}(\cdot, \cdot)  & \bmSIGMA_{kk}(\cdot, \cdot) \bmB^T_{lk}(\cdot, \cdot) \\
       \bmB_{lk}(\cdot, \cdot) \bmSIGMA_{kk} (\cdot, \cdot) & \bmB_{lk}(\cdot, \cdot) \bmSIGMA_{kk} (\cdot, \cdot) \bmB_{lk}^T(\cdot, \cdot) + \bmD_{ll} (\cdot, \cdot)
    \end{array}
    \right]    
\end{align*}

To obtain the updating formula, let $l = \{j+1 \}$, $k = \{1, 2, \ldots, j \}$, $k^c \in k$, 
\begin{scriptsize}
\begin{align*}
    \HSigma_{np \times np} \triangleq 
    \HSigma_{(j+1)n \times (j+1)n} = 
    \left[ 
    \begin{array}{cc}
       \bmSIGMA_{\{1, \ldots, j\}\{1, \ldots, j\}}(\cdot, \cdot)  & \bmSIGMA_{\{1, \ldots, j\}\{1, \ldots, j\}}(\cdot, \cdot) \bmB_{\{j+1\}\{1, \ldots, j\}}^T (\cdot, \cdot) \\
       \bmB_{\{j+1\}\{1, \ldots, j\}}\bmSIGMA_{\{1, \ldots, j\}\{1, \ldots, j\}}(\cdot, \cdot)  & \bmB_{\{j+1\}\{1, \ldots, j\}}\bmSIGMA_{\{1, \ldots, j\}\{1, \ldots, j\}} \bmB_{\{j+1\}\{1, \ldots, j\}}^T + \bmD_{\{j+1\}\{j+1\}}(\cdot, \cdot)
    \end{array}
    \right], 
\end{align*} 
\end{scriptsize}
where $\bmB_{(j+1)k^c} \neq \bm{0}$ if $k^c \in Pa(j+1)$.
\end{proof}

\section{Proof of Proposition 1}
\label{app:proof_prop1}
\begin{proof}
Theorem \ref{THRM:INDUCTION} implies that
$\HSigma_{np \times np}$ of any size can always be divided into four different blocks: the leading diagonal blocks SG,  the row block $\bmR$ beneath SG, the column block $\bmC$ to the right of SG, and the bottom-right block $\bmSIGMA_{rr}$. That is,
\begin{scriptsize}
$   \left[ \begin{array}{cc} 
  \left[SG\right]   & \left[ \bmC \right] \\
  \left[ \bmR\right]   & \left[ \bmSIGMA_{rr} \right] 
\end{array} \right] $
\end{scriptsize}. 

By \citet[p.~244, Theorem 14.8.5; Corollary 14.8.6]{harville1998matrix}, when the leading diagonal block $\bmSIGMA_{\{1, \cdots, j \}\{1, \cdots, j \}} (\cdot,\cdot) \triangleq SG $ in the induction formula in Theorem \ref{THRM:INDUCTION} is positive definite, and the Schur Complement of SG, denoted as $\HSigma_{np \times np}/SG = \bmSIGMA_{\{ j+1\}\{ j+1\}} - \bmR SG^{-1} \bmC $, i.e., 
    \begin{small}
    \begin{align}
        \nonumber
        &=\bmB_{ \{ j+1\} \{1, \cdots, j \}} \bmSIGMA_{\{1, \cdots, j \}\{1, \cdots, j \}} \bmB_{ \{ j+1\}\{1, \cdots, j \}}^T  + \bmD_{ \{ j+1\}\{ j+1\}} \\ \nonumber
        &- \bmB_{ \{ j+1\} \{1, \cdots, j \}}\bmSIGMA_{\{1, \cdots, j \}\{1, \cdots, j \}} \bmSIGMA^{-1}_{\{1, \cdots, j \}\{1, \cdots, j \}} \bmSIGMA_{\{1, \cdots, j \}\{1, \cdots, j \}} \bmB_{ \{ j+1\} \{1, \cdots, j \}}^T \\ 
        &= \bmD_{ \{ j+1\}\{ j+1\}} \triangleq \bmD_{rr}
    \end{align}
    \end{small}
    is positive definite, then $\HSigma_{np \times np}$ is positive definite.
\end{proof}


\section{Proof of Theorem 2}
\label{app:proof_thrm2}
\begin{proof}
The induction formula in Theorem \ref{THRM:INDUCTION} also implies that $\HSigma_{np \times np}$ of any size can always be divided into four different blocks: the leading diagonal blocks SG,  the row block $\bmR$ beneath SG, the column block $\bmC$ to the right of SG, and the bottom-right block $\bmSIGMA_{rr}$, 
that is,
\begin{scriptsize}
$   \left[ \begin{array}{cc} 
  \left[SG\right]   & \left[ \bmC \right] \\
  \left[ \bmR\right]   & \left[ \bmSIGMA_{rr} \right] 
\end{array} \right] $
\end{scriptsize}. 
For instance, 
when j = 2, $r = j+1 = 3$, \vspace{-7mm}
\begin{scriptsize}
\begin{align*}
    \HSigma_{3n \times 3n} = 
    \left[ \begin{array}{cc} 
  \left[SG\right]   & \left[ \bmC \right] \\
  \left[ \bmR\right]   & \left[ \bmSIGMA_{rr} \right] 
\end{array}  \right]
= \left[ \begin{array}{cc} 
  \left[SG = \bmSIGMA_{2n \times 2n} \right]   & \left[ \bmC \right] \\
  \left[ \bmR\right]   & \left[ \bmSIGMA_{33} \right] 
\end{array} \right] 
= \left[ \begin{array}{cc} 
    \left[\begin{array}{cc}
    \bmSIGMA_{11}   &  \bmSIGMA_{12}  \\
    \bmSIGMA_{21}  &  \bmSIGMA_{22} 
    \end{array} \right]
     & \left[ \begin{array}{c}
         \bmSIGMA_{13}  \\
         \bmSIGMA_{23}   
     \end{array} \right] \\
   \left[ \begin{array}{cc}
    \bmSIGMA_{31} & \bmSIGMA_{32} 
   \end{array} \right] & \left[ \bmSIGMA_{33} = \blacksquare + \bmD_{33} \right] 
\end{array} \right],
\end{align*}  
\end{scriptsize}
where the ``$\blacksquare$" denotes the $\bmB_{ \{ j+1\} \{1, \cdots, j \}} \bmSIGMA_{\{1, \cdots, j \}\{1, \cdots, j \}} \bmB_{ \{ j+1\}\{1, \cdots, j \}}^T$ in Theorem \ref{THRM:INDUCTION}.

    By \citet[p.~25, equation (0.8.5.6)]{horn2012matrix}, for a given $\HSigma_{np \times np}$, when the inverse of its leading diagonal block SG, i.e., $SG^{-1}$, and the inverse of the Schur complement of $SG$, i.e., $(\HSigma_{np \times np}/SG)^{-1} = \bmD_{rr}^{-1}$ (see the proof in Proposition \ref{prop:pd}) are provided, the $\HSigma^{-1}_{np \times np}$ can be obtained using formula 
    \begin{small}
    \begin{align}
    \label{eq:SIGMA_inv}
        \HSigma^{-1}_{np \times np}  =
        \left[ \begin{array}{cc}
           SG^{-1}[SG + \bmC \bmD_{rr}^{-1} \bmR]SG^{-1}  & -SG^{-1} \bmC \bmD_{rr}^{-1}  \\
           - \bmD_{rr}^{-1} \bmR SG^{-1}  &  \bmD_{rr}^{-1}
        \end{array}  \right].
    \end{align}     
    \end{small}

So, the joint precision matrix at the $r^{th}$ step, denoted as $\HSigma^{-1}_{rn \times rn}$, can be obtained using 
\begin{small}
    \begin{align}
    \label{eq:SIGMA_inv_r}
        \HSigma^{-1}_{rn \times rn} = 
        \left[ \begin{array}{cc}
           \Sigma_{(r-1)n \times (r-1)n}^{-1}[\Sigma_{(r-1)n \times (r-1)n} + \bmC \bmD_{rr}^{-1} \bmR]\Sigma_{(r-1)n \times (r-1)n}^{-1}  & -\Sigma_{(r-1)n \times (r-1)n}^{-1} \bmC \bmD_{rr}^{-1}  \\
           - \bmD_{rr}^{-1} \bmR \Sigma_{(r-1)n \times (r-1)n}^{-1}  &  \bmD_{rr}^{-1}
        \end{array}  \right].
    \end{align}     
    \end{small}

Therefore, when $j = 1$, $r = j + 1 = 2$,
\begin{small}
\begin{align*}
      \HSigma_{2n \times 2n}^{-1} = \left[ 
\begin{array}{cc}
    \bmSIGMA_{11}^{-1}[\bmSIGMA_{11} + \bmC \bmD_{22}^{-1} \bmR]\bmSIGMA_{11}^{-1} & -\bmSIGMA_{11}^{-1} \bmC \bmD_{22}^{-1} \\
    -\bmD_{22}^{-1} \bmR \bmSIGMA_{11}^{-1}  & \bmD_{22}^{-1} 
\end{array}
\right],  
\end{align*}   
\end{small}
where computing $\bmSIGMA_{11}^{-1}, \bmD_{22}^{-1}$ is $\mcalO(n^3)$;
when $j = 2$, $r = j + 1 = 3$, 
\begin{small}
\begin{align*}
      \HSigma_{3n \times 3n}^{-1} = \left[ 
\begin{array}{cc}
    \bmSIGMA_{2n\times 2n}^{-1}[\bmSIGMA_{2n \times 2n} + \bmC \bmD_{33}^{-1} \bmR]\bmSIGMA_{2n\times 2n}^{-1} & -\bmSIGMA_{2n\times 2n}^{-1} \bmC \bmD_{33}^{-1} \\
    -\bmD_{33}^{-1} \bmR \bmSIGMA_{2n\times 2n}^{-1}  & \bmD_{33}^{-1} 
\end{array}
\right],  
\end{align*} 
\end{small}
where computing $\bmD_{33}^{-1}$ is $\mcalO(n^3)$ and $\bmSIGMA_{2n \times 2n}^{-1}$ is already obtained from the last step, stored and can be easily fetched from memory cache; 
And when $j = (p - 1)$, $r = j+1 = p$,
\begin{small}
\begin{align}
\label{eq:SG_inv_p_iteration}
      \HSigma_{np \times np}^{-1} = \left[ 
\begin{array}{cc}
    \bmSIGMA_{(p-1)n \times (p-1)n}^{-1}[\bmSIGMA_{(p-1)n \times (p-1)n} + \bmC \bmD_{pp}^{-1} \bmR]\bmSIGMA_{(p-1)n \times (p-1)n}^{-1} & -\bmSIGMA_{(p-1)n \times (p-1)n}^{-1} \bmC \bmD_{pp}^{-1} \\
    -\bmD_{pp}^{-1} \bmR \bmSIGMA_{(p-1)n\times (p-1)n}^{-1}  & \bmD_{pp}^{-1} 
\end{array}\right],  
\end{align}   
\end{small}
where computing $\bmD_{pp}^{-1}$ is $\mcalO(n^3) $ and $\bmSIGMA_{(p-1)n\times (p-1)n}^{-1}$ is already obtained from the last step, stored and easily fetched from the memory cache.

Therefore, one only needs to compute $\bmSIGMA_{11}^{-1}$, $\bmD_{22}^{-1}$, $\ldots$, $\bmD_{pp}^{-1}$, and each of these $n \times n$ matrix inversion is $\mcalO(n^3)$, so the total computational complexity is $p * \mcalO(n^3)$, which is linear in $p$. 

Additionally, the construction of the joint precision matrix at step $r$, i.e., $\HSigma^{-1}_{rn \times rn}$ only relies on the joint covariance matrix at $(r-1)$ step, i.e. $\Sigma_{(r-1)n \times (r-1)n}$, not on the joint covariance matrix at $r$ step, therefore, the construction of joint precision matrix at step $r$ can be in parallel to the construction of joint covariance matrix at step $r$.
\end{proof}


\section{Proof of Proposition 2}
\label{app: proof_prop2}
\begin{proof}
       By \citet[Corollary 14.8.6]{harville1998matrix} and denote the $\HSigma^{-1}$ at step $r$ in the form of equation \eqref{eq:SIGMA_inv_r} as 
    \begin{scriptsize}
    $\left[ \begin{array}{cc}
        BK1 & BK2 \\
        BK3 & BK4
    \end{array} \right]$ 
    \end{scriptsize},
then $\HSigma^{-1}$ at step $r$ is PD iff
$BK4 = \bmD_{rr}^{-1}$ and its Schur Complement $\HSigma^{-1} / BK4 = BK1 - (BK2) (BK4^{-1}) (BK3) =
\HSigma^{-1}_{(r-1)n \times (r-1)n}$ are PD.
\end{proof}

 

\section{Algorithm 1}
\label{app:algo1}
\begin{small}
\begin{algorithm}
\caption{Algorithm for the generation of the desired $\Sigma$ and $\Sigma^{-1}$}
\label{alg:my-algorithm}
\KwData{The number of components $p$ and the data structure indicating parent and child relationship among $p$ component fields}
\KwResult{The desired $\HSigma_{np \times np}(\cdot, \cdot)$ and $\HSigma^{-1}_{np \times np}(\cdot, \cdot)$}

$\bmSIGMA_{11} \leftarrow $ known (e.g., exponential, \matern{}) \;
$\HSigma \leftarrow \bmSIGMA_{11}$; $n \leftarrow nrow(\bmSIGMA_{11})$ \;

\For{$r = 2$ \KwTo $p$}{
    PN = Pa(r) \;
    $\bmR = \bmC = NULL$ \;
    \For{$c = 1$ \KwTo $(r-1)$}{
        $\bmBT \leftarrow NULL$ \;
        $\bmSIGMA_{rc} \leftarrow \bm{0}$\;
        \For{t $\in$ PN} {
            $\bmB_{rt} \leftarrow f(h; \Delta_{rt}; A_{rt})$ \;
            $\bmBT \leftarrow rbind(\bmBT, \bmB_{rt}^T)$\;
            $\bmSIGMA_{rc} \leftarrow \bmSIGMA_{rc} + \bmB_{rt} \bmSIGMA[((t - 1)n + 1) : (tn), ((c-1)n + 1) :(cn)]$ \;   
        } 
        
        $\bmR \leftarrow cbind(\bmR, \bmSIGMA_{rc})$ \;
        $\bmSIGMA_{cr} \leftarrow \bmSIGMA_{rc}^T$ \;
        $\bmC \leftarrow rbind(\bmC, \bmSIGMA_{cr}) $ \;   
    }
    
    $\bmD_{rr} \leftarrow $ known (e.g., exponential, \matern{}) \;
    $\bmSIGMA_{rr} \leftarrow \bmR[, (t-1)n + 1 : (tn)]  \bmBT + \bmD_{rr}$ \;
    
    $SG \leftarrow \HSigma$  
    
    $Col \leftarrow rbind(\bmC, \bmSIGMA_{rr})$ \;
    $Row \leftarrow rbind(SG, \bmR)$ \;
    $\HSigma \leftarrow  cbind(Row, Col)$ \;  

    $\bmD_{rr}^{-1} \leftarrow $ Cholesky inversion ($\bmD_{rr}$)\;

    \If{ r== 2}{SG$^{-1} \leftarrow$ Cholesky inversion (SG)}
    $BK_1 \leftarrow SG^{-1}[SG + \bmC \bmD_{rr}^{-1} \bmR] SG^{-1}$ \;
    $BK_2 \leftarrow -SG^{-1} \bmC \bmD_{rr}^{-1}$ \;
    $BK_3 \leftarrow - \bmD_{rr}^{-1} \bmR SG^{-1}$ \;
    $BK_4 \leftarrow \bmD_{rr}^{-1}$ \;

    $\HSigma^{-1} \leftarrow rbind(cbind(BK1, BK2), cbind(BK3, BK4))$ \;
    $SG^{-1} \leftarrow \HSigma^{-1}$

    \If{r == p}{return $\HSigma$, $\HSigma^{-1}$ }
}
\end{algorithm}
\end{small}

\clearpage

\section{Details of 1D Simulation of First-stage Model}
\label{app:1dsimu}

The spatial domain $\mathcal{D}$ for this simulation is [-1, 1], with a grid size of 0.05. 

We use the special \matern{} function ($\nu$ = 3/2)
to model all the univariate conditional covariance $Cov((Y_l(s_i), Y_l(s_j) )| \bmY_{k^{sub}}(\cdot))$ for every single component $l$, 
and fix all the marginal variances $\sigma^2$ of each \matern{} 
to 1 and all the $\kappa$ to 2.

We propose a \textit{modified triangular wave} function 
with both positive and negative functional values for the $b_{rt}(\cdot, \cdot)$ to capture the cross-correlation.

The modified triangular wave has a functional form as follows, 
\begin{small}
\begin{align}
    Tri-Wave(h) = \left \{ \begin{array}{rcl}
A \{1 - \phi (\frac{/h - \Delta/}{/\Delta/})^2 \} & \mbox{for} & /h - \Delta/ \leq \rho /\Delta/ \\ 
0 & \mbox{for} & /h - \Delta/ >  \rho /\Delta/
\end{array}  
\right.,
\end{align}    
\end{small}
$A$ (amplitude) and $\Delta$ (horizontal translation) are two parameters obtained from inference, while $\phi$ and $\rho$ are two manually-set factors. 
$\phi$ controls the decay speed of the function value, and $\rho$ decides the compact support radius beyond which the function values are set to exact zero.

Asymmetric cross-correlation is accommodated via a shift parameter $\Delta$ \citep{li2011approach}
in the \textit{modified triangular wave} function, resulting in 
the $(i, j)^{th}$ element of the matrix $\bmB_{rt}$ not equal to its $(j, i)^{th}$ element.
Hence, the off-diagonal blocks of $\HSigma_{np \times np}$ are asymmetric.

Iterative $\HSigma_{np \times np}$ and $\HSigma^{-1}_{np \times np}$ construction, involving repeated multiplications of $\bmB_{rt}$, may amplify unavoidable tiny errors (e.g., rounding), leading to numerical instability if $\bmB_{rt}$ has a large spectral norm and/or condition number.

We apply spectral normalisation (SpN) and regularisation (Reg) to control 
$\bmB_{rt}$'s spectral norm and condition number, a technique from deep neural network training \citep{miyato2018spectral}. 
Test results confirm the robustness of the positive definiteness of $\HSigma_{np \times np}$ and $\HSigma^{-1}_{np \times np}$ using $\bmB_{rt}^{SpN + Reg}$. For the full test report, 
see Supplementary Material \ref{app:Experiemnt_repo}.



\section{Wendland function}
\label{app:Wendland}
\begin{figure}[htbp]
    \centering
    \includegraphics[width = 0.5\textwidth]{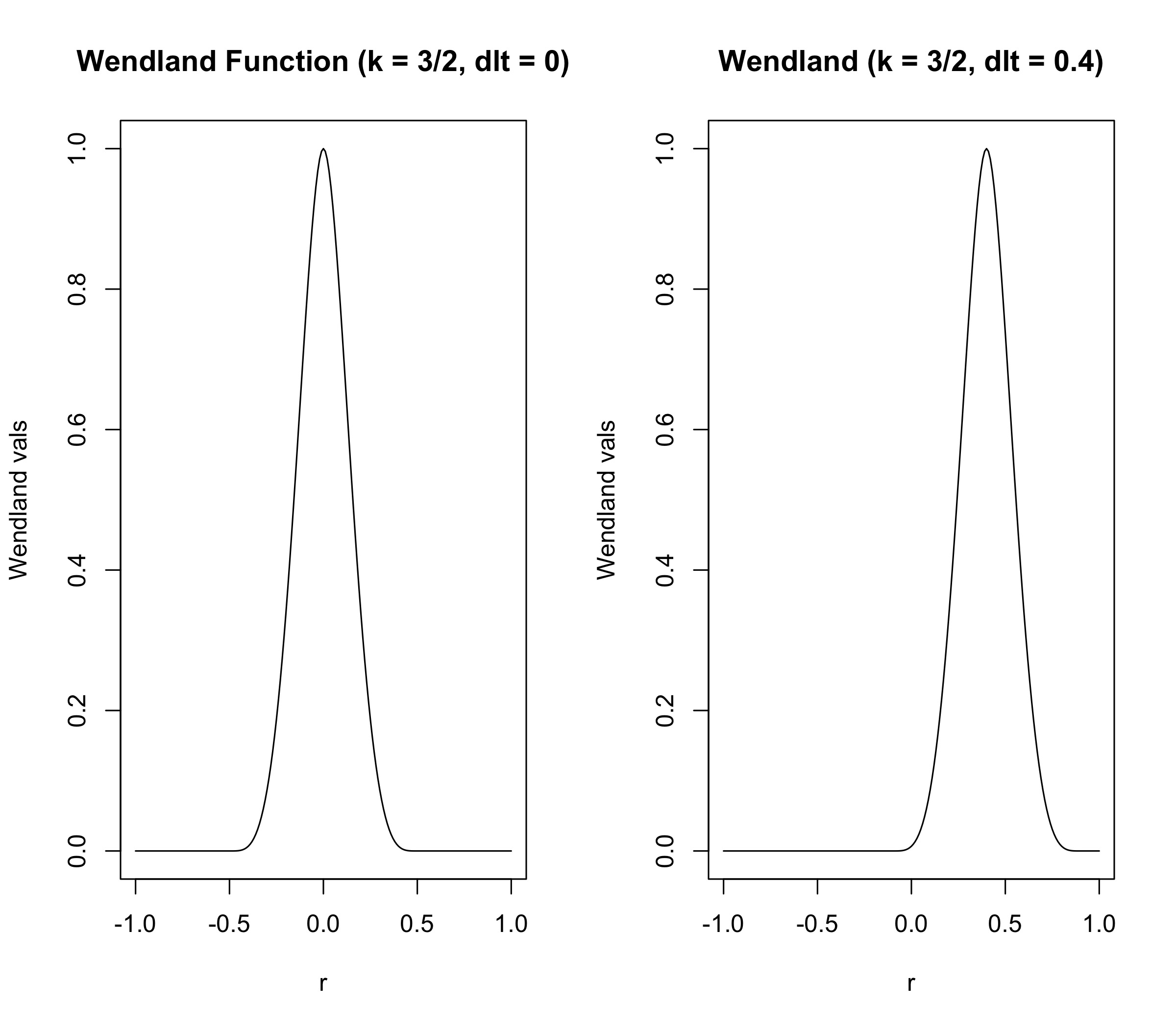}
    \caption{Wendland function where smoothness $k = 3/2$. The spatial domain is a sequence from -1 to 1 by 0.01, and the $R$ is set to 0.5. The left figure corresponds to translation $\Delta = 0$, and the right corresponds to $\Delta = 0.4$.}
    \label{fig:Wendland}
\end{figure}

Wendland function \citep{wendland1995piecewise}
\begin{align}
    Wendland_k(r) = \left \{ \begin{array}{rcl}
A \{1 - (\frac{/r - \Delta/}{R})^{2k + 1}(1 + (2k + 1) \frac{/r - \Delta/}{R}) \} 
& \mbox{for} & /h - \Delta/ \geq  R \\ 
0 & \mbox{for} & /h - \Delta/ >  R
\end{array}  
\right.,
\end{align}
in which $A$ (amplitude) and $\Delta$ (translation) are parameters obtained from inference, while $k$ and $R$ are manually set factors. $k$ controls the smoothness (continuously differentiability) of the function and is set to the same value as the smoothness $\nu$ in \matern{}, e.g., 3/2 for special \matern{}. $R$ is the compact support radius beyond which the function values are set to exact zero. 
If this Wendland function is used to model the covariance, it 
also embodies the meaning of effective range, at which the spatial correlation drops to a negligible 0.05, see \citet{furrer2006covariance}; For the shape of the Wendland function (k = 3/2), see below Figure \ref{fig:Wendland}.

\section{Test Report on the Robustness of the Positive Definiteness of $\Sigma$ and $\Sigma^{-1}$}
\label{app:Experiemnt_repo}
This section aims to test the robustness of the positive definiteness (PD) of $\HSigma$ and $\HSigma^{-1}$ constructed using Algorithm \ref{alg:my-algorithm} with original $\bmB_{rt}$ ($\bmB_{rt}^{Original}$) functions and $\bmB_{rt}$ undergone spectral normalisation together with regularisation ($\bmB_{rt}^{SpN + Reg}$). 

\begin{figure}[htbp]
  \centering
  \begin{subfigure}[b]{0.32\textwidth}
    \centering
    \includegraphics[width=\textwidth]{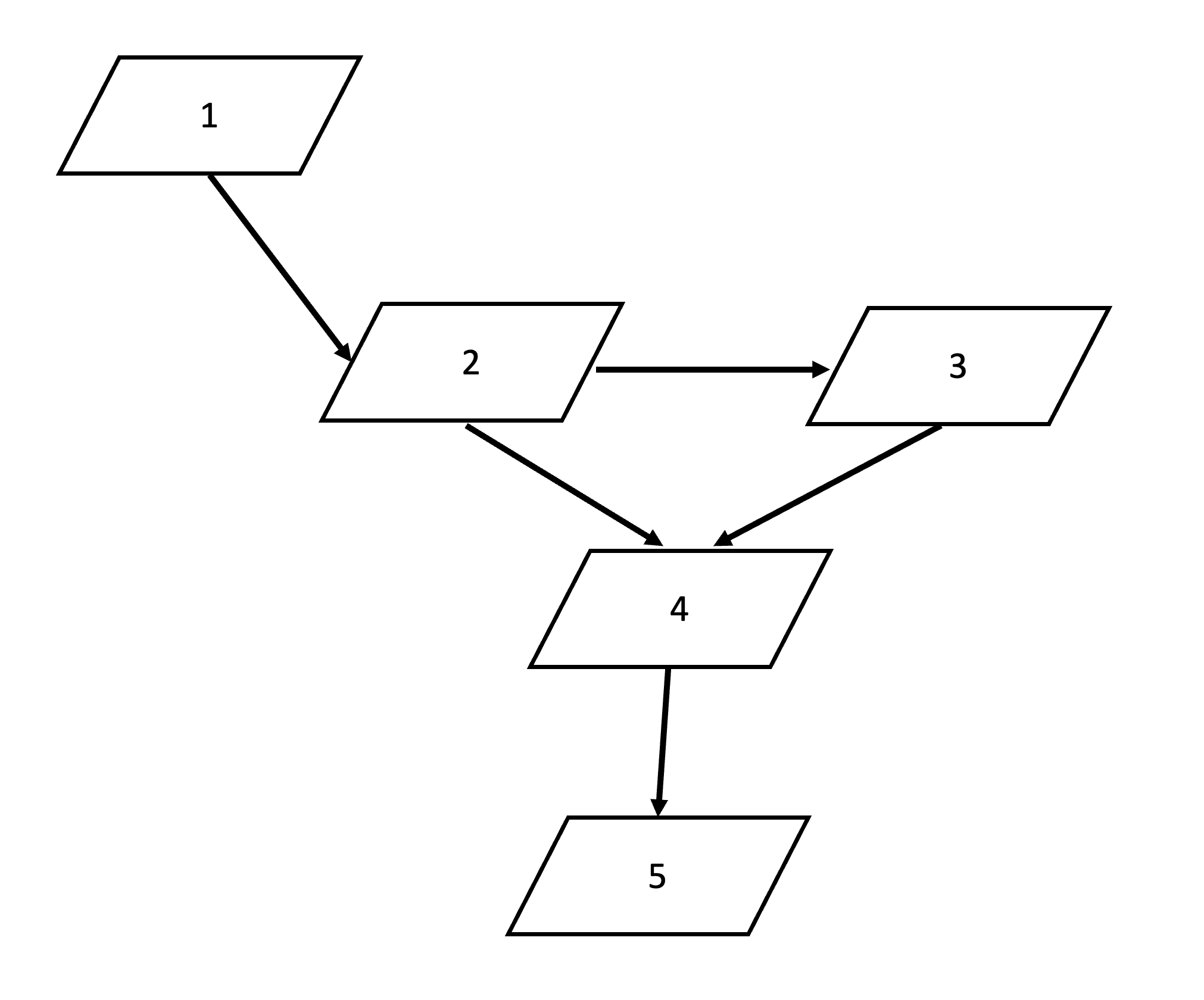}
    \caption{A randomly organised acyclic graph for five fields}
    \label{fig:sub1}
  \end{subfigure}
  \hfill
  \begin{subfigure}[b]{0.34\textwidth}
    \centering
    \includegraphics[width=\textwidth]{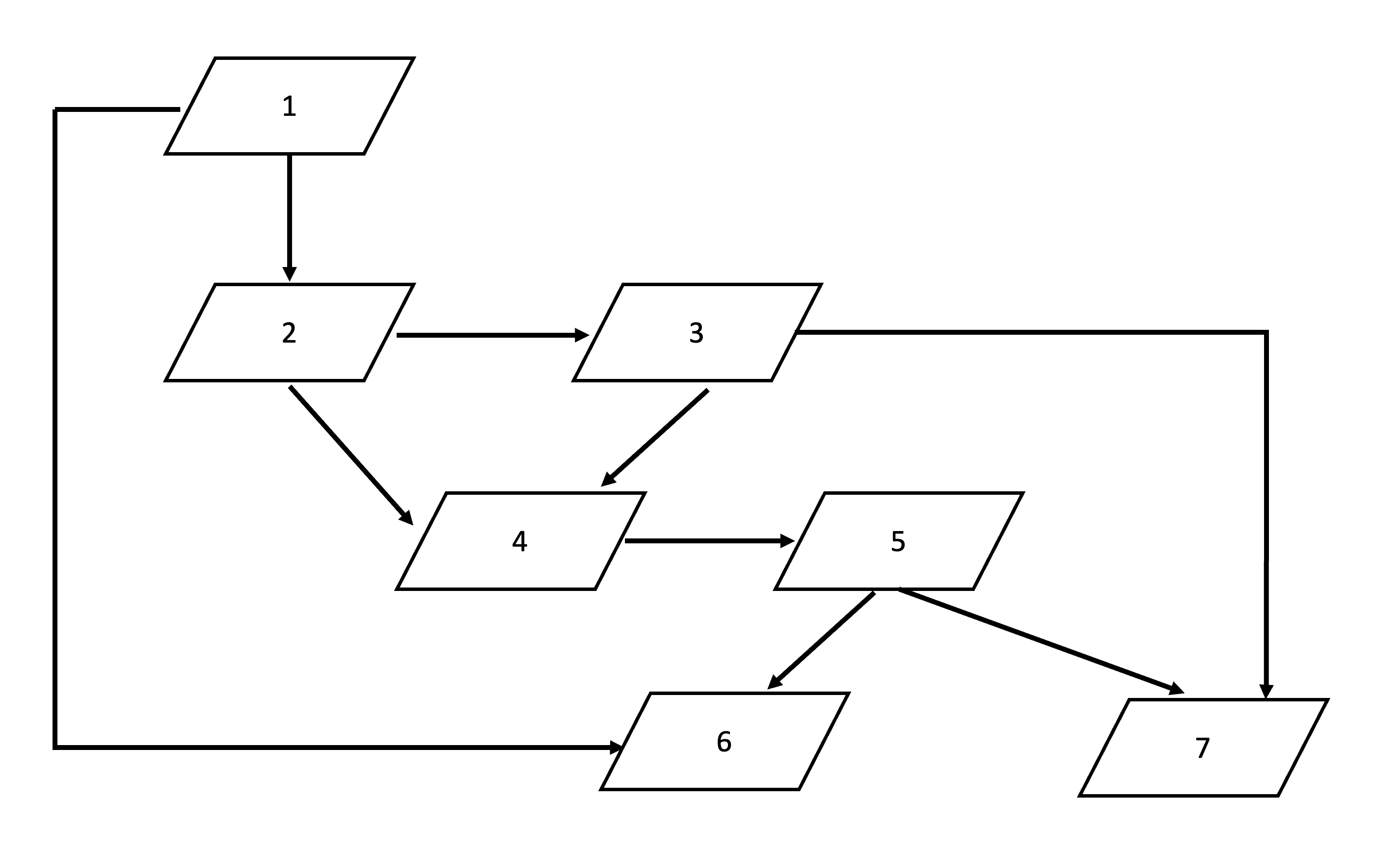}
    \caption{A randomly organised acyclic graph for seven fields}
    \label{fig:sub2}
  \end{subfigure}
  \caption{Randomly drawn acyclic graphs for tests}
  \label{fig:2_tst_graph}
\end{figure}

The test settings are three versions of the modified Tri-Wave functions and one version of the Wendland function (k = 3/2) for $\bmB_{rt}$ on three combinations of grid size (ds) and domain ($\mcalD$), i.e., $\{ds = 0.1, \mcalD = [-1, 1] \}$, 
$\{ds = 0.05, \mcalD = [-1, 1] \}$, $\{ds = 0.1, \mcalD = [-10, 10] \}$ under two graph structures corresponds to two fields (p = 5, p = 7), see Figure \ref{fig:2_tst_graph}. 

The modified Tri-Wave function is 
\begin{align*}
    Tri-Wave(h) = \left \{ \begin{array}{rcl}
A \{1 - \phi (\frac{/h - \Delta/}{/\Delta/})^2 \} & \mbox{for} & /h - \Delta/ \leq \rho /\Delta/ \\ 
0 & \mbox{for} & /h - \Delta/ >  \rho /\Delta/
\end{array}  
\right.
\end{align*}
and the three versions are V4 ($\phi = 1/2, \rho = 2$), V5 ($\phi = 2, \rho = 1$), and V7 ($\phi = 2, \rho = 2$).

The Wendland function used in tests adopts $ k = 3/2$ and $R = 0.5$, i.e.,
\begin{align}
    Wendland_k(r) = \left \{ \begin{array}{rcl}
A \{1 - (\frac{/r - \Delta/}{R})^{4}(1 + 4 \frac{/r - \Delta/}{R}) \} 
& \mbox{for} & /h - \Delta/ \leq R \\ 
0 & \mbox{for} & /h - \Delta/ >  R
\end{array}  
\right.
\end{align}

The PD is tested using 100 parameter combinations, where $A$ ranges from 0.1 $\sim$ 1, and $\Delta$ ranges from 0.1 $\sim$ 1, both by 0.1.

The test results are presented below in Table \ref{Tab:Exp_triwave} and \ref{Tab:Exp_wendland}.

\renewcommand{\arraystretch}{0.8}
\begin{table}[htbp]
\centering
\caption{Test Report on the Positive Definiteness of $\Sigma$ and $\Sigma^{-1}$ using $\bmB_{rt}^{Original}$ and $\bmB_{rt}^{SpN + Reg}$ with Different Versions of Modified Tri-Wave}
\label{Tab:Exp_triwave}
\begin{tabular}{llllll}
 \toprule
   \begin{tabular}[c]{@{}l@{}} Version \\ $(\phi$, $\rho)$ \end{tabular} &  Grid size &  Domain $\mcalD$   & \begin{tabular}[c]{@{}l@{}} Field No. \\ p  \end{tabular} & $\bmB_{rt}^{Original}$  & $\bmB_{rt}^{SpN + Reg}$ \\ \midrule
      \begin{tabular}[c]{@{}l@{}}  V4 \\ (1/2, 2) \end{tabular}  & 
      ds = 0.1 & $[-1, 1]$   &  7 & \begin{tabular}[c]{@{}l@{}} few $\HSigma^{-1}$ PD \end{tabular} & \begin{tabular}[c]{@{}l@{}} all $\HSigma^{-1}$, $\HSigma$ PD \\ no perturb required  \end{tabular} \\ \\ 
        \begin{tabular}[c]{@{}l@{}}  V4 \\ (1/2, 2) \end{tabular} & ds = 0.05 & $[-1, 1]$  &  7 & \begin{tabular}[c]{@{}l@{}} 6 $\HSigma$ not PD; \\ none of $\HSigma^{-1}$ PD  \end{tabular}  & \begin{tabular}[c]{@{}l@{}} all $\HSigma^{-1}$, $\HSigma$ PD \\ no perturb required \end{tabular}  \\ \\
    \begin{tabular}[c]{@{}l@{}}  V4 \\ (1/2, 2) \end{tabular} & ds = 0.1 & $[-10, 10]$ &  5 &  none of $\HSigma^{-1}$ PD   & 
    \begin{tabular}[c]{@{}l@{}} all $\HSigma^{-1}$, $\HSigma$ PD \\ no perturb required  \end{tabular} \\ \\ 
    \begin{tabular}[c]{@{}l@{}}  V4 \\ (1/2, 2) \end{tabular} & ds = 0.1 & $[-10, 10]$ &  7 & \begin{tabular}[c]{@{}l@{}}  none of $\HSigma^{-1}$ PD; \\ few $\HSigma$ PD; \\ only field 1-3 PD; \\ field 4$\sim$7 all non-PD   \end{tabular}  & \begin{tabular}[c]{@{}l@{}}  20 $\HSigma^{-1}$ not PD; \\ the rest require perturb \\ 0.001 $\sim$ 0.2; \\ $\HSigma$, $\HSigma^{-1}$ PD in field 1$\sim$6  \end{tabular} \\ \\
    \begin{tabular}[c]{@{}l@{}}  V5 \\ (2, 1) \end{tabular}  &  ds = 0.1 &  $[-1, 1]$ &  7 & \begin{tabular}[c]{@{}l@{}} none of $\HSigma^{-1}$ PD \end{tabular} & \begin{tabular}[c]{@{}l@{}} all $\HSigma^{-1}$, $\HSigma$ PD \\ no perturb required  \end{tabular}  \\  \\
    \begin{tabular}[c]{@{}l@{}}  V5 \\ (2, 1) \end{tabular} & ds = 0.05 & $[-1, 1]$  &  7 & \begin{tabular}[c]{@{}l@{}} none of $\HSigma^{-1}$ PD; \end{tabular}  & \begin{tabular}[c]{@{}l@{}} 3 $\HSigma^{-1}$ not PD \\ all $\HSigma$ PD; \\ $\HSigma$, $\HSigma^{-1}$ PD in field 1$\sim$6 \end{tabular} \\ \\
    \begin{tabular}[c]{@{}l@{}}  V5 \\ (2, 1) \end{tabular}  &  ds = 0.1 & $[-10, 10]$   &  5 &  12 $\HSigma^{-1}$ not PD  & \begin{tabular}[c]{@{}l@{}} all $\HSigma^{-1}$, $\HSigma$ PD \\ no perturb required \end{tabular} \\ \\
    \begin{tabular}[c]{@{}l@{}}  V5 \\ (2, 1) \end{tabular} & ds = 0.1 & $[-10, 10]$  &  7 & \begin{tabular}[c]{@{}l@{}} none of $\HSigma^{-1}$ PD; \\ only field 1$\sim$4 PD; \\ field 5$\sim$7 non-PD  \end{tabular} & \begin{tabular}[c]{@{}l@{}} 20 $\HSigma^{-1}$ not PD;\\ all $\HSigma$ PD;  \\$\HSigma$, $\HSigma^{-1}$ PD in field 1$\sim$6 \end{tabular} \\ \\
   \begin{tabular}[c]{@{}l@{}}  V7 \\ (2, 2) \end{tabular}  &  ds = 0.1 & $[-1, 1]$ &  7 & \begin{tabular}[c]{@{}l@{}} none of $\HSigma^{-1}$ PD  \\ none of $\HSigma$ PD  \end{tabular} & \begin{tabular}[c]{@{}l@{}} all $\HSigma^{-1}$, $\HSigma$ PD  \\ no perturb required \end{tabular} \\ \\ 
   \begin{tabular}[c]{@{}l@{}}  V7 \\ (2, 2) \end{tabular} &  ds = 0.05 & $[-1, 1]$  &  7 & \begin{tabular}[c]{@{}l@{}} none of $\HSigma^{-1}$ PD  \\ none of $\HSigma$ PD   \end{tabular}  &  \begin{tabular}[c]{@{}l@{}} all $\HSigma^{-1}$, $\HSigma$ PD \\ no perturb required \end{tabular} \\ \\ 
   \begin{tabular}[c]{@{}l@{}}  V7 \\ (2, 2) \end{tabular} & ds = 0.1 & $[-10, 10]$  &  5 & \begin{tabular}[c]{@{}l@{}} none of $\HSigma^{-1}$ PD  \\ none of $\HSigma$ PD   \end{tabular}  &  \begin{tabular}[c]{@{}l@{}} all $\HSigma^{-1}$, $\HSigma$ PD \\ no perturb required \end{tabular} \\ \\  
   \begin{tabular}[c]{@{}l@{}}  V7 \\ (2, 2) \end{tabular} & ds = 0.1 & $[-10, 10]$  &  7 & \begin{tabular}[c]{@{}l@{}} none of $\HSigma^{-1}$ PD  \\ none of $\HSigma$ PD   \end{tabular}  & \begin{tabular}[c]{@{}l@{}} 5 $\HSigma^{-1}$ not PD; \\ the rest perturb 0.1 $\sim$ 0.4; \\ $\HSigma$, $\HSigma^{-1}$ PD in field 1$\sim$6 \end{tabular} \\ \\       
\bottomrule
\end{tabular}
\end{table}
\renewcommand{\arraystretch}{1}

\renewcommand{\arraystretch}{0.5}
\begin{table}[htbp]
\centering
\caption{Test Report on the Positive Definiteness of $\Sigma$ and $\Sigma^{-1}$ using $\bmB_{rt}^{Original}$ and $\bmB_{rt}^{SpN + Reg}$ with Wendland$_{32}$ function}
\label{Tab:Exp_wendland}
\begin{tabular}{cccccc}
 \bottomrule
   Function &  Grid size &  Domain $\mcalD$   & Field No. p  & $\bmB_{rt}^{Original}$  & $\bmB_{rt}^{SpN}$ \\ \midrule
      WL32  & 
      ds = 0.1 & $[-1, 1]$   &  7 & all $\HSigma^{-1}$, $\HSigma$ PD  & \begin{tabular}[c]{@{}l@{}} all $\HSigma^{-1}$, $\HSigma$ PD \\ no perturb required  \end{tabular} \\ \\ 
        WL32 & ds = 0.05 & $[-1, 1]$  &  7 & \begin{tabular}[c]{@{}l@{}} 6 $\HSigma^{-1}$ not PD; \\ none of $\HSigma^{-1}$ PD \end{tabular} & \begin{tabular}[c]{@{}l@{}} all $\HSigma$, $\HSigma^{-1}$ PD \\ perturb 1e-4 $\sim$ 0.001  \end{tabular}  \\ \\
    WL32 & ds = 0.1 & $[-10, 10]$ &  5 & all $\HSigma^{-1}$, $\HSigma$ PD  & 
    \begin{tabular}[c]{@{}l@{}} all $\HSigma^{-1}$, $\HSigma$ PD \\ no perturb required  \end{tabular} \\ \\ 
   WL32 & ds = 0.1 & $[-10, 10]$ &  7 & \begin{tabular}[c]{@{}l@{}}  9 $\HSigma^{-1}$ not PD \\ the rest perturb \\ 1e-5 $\sim$ 0.3  \end{tabular}  & \begin{tabular}[c]{@{}l@{}}  all $\HSigma$, $\HSigma^{-1}$ PD; \\ perturb 1e-7 $\sim$ 1e-4  \end{tabular} \\ \\
    \bottomrule
\end{tabular}
\end{table}
\renewcommand{\arraystretch}{1}

Several conclusions are obtained below based on the results in both tables.
\begin{enumerate}
    \item In general, $\bmB_{rt}^{SpN + Reg}$ provides robust PD for $\HSigma$ and $\HSigma^{-1}$ generation; 
    \item There exist functions, such as Tri-Wave V7, where if $\bmB_{rt}^{Original}$ is used, none of the $\HSigma$ and $\HSigma^{-1}$ is PD under any of the grid size and domain combinations, and any graph structures for any parameter combinations; 
    \item For two versions of the Tri-Wave function, V4 and V5, there are scenarios, e.g., ds = 0.1, $\mcalD$ = [-10, 10], p = 7, even using $\bmB_{rt}^{SpN + Reg}$ whose condition number is less than 3, yet still encounter 20 out of 100 non-PD $\HSigma^{-1}$. This could be due to numerical issues when multiple \textit{for} loops for different parameter combinations
    run together. They all become PD when checking each of these non-PD parameter combinations individually.
    
    
\end{enumerate}

See GitHub R scripts 027-031.
\clearpage

\section{Proof of Extension of the Hammersley-Clifford Theorem for Multivariate Spatial Stochastic Processes (Observation 1)}
\label{app:Proof_HC_col}
As mentioned briefly by \citet[Section~3]{besag1974spatial},
replace the univariate site with $p$ notational sites, each of which corresponds to a single component of p components. $l \in \{1, \ldots, p\}, k = \{1, \ldots, p \} \backslash \{ l\}$.
$l \cup k = \{ 1, \ldots, p \}$,  $i, j = 1, \ldots, n$, $i \neq j$. 
Partition the collection of $p$ random vectors 
$(\bmY_1(\cdot), \ldots, \bmY_p(\cdot))$ into $(\bmY_k(\cdot), \bmY_l(\cdot))$, where 
$\bmY_l(\cdot) \in \mbbR^n$ indicates one random vector corresponding to index $l$, and $\bmY_k(\cdot)$ indicates a collection of random vectors corresponding to the remaining collection of indices except $l$.

\begin{dmath*}
   pr(\bmY) = pr(\bmY_k(\cdot), \bmY_l(\cdot))  
    = \frac{pr(Y_l(s_n) \mid Y_l(s_1), \cdots, Y_l(s_{n-1}), \bmY_k(\cdot))}{pr(X_l(s_n) \mid Y_l(s_1), \cdots, Y_l(s_{n-1}), \bmY_k(\cdot))} \times pr( Y_l(s_1), \cdots, Y_l(s_{n-1}), \bmY_k(\cdot), X_l(s_n)) \\ 
    = \frac{pr(Y_l(s_n) \mid Y_l(s_1), \cdots, Y_l(s_{n-1}), \bmY_k(\cdot))}{pr(X_l(s_n) \mid Y_l(s_1), \cdots, Y_l(s_{n-1}), \bmY_k(\cdot))} \times \frac{pr(Y_l(s_{n-1}) \mid Y_l(s_1), \cdots, Y_l(s_{n-2}), \bmY_k(\cdot), X_l(s_n))}{pr(X_l(s_{n-1})\mid Y_l(s_1), \cdots, Y_l(s_{n-2}), \bmY_k(\cdot), X_l(s_n))} \times pr(Y_l(s_1), \cdots, Y_l(s_{n-2}), \bmY_k(\cdot), X_l(s_{n-1}), X_l(s_{n}))) \\ 
    = \frac{pr(Y_l(s_n) \mid Y_l(s_1), \cdots, Y_l(s_{n-1}), \bmY_k(\cdot))}{pr(X_l(s_n) \mid Y_l(s_1), \cdots, Y_l(s_{n-1}), \bmY_k(\cdot))} \times \frac{pr(Y_l(s_{n-1}) \mid Y_l(s_1), \cdots, Y_l(s_{n-2}), \bmY_k(\cdot), X_l(s_n))}{pr(X_l(s_{n-1})\mid Y_l(s_1), \cdots, Y_l(s_{n-2}), \bmY_k(\cdot), X_l(s_n))} \\ 
    \times \frac{pr(Y_l(s_{n-2}) \mid Y_l(s_1), \cdots, Y_l(s_{n-3}), \bmY_k(\cdot), X_l(s_{n-1}), X_l(s_n))}{pr(X_l(s_{n-2}) \mid Y_l(s_1), \cdots, Y_l(s_{n-3}), \bmY_k(\cdot), X_l(s_{n-1}), X_l(s_n))} \\
    \times \cdots \times \frac{pr(Y_l(s_1) \mid \bmY_k(\cdot), X_l(s_2), X_l(s_3), \cdots, X_l(s_n))}{pr(X_l(s_1) \mid \bmY_k(\cdot), X_l(s_2), X_l(s_3), \cdots, X_l(s_n))} \times pr(\bmY_k(\cdot), X_l(s_1), X_l(s_2), \cdots, X_l(s_n)). 
\end{dmath*}
Then divide both sides by $pr(\bmY_k(\cdot), X_l(s_1), X_l(s_2), \cdots, X_l(s_n))$.







\section{Verification of the Proportionality of Probability Ratio to a Product of Terms }
\label{app:verification_prop_product}
Following Assumption 1 in \citet[Sec.~4]{besag1974spatial} that involves only up to pairwise interaction terms. Then the equation $Q(\bmY) - Q(\bmY_{-(li)})$ is expanded as 
\begin{small}
\begin{dmath*}
\label{Q-Qil_pairwise}
    Q(\bmY) - Q(\bmY_{-(li)}) =  Y_l(s_i)G_i^l(Y_l(s_i)) + 
     \sum_{1 \leq j \leq n} Y_l(s_i) Y_l(s_j) G_{ij}^{ll} (Y_l(s_i), Y_l(s_j))  
    + \sum_{1 \leq k^c \leq p}  Y_l(s_i) Y_{k^c}(s_i) G_{ii}^{l k^{c}}(Y_l(s_i), Y_{k^c}(s_i))  
     +  \sum_{1 \leq k^c \leq p} \sum_{1 \leq  j \leq n} Y_l(s_i) Y_{k^c}(s_j)G_{i j}^{l {k}^c} (Y_l(s_i), Y_{k^c}(s_j)) 
\end{dmath*}    
\end{small}

Substitute the G-functions using the equation \eqref{eq:sinG}, \eqref{eq:autoG_samevar}, \eqref{eq:autoG_sameloc}, \eqref{eq:crossG} and obtain 

\begin{small}
\begin{dmath*}
    Q(\bmY) - Q(\bmY_{-(li)}) =  Y_l(s_i)G_i^l(Y_l(s_i)) + 
     \sum_{1 \leq j \leq n} Y_l(s_i) Y_l(s_j) G_{ij}^{ll} (Y_l(s_i), Y_l(s_j))  
    + \sum_{1 \leq k^c \leq p}  Y_l(s_i) Y_{k^c}(s_i) G_{ii}^{l k^{c}}(Y_l(s_i), Y_{k^c}(s_i))  
     +  \sum_{1 \leq k^c \leq p} \sum_{1 \leq  j \leq n} Y_l(s_i) Y_{k^c}(s_j)G_{i j}^{l {k}^c} (Y_l(s_i), Y_{k^c}(s_j)) \\
    = log \frac{pr(Y_l(s_i) \mid \{ 0_r(s_h): r \neq l, h \neq i \})}{pr(0_l(s_i) \mid \{ 0_r(s_h): r \neq l, h \neq i \})}  
    + \sum_{1 \leq j \leq n} log \frac{pr(Y_l(s_i) \mid \{Y_l(s_j): j \in \mcalN(i) \}, \{ 0_r(s_h): r \neq l, h \neq i, j  \})}{ pr(0_l(s_i) \mid \{Y_l(s_j): j \in \mcalN(i) \}, \{0_r(s_h) : r \neq l, h \neq i, j \})} - \sum_{1 \leq j \leq n} log \frac{pr(Y_l(s_i) \mid \{0_r(s_h):  r \neq l, h \neq i \})}{pr(0_l(s_i) \mid \{0_r(s_h):  r \neq l , h \neq i\})}
    + \sum_{1 \leq k^c \leq p} log \frac{pr(Y_l(s_i) \mid \{Y_{k^c}(s_i): k^c \text{ --- } l , k^c \in \{1, \ldots, p\} \backslash \{ l\} \}, \{ 0_r(s_h): r \neq k^c, l, h \neq i  \})}{ pr(0_l(s_i) \mid \{Y_{k^c}(s_i): k^c \text{ --- } l , k^c \in \{1, \ldots, p\} \backslash \{ l\} \}, \{ 0_r(s_h): r \neq k^c, l, h \neq i  \})} - \sum_{1 \leq k^c \leq p} log \frac{pr(Y_l(s_i) \mid \{0_r(s_h):  r \neq l, h \neq i \})}{pr(0_l(s_i) \mid \{0_r(s_h):  r \neq l , h \neq i\})}
    + \sum_{1 \leq k^c \leq p} \sum_{1 \leq  j \leq n} log \frac{pr(Y_l(s_i) \mid \{Y_{k^c}(s_j): k^c \text{ --- } l , k^c \in \{1, \ldots, p\} \backslash \{ l\}, j \in \mcalN(i) \}, \{ 0_r(s_h): r \neq k^c, l, h \neq i,j  \})}{ pr(0_l(s_i) \mid \{Y_{k^c}(s_j): k^c \text{ --- } l , k^c \in \{1, \ldots, p\} \backslash \{ l\}, j \in \mcalN(i) \}, \{ 0_r(s_h): r \neq k^c, l, h \neq i,j  \})} - \sum_{1 \leq k^c \leq p} \sum_{1 \leq  j \leq n} log \frac{pr(Y_l(s_i) \mid \{0_r(s_h):  r \neq l, h \neq i \})}{pr(0_l(s_i) \mid \{0_r(s_h):  r \neq l , h \neq i\})}\\
    = log \left[ \frac{pr(Y_l(s_i) \mid \{ 0_r(s_h): r \neq l, h \neq i \})}{pr(0_l(s_i) \mid \{ 0_r(s_h): r \neq l, h \neq i \})} \right] ^{(1-n-p-np)} + \\
    log \prod_{1 \leq j \leq n} \frac{pr(Y_l(s_i) \mid \{Y_l(s_j): j \in \mcalN(i) \}, \{ 0_r(s_h): r \neq l, h \neq i, j  \})}{ pr(0_l(s_i) \mid \{Y_l(s_j): j \in \mcalN(i) \}, \{0_r(s_h) : r \neq l, h \neq i, j \})}  \\
    + log \prod_{1 \leq k^c \leq p} \frac{pr(Y_l(s_i) \mid \{Y_{k^c}(s_i): k^c \text{ --- } l , k^c \in \{1, \ldots, p\} \backslash \{ l\} \}, \{ 0_r(s_h): r \neq k^c, l, h \neq i  \})}{ pr(0_l(s_i) \mid \{Y_{k^c}(s_i): k^c \text{ --- } l , k^c \in \{1, \ldots, p\} \backslash \{ l\} \}, \{ 0_r(s_h): r \neq k^c, l, h \neq i  \})} \\
    + log \prod_{1 \leq k^c \leq p} \prod_{1 \leq j \leq n} \frac{pr(Y_l(s_i) \mid \{Y_{k^c}(s_j): k^c \text{ --- } l , k^c \in \{1, \ldots, p\} \backslash \{ l\}, j \in \mcalN(i) \}, \{ 0_r(s_h): r \neq k^c, l, h \neq i,j  \})}{ pr(0_l(s_i) \mid \{Y_{k^c}(s_j): k^c \text{ --- } l , k^c \in \{1, \ldots, p\} \backslash \{ l\}, j \in \mcalN(i) \}, \{ 0_r(s_h): r \neq k^c, l, h \neq i,j  \})} 
\end{dmath*}    
\end{small}

Therefore, 
\begin{small}
\begin{align*}
    \frac{pr(\bmY)}{pr(\bmY_{-(li)})} &= exp \{ Q(\bmY) - Q(\bmY_{-(li)}) \} \\
    &\varpropto \left[ [Y_l(s_i)] \right]^{(1-n-p-np)}
    \times \prod_{1 \leq j \leq n} [Y_l(s_i) \mid \{Y_l(s_j): j \in \mcalN(i) \}] \\
    &\times \prod_{1 \leq k^c \leq p} [Y_l(s_i) \mid \{Y_{k^c}(s_i): k^c \text{ --- } l , k^c \in \{1, \ldots, p\} \backslash \{ l\} \}] \\
    &\times \prod_{1 \leq k^c \leq p} \prod_{1 \leq j \leq n} [Y_l(s_i) \mid \{ Y_{k^c}(s_j):  k^c \text{ --- } l , k^c \in \{1, \ldots, p\} \backslash \{ l\}, j \in \mcalN(i) \}],  
\end{align*} 
\end{small}
where ``$[\cdot]$'' represents some distribution.

Therefore, the probability ratio $\frac{pr(\bmY)}{pr(\bmY_{-(li)})}$ is proportionate to the product of the distribution of the random quantity $Y_l(s_i)$ (raised to a power) along with its auto-conditional distributions (same-component, same-location) and cross-conditional distributions.

\section{Proof of Lemma \ref{thrm:alt_H-C}}
\label{app:proof_lemma1}
\begin{proof}

By the definition of $Q(\bmY)$ expansion in equation \eqref{Q-G}, different G-functions can be written in terms of $Q(\cdot)$. Specifically, 
singleton G-function $Y_l(s_i) G_i^l(Y_l(s_i))$ can be written as
\begin{small}
\begin{align}
\label{eq:sinG}
Y_l(s_i) G_i^l (Y_l(s_i)) &= Q(0, \cdots, 0, Y_l(s_i), 0, \cdots, 0, \underline{0}, 0, \cdots, 0, \underline{0}, 0, \cdots, 0, \underline{0}, 0, \cdots, 0) \nonumber \\
    &= log \frac{pr(Y_l(s_i) \mid \{ 0_r(s_h): r \neq l, h \neq i \})}{pr(0_l(s_i) \mid \{ 0_r(s_h): r \neq l, h \neq i \})}  \quad \mbox{(by definition of $Q(\bmY)$ in equation \eqref{def:Q}) };
\end{align}   
\end{small}

same-component auto G-function $Y_l(s_i)Y_l(s_j) G_{ij}^{ll}(Y_l(s_i), Y_l(s_j))$ can be written as
\begin{small}
\begin{dmath}
\label{eq:autoG_samevar}
    Y_l(s_i) Y_l(s_j) G_{ij}^{l l} (Y_l(s_i), Y_l(s_j))
= Q(0, \cdots, 0, Y_l(s_i), 0, \cdots, 0, Y_l(s_j), 0, \cdots, 0, \underline{0}, 0, \cdots, 0, \underline{0}, 0, \cdots, 0) \nonumber \\
- Q(0, \cdots, 0, \underline{0}, 0, \cdots, 0, Y_l(s_j), 0, \cdots, 0, \underline{0}, 0, \cdots, 0, \underline{0}, 0, \cdots, 0) \nonumber \\
- Q(0, \cdots, 0, Y_l(s_i), 0, \cdots, 0, \underline{0}, 0, \cdots, 0, \underline{0}, 0, \cdots, 0, \underline{0}, 0, \cdots, 0) \\
= log \frac{pr(Y_l(s_i) \mid \{Y_l(s_j): j \in \mcalN(i) \}, \{ 0_r(s_h): r \neq l, h \neq i, j  \})}
{ pr(0_l(s_i) \mid \{Y_l(s_j): j \in \mcalN(i) \}, \{0_r(s_h) : r \neq l, h \neq i, j \})}
- log \frac{pr(Y_l(s_i) \mid \{0_r(s_h):  r \neq l, h \neq i \})}{pr(0_l(s_i) \mid \{0_r(s_h):  r \neq l , h \neq i\})};
\end{dmath}   
\end{small}

same-location auto G-function $Y_l(s_i)Y_{k^c}(s_i) G_{ii}^{lk^c}(Y_l(s_i), Y_{k^c}(s_i))$ can be written as
\begin{small}
\begin{dmath}
\label{eq:autoG_sameloc}
    Y_l(s_i)Y_{k^c}(s_i) G_{ii}^{lk^c}(Y_l(s_i), Y_{k^c}(s_i))
= Q(0, \cdots, 0, Y_l(s_i), 0, \cdots, 0, \underline{0}, 0, \cdots, 0, Y_{k^c}(s_i), 0, \cdots, 0, \underline{0}, 0, \cdots, 0) \nonumber \\
- Q(0, \cdots, 0, \underline{0}, 0, \cdots, 0, \underline{0}, 0, \cdots, 0, Y_{k^c}(s_i), 0, \cdots, 0, \underline{0}, 0, \cdots, 0) \nonumber \\
- Q(0, \cdots, 0, Y_l(s_i), 0, \cdots, 0, \underline{0}, 0, \cdots, 0, \underline{0}, 0, \cdots, 0, \underline{0}, 0, \cdots, 0) \\
= log \frac{pr(Y_l(s_i) \mid \{Y_{k^c}(s_i): k^c \text{ --- } l , k^c \in \{1, \ldots, p\} \backslash \{ l\} \}, \{ 0_r(s_h): r \neq k^c, l, h \neq i  \})}
{ pr(0_l(s_i) \mid \{Y_{k^c}(s_i): k^c \text{ --- } l , k^c \in \{1, \ldots, p\} \backslash \{ l\} \}, \{ 0_r(s_h): r \neq k^c, l, h \neq i  \})}
- log \frac{pr(Y_l(s_i) \mid \{0_r(s_h):  r \neq l, h \neq i \})}{pr(0_l(s_i) \mid \{0_r(s_h):  r \neq l , h \neq i\})};
\end{dmath}   
\end{small}

and cross G-function $Y_l(s_i)Y_{k^c}(s_j) G_{ij}^{lk^c}(Y_l(s_i), Y_{k^c}(s_j))$ can be written as
\begin{small}
\begin{dmath}
\label{eq:crossG}
    Y_l(s_i)Y_{k^c}(s_j) G_{ij}^{lk^c}(Y_l(s_i), Y_{k^c}(s_j))
= Q(0, \cdots, 0, Y_l(s_i), 0, \cdots, 0, \underline{0}, 0, \cdots, 0, \underline{0}, 0, \cdots, 0, Y_{k^c}(s_j), 0, \cdots, 0) \nonumber \\
- Q(0, \cdots, 0, \underline{0}, 0, \cdots, 0, \underline{0}, 0, \cdots, 0, \underline{0}, 0, \cdots, 0, Y_{k^c}(s_j), 0, \cdots, 0) \nonumber \\
- Q(0, \cdots, 0, Y_l(s_i), 0, \cdots, 0, \underline{0}, 0, \cdots, 0, \underline{0}, 0, \cdots, 0, \underline{0}, 0, \cdots, 0) \\
= log \frac{pr(Y_l(s_i) \mid \{Y_{k^c}(s_j): k^c \text{ --- } l , k^c \in \{1, \ldots, p\} \backslash \{ l\}, j \in \mcalN(i) \}, \{ 0_r(s_h): r \neq k^c, l, h \neq i,j  \})}
{ pr(0_l(s_i) \mid \{Y_{k^c}(s_j): k^c \text{ --- } l , k^c \in \{1, \ldots, p\} \backslash \{ l\}, j \in \mcalN(i) \}, \{ 0_r(s_h): r \neq k^c, l, h \neq i,j  \})}
- log \frac{pr(Y_l(s_i) \mid \{0_r(s_h):  r \neq l, h \neq i \})}{pr(0_l(s_i) \mid \{0_r(s_h):  r \neq l , h \neq i\})}.
\end{dmath}   
\end{small}

The above equations \eqref{eq:sinG}--\eqref{eq:crossG} indicate that from local conditional distributions reflecting auto-/cross-neighbourhood structures, we could obtain corresponding auto-G and cross-G functions; from these G-functions, we could obtain $Q(\bmY)$ (by equation \eqref{Q-G}); and from the $Q(\bmY)$, we could arrive at the desired joint $pr(\bmY)$ on the global level (by equation \eqref{def:Q}). 

\end{proof}

\section{Verification of Proportionality of $pr(\mathbf{Y})$ to the Product of Terms }
\label{app:verify_p(y)}
\begin{proof}
By equation \eqref{def:Q}, $pr(\bmY) \varpropto exp(Q(\bmY))$. 
By equation \eqref{Q-G}, $Q(\bmY)$ can be expressed in terms of G-functions, and following the Assumption 1 in \citet[Sec.~4.1]{besag1974spatial},
we involve G-functions only up to pairwise interactions, then
\begin{dmath*}
  Q(\bmY) = \sum_{l=1 }^p \sum_{i = 1}^n Y_l(s_i) G_i^l(Y_l(s_i)) + 
\sum_{l= 1}^p \sum_{1 \leq i, j \leq n} Y_l(s_i) Y_l(s_j) G_{ij}^{ll} (Y_l(s_i), Y_l(s_j)) +
\sum_{1 \leq k^c, l \leq p} \sum_{i = 1}^n Y_l(s_i) Y_{k^c}(s_i) G_{ii}^{lk^c} (Y_l(s_i), Y_{k^c}(s_i)) +    
\sum_{1 \leq k^c, l \leq p }\sum_{1 \leq i, j \leq n}  Y_l(s_i) Y_{k^c}(s_j) G_{ij}^{lk^c} (Y_l(s_i), Y_{k^c}(s_j))   
\end{dmath*}

By the relationship between G-functions and conditional distributions in equation \eqref{eq:sinG}, \eqref{eq:autoG_samevar}, \eqref{eq:autoG_sameloc}, \eqref{eq:crossG}, we re-write the above equation as follows
\begin{small}
\begin{dmath}
    Q(\bmY) = (1-n-p-np)log \prod_{l=1 }^p \prod_{i = 1}^n \frac{pr(Y_l(s_i) \mid \{ 0_r(s_h): h \neq i; r \neq l \})}{pr(0_l(s_i) \mid \{ 0_r(s_h): h \neq i; r \neq l \})} \\
    + log \prod_{l=1 }^p \prod_{1 \leq i,j \leq n} \frac{pr(Y_l(s_i) \mid \{Y_l(s_j): j \in \mcalN(i) \}, \{ 0_r(s_h): h \neq i, j; r \neq l \})}{ pr(0_l(s_i) \mid \{Y_l(s_j): j \in \mcalN(i) \}, \{0_r(s_h) : h \neq i, j; r \neq l \})} \\
    + log \prod_{1 \leq k^c, l \leq p} \prod_{i = 1}^n  \frac{p(Y_l(s_i) \mid \{Y_{k^c}(s_i): k^c \text{ --- } l , k^c \in \{1, \ldots, p\} \backslash \{ l\} \}, \{ 0_r(s_h) : h \neq i, j; 
    r \neq l, k^c \}) }{pr(0_l(s_i) \mid \{Y_{k^c}(s_i): k^c \text{ --- } l , k^c \in \{1, \ldots, p\} \backslash \{ l\} \}, \{0_r(s_h) : h \neq i, j; r \neq l, k^c \})} \\
    + log \prod_{1 \leq k^c, l \leq p} \prod_{1 \leq i,j \leq n}
    \frac{p(Y_l(s_i) \mid \{Y_{k^c}(s_j): k^c \text{ --- } l , k^c \in \{1, \ldots, p\} \backslash \{ l\}, j \in \mcalN(i) \}, \{ 0_r(s_h) : h \neq i, j; r \neq l, k^c \}) }{pr(0_l(s_i) \mid \{Y_{k^c}(s_j): k^c \text{ --- } l , k^c \in \{1, \ldots, p\} \backslash \{ l\}, j \in \mcalN(i) \}, \{0_r(s_h) : h \neq i, j; r \neq l, k^c \})}.
\end{dmath} 
\end{small}

Therefore, the $pr(\bmY) \varpropto exp(Q(\bmY))$ can be written as directly proportionate to a product of various auto-conditionals and cross-conditionals.
\begin{small}
\begin{dmath*}
    pr(\bmY) \varpropto exp(Q(\bmY)) \varpropto \\
    \left( \prod_{l=1 }^p \prod_{i =}^n \frac{pr(Y_l(s_i) \mid \{0_r(s_h): h \neq i; r \neq l \})}{pr(0_l(s_i) \mid \{ 0_r(s_h): h \neq i; r \neq l \})} \right)^{(1-n-p-np)} \\ 
    \times \prod_{l = 1}^p \prod_{1 \leq i, j \leq n} \frac{pr(Y_l(s_i) \mid \{Y_l(s_j): j \in \mcalN(i) \}, \{ 0_r(s_h): h \neq i, j; r \neq l \})}{ pr(0_l(s_i) \mid \{Y_l(s_j): j \in \mcalN(i) \}, \{0_r(s_h) : h \neq i, j; r \neq l \})} \\
    \times \prod_{1 \leq k^c, l \leq p} \prod_{i = 1}^n  \frac{pr(Y_l(s_i) \mid \{Y_{k^c}(s_i): 
    k^c \text{ --- } l , k^c \in \{1, \ldots, p\} \backslash \{ l\} \}, \{ 0_r(s_h) : h \neq i, j; r \neq l, k \}) }{pr(0_l(s_i) \mid \{Y_{k^c}(s_i): k^c \text{ --- } l , k^c \in \{1, \ldots, p\} \backslash \{ l\} \}, \{ 0_r(s_h) : h \neq i, j; r \neq l, k \})} \\
    \times \prod_{1 \leq k^c, l \leq p} \prod_{1 \leq i, j \leq n}
    \frac{pr(Y_l(s_i) \mid \{Y_{k^c}(s_j): 
    k^c \text{ --- } l , k^c \in \{1, \ldots, p\} \backslash \{ l\}, j \in \mcalN(i) \}, \{ 0_r(s_h) : h \neq i, j; r \neq l, k \}) }{pr(0_l(s_i) \mid \{Y_{k^c}(s_j): k^c \text{ --- } l , k^c \in \{1, \ldots, p\} \backslash \{ l\}, j \in \mcalN(i)\}, \{ 0_r(s_h) : h \neq i, j; r \neq l, k \})} \\
    \varpropto 
    \left( \prod_{l =1 }^p \prod_{i = 1}^n [Y_l(s_i)] \right)^{(1-n-p-np)} \times \prod_{l=1}^p \prod_{1 \leq i, j \leq n}
    [Y_l(s_i) \mid \{Y_l(s_j): j \in \mcalN(i) \} ] \\
    \times \prod_{1 \leq k^c, l \leq p}  \prod_{i = 1}^n [Y_l(s_i) \mid \{ Y_{k^c}(s_i): k^c \text{ --- } l , k^c \in \{1, \ldots, p\} \backslash \{ l\} \}] \\
    \times \prod_{1 \leq k^c, l \leq p} \prod_{1 \leq i, j \leq n} [Y_l(s_i) \mid \{ Y_{k^c}(s_j): k^c \text{ --- } l , k^c \in \{1, \ldots, p\} \backslash \{ l\}, j \in \mcalN(i) \} ]
\end{dmath*}
\end{small}
Here, ``$[\cdot]$'' represents some distribution. 
\end{proof}

\section{Comparision Table between MRF and Cross-MRF}
\label{app:compare_tab}
Table \ref{Tab:compare_MRF_crossMRF} compares MRF for univariate spatial processes and cross-MRF for multivariate spatial processes.

The content of MRF for univariate spatial processes can be found in \citet[Sec.~3]{besag1974spatial} and \citet[pp.~174-181]{cressie2011statistics}.
\begin{table}[htpb]
\centering
\caption{A Comparison between MRF and Cross-MRF}
\label{Tab:compare_MRF_crossMRF}
\begin{tabular}{lll}
 \toprule
   \multicolumn{1}{c}{\textbf{Facets}} & \qquad \textbf{Univariate ($p = 1$)} & \qquad \textbf{Multivariate ($p \gg 2$)} \\ \hline
   \begin{tabular}[c]{@{}l@{}} Spatial process \end{tabular} & \begin{tabular}[c]{@{}l@{}} $\{Y(s_i): i = 1, \ldots, n \}$  \end{tabular} &  \begin{tabular}[c]{@{}l@{}} $\{(Y_1(s_i), \ldots, Y_p(s_i)): $ $i = 1, \ldots, n \}$ \end{tabular}  \\  \\
    \begin{tabular}[c]{@{}l@{}} Joint \\ distribution \end{tabular} & $pr(Y(s_1), \ldots, Y(s_n))$ & \begin{tabular}[c]{@{}l@{}} 
    $pr(\bmY_1,\ldots, \bmY_l, \ldots, \bmY_p) =$ \\ $pr(\bmY_1, \ldots, \bmY_i, \ldots, \bmY_n) $, \\ where  $\bmY_l\in \mbbR^n$,  $\bmY_i \in \mbbR^p$\\ 
    \end{tabular}  \\ \\
    \begin{tabular}[c]{@{}l@{}} Neighbourhood \\ structure of \\ $Y_l(s_i)$ \end{tabular} & \begin{tabular}[c]{@{}l@{}}  
    $ \{Y(s_j): j \in \mcalN(i)\} $, \\ $i = 1, \ldots, n$  \end{tabular} & \begin{tabular}[c]{@{}l@{}} Same-component auto-neighbourhood \\ $\{Y_l(s_j): j \in \mcalN(i)\} $ 
    \\ Same-location auto-neighbourhood \\ $\{Y_{k^c}(s_i): k^c \text{ --- } l , k^c \in \{1, \ldots, p\} \backslash \{ l\} \}$ \\
    Cross-neighbourhood \\ $ \{Y_{k^c}(s_j): k^c \text{ --- } l , k^c \in \{1, \ldots, p\} \backslash \{ l\}, j \in \mcalN(i) \} $,   \\ $l = 1, \ldots, p$, $i = 1, \ldots, n $ \end{tabular}   \\ \\
    MRF &  \begin{tabular}[c]{@{}l@{}} when \\ $\{[Y(s_i) \mid \{Y(s_j): j \in \mcalN(i) \}]$ \\ $ :i = 1, \ldots, n\}$ \\ can define the $pr(\bmY)$, \\ then $\{Y(s_i): i = 1, \ldots, n \}$ \\ is a MRF, $\bmY \in \mbbR^{n}$ \end{tabular} & \begin{tabular}[c]{@{}l@{}} when conditions in Sec. 5.2 are satisfied, \\ and same-component auto-conditionals \\ $\{[Y_l(s_i)\mid \{Y_l(s_j): j \in \mcalN(i) \}] :$ \\ $l = 1, \ldots, p$, $i = 1, \ldots, n \}$, \\ same-location auto-conditionals \\
     $\{[Y_l(s_i)\mid \{Y_{k^c}(s_i): k^c \text{ --- } l , k^c \in \{1, \ldots, p\} \backslash \{ l\} \}] :$ \\ $l = 1, \ldots, p$, $i = 1, \ldots, n \}$, \\ and cross-conditionals \\
    $\{[Y_l(s_i)\mid \{Y_{k^c}(s_j): k^c \text{ --- } l , k^c \in \{1, \ldots, p\} \backslash \{ l\}$,\\ $ j \in \mcalN(i) \}] :$  $l = 1, \ldots, p$, $i = 1, \ldots, n \}$ \\ can define the joint distribution $pr(\bmY)$, $\bmY \in \mbbR^{np}$, \\ then
    $\{(Y_1(s_i), \ldots, Y_p(s_i)): $ $i = 1,  \ldots, n \}$ \\ is a cross-MRF \end{tabular}  \\ \\
    \begin{tabular}[c]{@{}l@{}} $Q(\bmY)$ expansion \\ for alternative \\ H-C Theorem \\ (up to pairwise \\ interaction \\ terms) \end{tabular} &  \begin{tabular}[c]{@{}l@{}} $Q(\bmY) $ = \\ $\sum_{i = 1}^n Y(s_i)G_i(Y(s_i))$ + \\ 
    $ \sum_{1 \leq i,j \leq n} Y(s_i)Y(s_j)$ \\$ G_{ij}(Y(s_i), Y(s_j))$ \end{tabular} & 
    \begin{tabular}[c]{@{}l@{}} $Q(\bmY) $ = \\ $\sum_{l = 1}^p \sum_{i = 1}^n Y_l(s_i)G_i^l(Y_l(s_i))$ + \\ 
    $ \sum_{l = 1}^p \sum_{1 \leq i,j \leq n} Y_l(s_i) Y_l(s_j) G_{ij}^{ll} (Y_l(s_i), Y_l(s_j))$ + \\ $\sum_{1 \leq l, k^c \leq p} \sum_{i = 1}^n Y_l(s_i) Y_{k^c}(s_i) G_{ii}^{l k^c}(Y_l(s_i), Y_{k^c}(s_i))$ + \\
    $\sum_{1 \leq l, k^c \leq p} \sum_{1 \leq i,j \leq n} Y_l(s_i) Y_{k^c}(s_j)G_{ij}^{lk^c} (Y_l(s_i), Y_{k^c}(s_j)) $  \end{tabular}  \\ \\ 
    $pr(\bmY) \varpropto$ & \begin{tabular}[c]{@{}l@{}} $\left( \prod_{i} [Y(s_i)] \right)^{(1-n)} \times \prod \limits_{1 \leq i, j \le n}$  \\
    $ [Y(s_i) \mid \{Y(s_j): j\in \mcalN(i) \}]$ \end{tabular}  & 
    \begin{tabular}[c]{@{}l@{}} $\left( \prod_{l=1}^p \prod_{i = 1}^n [Y_l(s_i)] \right)^{(1-n-p-np)} \times  $ \\ 
    $\prod_{l=1}^p \prod_{1 \leq i, j \leq n}[Y_l(s_i) \mid \{Y_l(s_j): j\in \mcalN(i) \}] \times  $ \\ $\prod_{1 \leq k^c, l \leq p} \prod_{i = 1}^n [Y_l(s_i) \mid \{ Y_{k^c}(s_i): k^c \text{ --- } l$ , \\ $k^c \in \{1, \ldots, p\} \backslash \{ l\} \}] \times $ \\
    $\prod_{1 \leq k^c, l \leq p} \prod_{1 \leq i, j \leq n} [Y_l(s_i) \mid \{ Y_{k^c}(s_j): k^c \text{ --- } l$ , \\ $k^c \in \{1, \ldots, p\} \backslash \{ l\}, j \in \mcalN(i) \}]$    \end{tabular}  \\ \\ 
    \bottomrule
\end{tabular}
\end{table}

\section{Updated Algorithm Linking Cross-MRF and the Mixed Spatial Graphical Model Framework}
\label{app:updated_algo}
Below Algorithm \ref{alg:my-algorithm_update} is the updated algorithm that links the cross-MRF with the mixed spatial graphical model framework, in which steps 1, 2, 17, and 18 reflect the $s_j \in \mcalN(s_i)$ using uni-variate CAR model.

\begin{spacing}{0.5}
\begin{algorithm}
\caption{Algorithm for the generation of the desired $\Sigma$ and $\Sigma^{-1}$}
\label{alg:my-algorithm_update}
\KwData{The number of components $p$ and the data structure indicating parent and child relationship among $p$ component fields}
\KwResult{The desired $\HSigma_{np \times np}(\cdot, \cdot)$ and $\HSigma^{-1}_{np \times np}(\cdot, \cdot)$}

$\bmSIGMA_{11}^{-1} \leftarrow $ uni-variate CAR  \;
$\bmSIGMA_{11} \leftarrow $ Cholesky inversion($\bmSIGMA_{11}^{-1}$) \;
$\HSigma \leftarrow \bmSIGMA_{11}$; $n \leftarrow nrow(\bmSIGMA_{11})$ \;

\For{$r = 2$ \KwTo $p$}{
    PN = Pa(r) \;
    $\bmR = \bmC = NULL$ \;
    \For{$c = 1$ \KwTo $(r-1)$}{
        $\bmBT \leftarrow NULL$ \;
        $\bmSIGMA_{rc} \leftarrow \bm{0}$\;
        \For{t $\in$ PN} {
            $\bmB_{rt} \leftarrow f(h; \Delta_{rt}; A_{rt})$ \;
            $\bmBT \leftarrow rbind(\bmBT, \bmB_{rt}^T)$\;
            $\bmSIGMA_{rc} \leftarrow \bmSIGMA_{rc} + \bmB_{rt} \bmSIGMA[((t - 1)n + 1) : (tn), ((c-1)n + 1) :(cn)]$ \;   
        } 
        
        $\bmR \leftarrow cbind(\bmR, \bmSIGMA_{rc})$ \;
        $\bmSIGMA_{cr} \leftarrow \bmSIGMA_{rc}^T$ \;
        $\bmC \leftarrow rbind(\bmC, \bmSIGMA_{cr}) $ \;   
    }
    
    $\bmD_{rr}^{-1} \leftarrow $ uni-variate CAR \;
    $\bmD_{rr} \leftarrow$ Cholesky inversion($\bmD_{rr}^{-1}$) \;
    
    $\bmSIGMA_{rr} \leftarrow \bmR[, (t-1)n + 1 : (tn)]  \bmBT + \bmD_{rr}$ \;
    
    $SG \leftarrow \HSigma$  
    
    $Col \leftarrow rbind(\bmC, \bmSIGMA_{rr})$ \;
    $Row \leftarrow rbind(SG, \bmR)$ \;
    $\HSigma \leftarrow  cbind(Row, Col)$ \;  


    \If{ r == 2}{SG$^{-1} \leftarrow \Sigma_{11}^{-1}$}
    
    $BK_1 \leftarrow SG^{-1} + (SG^{-1} \bmC \bmD_{rr}^{-1}) (\bmR SG^{-1})$ \;
    $BK_2 \leftarrow -SG^{-1} (\bmC \bmD_{rr}^{-1})$ \;
    $BK_3 \leftarrow - (\bmD_{rr}^{-1} \bmR) SG^{-1}$ \;
    $BK_4 \leftarrow \bmD_{rr}^{-1}$ \;

    $\HSigma^{-1} \leftarrow rbind(cbind(BK1, BK2), cbind(BK3, BK4))$ \;
    $SG^{-1} \leftarrow \HSigma^{-1}$

    \If{r == p}{return $\HSigma$, $\HSigma^{-1}$ }
}
\end{algorithm}
\end{spacing}

\section{Microbenchmark Generation Time between Two $s_j \in \mcalN(s_i) $ Realisation Strategies: Univariate CAR and Tapering \matern}
\label{app:car_vs_matern}
The graph structure of the component fields is the first six fields in Figure \ref{fig:ten_fields}. 
The 1D simulation setting is grid size $ds = 0.1$, spatial domain $\mcalD = [-10, 10]$. 
Univariate CAR model using lag-3 neighbourhood structure. 
Parameters for the univariate Mat\'{e}rn covariance is $A = 0.1$, $\Delta = 0.5$, $\sigma^2 = 1$, $\kappa = 2$. 
Table \ref{Tab:timecompare_UniCARvsTaper} shows the benchmark of operational time of 100 randomly evaluated $\HSigma_{np \times np}$ and $\HSigma_{np \times np}^{-1}$ using two $s_j \in \mcalN(s_i)$ realisation strategies.
\begin{table}[ht]
\centering
\caption{Comparison of the Operational Time of 100 Randomly Evaluated $\Sigma_Y$ and $\Sigma_Y^{-1}$ Generation using Two $s_j \in \mcalN(s_i)$ Realisation Methods. Unit: Nanoseconds.}
\label{Tab:timecompare_UniCARvsTaper}
\begin{tabular}{llllll}
 \\ \toprule
  & min & lq & mean & up & max  \\ \cmidrule(lr){2-6}
CAR  & 2.778   & 2.897   & 2.898  & 2.912   &  3.043 \\
Taper \matern{} & 3.879 & 3.902 & 3.976 & 4.012 & 4.154
 \\ \bottomrule 
\end{tabular}
\end{table}

See GitHub R script 037.



\end{appendices}

\end{document}